# Anomalous Sodium Insertion in Highly Oriented Graphite: Thermodynamics, Kinetics and Evidence for Two-Sided Intercalation

Chuanhai Gan*[1], Chuanlian Xiao[1], Hongguang Wang, Peter A. van Aken, Rotraut Merkle, Sebastian Bette, Bettina V. Lotsch, Joachim Maier*

Max Planck Institute for Solid State Research, 70569 Stuttgart, Germany

ABSTRACT

The difficult intercalation of sodium (Na) into graphite is studied by systematic and long-time investigations (of up to 2 years) using highly oriented pyrolytic graphite (HOPG). By studying chemical insertion of solid, liquid and gaseous Na at low and high temperatures (LT, HT) as well as using electrochemical insertion at 25 °C into uncoated and coated HOPG, it became clear that insertion equilibrium requires HT. On decreasing chemical intercalation temperature from HT (500 °C) to LT (25 °C), thermodynamic control was found to change to diffusion control and finally to interfacial control. For the electrochemical insertion, coating ($TiO_2$) proved advisable (to avoid co-intercalation) and efficient in reducing the interfacial resistance. Measured saturation values were found to be not higher than ∼ 1 mol %. Towards room temperature higher equilibrium values cannot be excluded but would in view of the very low driving force kinetically be very difficult to reach. The reversible cell voltage of the saturated composition (versus alkali metal) is distinctly lower than for the analogous cells using lithium (Li) or potassium (K). Detailed transmission electron microscopy (TEM) studies reveal the unexpected fact that at HT Na predominantly enters HOPG in the form of two-sided intercalation sandwiching carbon layers (bilayers), while at LT more highly aggregated layers appear to a comparable degree, accompanied with the formation of higher-dimensional crystal imperfections. The reasons for this peculiar feature and the non-monotonic thermodynamics in the sequence Li-Na-K-Rb-Cs are discussed not only from an energetic but also from an entropic point of view.

## INTRODUCTION

Li based batteries have quickly penetrated our daily life owing to their large storage capacities combined with sufficient power densities. These features rely on the availability of high-performance electrode materials. Currently, the material of choice for the negative electrode is graphite. It is inexpensive and allows for rapid and extensive intercalation and deintercalation of Li. The significance of Li-based batteries ranges from powering commodities to enabling electromobility or efficient grid storage. This ubiquitous demand is in conflict with the limited resources[1]. From that point of view, Na-based batteries are advantageous, and their electrochemical properties have long been underestimated[2]. Various positive electrode materials are available that quickly store significant amounts of Na[2]. Nonetheless, the negative electrode remains a severe problem.

[1]These authors contributed equally.

*Corresponding authors, c.gan@fkf.mpg.de; office-maier@fkf.mpg.de

The observation of Na insertion into carbon materials dates back to as early as 1925[3]. Na-intercalated graphite compounds were firstly reported in 1958[4], while K- and Li- intercalated graphite compounds had been described earlier, in 1926[5] and 1955[6], respectively. Owing to the interest in Na-based batteries, there have more recently been numerous investigations of Na insertion in graphite[7-14]. All authors agree on observing very low if not immeasurably small insertion at room temperature. Higher solubilities have only been reported for high pressure and HT chemical insertion[15]. The larger solubility in "hard carbon", the canonical negative electrode material for Na-based batteries, appears to be largely due to stabilizing Na at higher-dimensional defects, going along with an open circuit voltage vs. Na that is close to zero and does not stabilize Na enough to avoid the danger of dendrite formation[16, 17]. The reason why Na intercalation in graphite is not favored has been inspected by various authors[15, 18-23]. In spite of the multitude of investigations, systematic long-time experiments have not been reported, so that the saturation concentrations are unclear. Even the decision between thermodynamic control and kinetic control could not be made, let alone a decision on diffusion or interfacial control. Beyond that, a microscopic picture of Na intercalates has not been given.

The present work distinguishes itself from works presented in the literature in essentially three respects: (i) It presents a systematic long-time study (up to 2 years) of chemical and electrochemical insertion into pure and coated HOPG at various temperatures, allowing us to discriminate between thermodynamic control, diffusion control and reaction control, and moreover to measure or derive equilibrium intercalation potentials. (ii) The combined use of advanced TEM, X-ray diffraction (XRD) and chemical analysis yields the surprising but clear information that the intercalation pattern is dominated by two-sided intercalation at HT rather than monolayers and also higher aggregates at LT. (iii) The thermodynamic considerations explicitly include entropy effects which proved to be key to understand the intercalation pattern.

CHEMICAL AND ELECTROCHEMICAL INSERTION IN HOPG: KINETIC PICTURE

Since in the literature neither the graphite samples used have been well characterized nor waiting times have been sufficiently long, (i) literature results are contradictory, (ii) Na concentrations reported are not based on reliable analyses, (iii) and to reiterate, no conclusions on thermodynamic or kinetic control could be drawn.

Here we report on systematic chemical and electrochemical (Fig. 1(a, b)) insertion into HOPG on time scales of hours, months and years accompanied by careful reference experiments (see Fig. SIII-1, Supplementary Information (SI)). The chemical insertion was done by direct contact to solid or liquid (or even gaseous) Na and characterized by XRD and chemical analysis. XRD investigation gives qualitative information on whether or not perceptible amounts of Na have entered HOPG (appearance of new peaks at lower and higher diffraction angles compared to original (002) peaks of graphite as depicted in Fig. 2(a)) or graphite powder (Fig. SIII-2, SI, also see detailed XRD data analysis in Part IV, SI). Chemical analyses by titration[24] and inductively coupled plasma optical emission spectroscopy (ICP-OES) deliver quantitative information on the average composition, provided the molar Na-content is higher than $5\times10^{-5}$ ($NaC_{20000}$). For chemical insertion, a wide temperature range from 25 to 500 °C was covered (Fig. SIII-3(a), SI), and waiting times of up to 2 years were allowed for (Fig. SIII-3(c), SI). For 25, 90 and 110 °C (below and above the Na melting point of 98 °C), also coatings with $TiO_2$ or amorphous carbon were employed (Fig. SIII-3(b, c, d), SI).

The electrochemical insertion experiments were performed (Fig. 1(b)) using the cell Na / liquid electrolyte (1 M sodium hexafluorophosphate ($NaPF_6$) (or sodium bis(fluorosulfonyl)imide (NaFSI))-EC/DMC) / HOPG, whereby pure and coated HOPG ($TiO_2$ as well as amorphous carbon) were used. Here the temperature was around 25 °C and the waiting times ranged from 1 week to 8 months (Figs. SIII-3(d), -4, SI).

Unsurprisingly, we found in targeted experiments (Fig. 1(c, d)) that insertion only occurs perpendicular to the c-axis (parallel to the carbon layers), so that the edge regions (side-surface) of our samples must be available for insertion as shown in Fig. 2(b) and Fig. SIII-5 (SI). The results justify the neglect of higher-dimensional defects such as dislocations as far as perpendicular long-range transport is concerned.

Let us first discuss the results of chemical insertion. At 400 and 300 °C waiting times of 2 days and 2 weeks, respectively, are enough to reach saturation (for Na activity $a_{\mathrm{Na}} = 1$ with compositions on the order of $NaC_{100}$ (Table SII-1, Fig. SIII-6, SI)). (Polycrystalline graphite powder behaves similarly according to XRD measurement, Fig. SIII-2, SI). In all these cases Na is present in liquid form. Na vapor intercalation into HOPG at 400 °C for 2 days yields significantly lower values, which is expected due to the lower Na activity (low Na vapor pressure, Fig. SIII-7, SI). Fig. 2(a) and Table SII-1 (SI) summarize the insertion results.

Insertion experiments at LT (< 100 °C) suffer from the typical problems of solid-solid contacts. 25 °C synthesized samples show distinctly lower Na contents (~$NaC_{13000}$) for a waiting time of 2 weeks, and not very different values (~$NaC_{14000}$) for a waiting time of 1 year (whilst Li and K can intercalate easily, see Fig. SIII-3(e, f), SI). Comparison with the electrochemical insertion (see below) indicates that it is still very far from saturation and that even 2 year-long waiting times are too short to reach saturation (also indicated by XRD data, Fig. SIII-3(c, d), SI). Chemical insertion at LT does not perceptibly benefit from coating (~$NaC_{13700}$, Table SII-1, Fig. SIV-3(c), SI) by $TiO_2$ or by amorphous carbon.

At 400 °C, $TiO_2$ coated and amorphous carbon coated HOPG cases show similar results to uncoated HOPG (whereas a thicker coating, e.g. 10 nm $TiO_2$, slows down Na intercalation (Fig. SIII-8, SI)). Whilst this is expected as at these temperatures interfacial kinetics is comparatively fast, the negligible influence of coating at LT should be ascribed to the difficult solid-solid contact (Na/solid storage medium).

As far as electrochemical insertion is concerned, we meet different interfacial problems. The difficult solid-solid contact is absent, but the Na-liquid electrolyte contact is chemically unstable resulting in SEI formation (characterized by XPS (Fig. SIII-9), EDX (Figs. SIII-10, 11, 12), SI and Fig.2 (c)). Accordingly, we face - for kinetic reasons - higher storage (at 25 °C) than for LT chemical insertion, but still distinctly smaller storage than observed at HT. For uncoated HOPG, the highest Na content was ~$NaC_{300}$ for electrochemical insertion (for a waiting time of 8 weeks, Table SII-1, SI). Yet, these values are not reliable as evidenced by the observation of severe sample expansion that we attribute to co-intercalation most probably of the solvent (Fig. SIII-13, SI). For these reasons we rely on $TiO_2$-coated samples for the electrochemical intercalation yielding ~$NaC_{300}$ (for a waiting time of 3.5 months), and higher Na content may be realizable at much longer time. At 40 °C, the electrochemically intercalated $TiO_2$-coated samples (4 weeks) show a similar storage as observed at HT when we analyze the outer half of the sample (as discussed below the inner half still shows low Na-content). Coating has various advantages: (i) The parasitic SEI formation is less marked[25] and the proper interfacial kinetics may be faster (EIS, Fig. SIII-14, SI), and (ii) very importantly, solvent co-intercalation is suppressed (Part II-2), SI). Impedance spectroscopy indeed shows a more stable behavior for the coated sample (Fig. SIII-14, SI). No further conclusions about the coating's impact on kinetics could be drawn.

Worthwhile information on the incorporation kinetics comes from cutting the samples and comparing inner (far from the Na source) and outer parts (close to the Na source, see Fig. SI-1(d), SI) using XRD or titration measurements. For the HT (400 and 300 °C) samples, after a waiting time of 2 days, no difference was observed which is in line with the fact that 2 days were sufficient to reach equilibrium (Fig. SIII-6(a, b), SI). For 200 °C, where a waiting time of 2 days was shorter than the equilibration time (Fig. SIII-6(c), SI), a significantly higher Na content in the outer part was found when compared to the inner part indicating diffusion control. At 25 °C, from 2 weeks to 2 years, again no perceptible difference could be observed (Fig. SIII-3(c, d), SI), indicating that now interfacial control has taken over. This transition from interfacial control to diffusion control and finally thermodynamic control on increasing temperature is not unexpected as it signifies positive effective activation energies and higher barrier values for interfacial kinetics than for diffusion kinetics. In agreement with this, coating does not change the HT values, while at LT it can have a significant impact on the kinetics (even at 150 °C, Table SII-1, SI). The comparison between electrochemical intercalation for coated and uncoated samples is very revealing in this context (Fig. SIII-3(d), SI)). While the results for the uncoated samples are the same as for the chemical intercalation and are definitely interfacially controlled (no profile), the values for electrochemically intercalated coated samples show a diffusion profile indicating that the coating could substantially reduce the interfacial resistance.

Cyclic voltammetry (CV) using voltage ramps of various rates performed on electrochemically inserted HOPG confirms the picture. For Na, CV clearly shows that insertion requires diffusion parallel to the planes (carbon layer) (Fig. 2(b)). The slope of logarithms of peak current versus sweep rate is less than 0.5 (Fig. 2(e)) indicating surface control for the Na insertion into uncoated samples (for more details, see Part II-3), SI). Li and K show under the same conditions diffusion control which changes to surface control only at high voltage-sweep rate; notably the interfacial kinetics worsen drastically from Li to Na and then improve slightly towards K (Fig. SIII-15, SI). The CV measurements indicate positive de-intercalation potentials (cf. intercalation peak potentials close to 0 V) for Na being as small as 0.05 V compared to 0.45 V for K, 0.36 V for Li (Fig. SIII-16, SI) as more accurately measured by cell potential discussed below. CV results on $TiO_2$ coated material confirm lower interfacial resistance and hence faster rates which eventually would be limited by diffusion (Fig. 2 (e) and Fig. SIII-17, SI).

More detailed information stems from targeted open circuit voltage (OCV) measurements on $(Na_yC)_{sat.}$/electrolyte/Na (sample $(Na_yC)_{sat.}$ synthesized at 400 °C). The transient in Fig. 2(f) shows approximately exponential decay toward an end value which we attribute to the reversible cell voltage vs. Na (OCV) for this special composition. These end values are small indeed but positive (10-15 mV), much smaller than for Li (≥ 0.20 V) or K (≥ 0.40 V) (see Fig. 2(f), Figs. SIII-18, 19, SI. The OCV values are consistently in between the intercalation and de-intercalation peak potentials of CV, Fig. SIII-16, SI). Such OCV values are not influenced by electrolyte characteristics, such as solvation free energies, as those effects cancel in the cell reaction. Thus, we can conclude that - in agreement with literature - the free energy of accommodating the alkali metal into HOPG (which should not significantly differ for usual graphite), shows a minimum for Na, when the atomic number of the alkaline metals is varied. A similar sequence is seen for the incorporation kinetics: Li can be easily intercalated at 25 °C electrochemically; this also holds for K but at significantly smaller rates (CV, Fig. SIII-16, and Fig. SIII-20, SI). Na intercalation had to be done at higher temperatures to obtain a significant alkaline amount inside. In the case of 25 °C storage, we can check the distance to equilibrium by following the decrease of OCV towards zero.

The OCV values in Fig. 2(f) show after a certain period (which decreases with decreasing sample mass) significant rises reflecting Na-loss by SEI formation. We see a qualitatively similar behavior with Li and K but connected with larger OCV values (Fig. SIII-19, SI). In all cases the voltage increase occurs the later the higher the sample mass. This we can interpret as sluggish alkali metal removal from the graphite by reaction with the electrolyte. The fact that the time until the rise starts depends roughly linearly rather than quadratically on the sample size, is consistent with interfacial kinetic control[26]. The mechanisms of interfacial control can be manifold and range from hindered phase transfer to phase formation (e.g. SEI in the electrochemical experiments)[27, 28] or may be simply governed by point contact problems (solid-solid contacts in the case of chemical experiments). The just mentioned onset of corrosion sets a time limit to our storage experiments.

If overvoltages (in fact negative voltages) are electrochemically imposed on polycrystalline graphite powder (diameter: ~11 µm), we observe already for -0.01 V, and even more so for -0.40 V, an acceleration of the kinetics. For -0.50 V up to -2.00 V no further increase of the values beyond Na values being on the order of the values obtained for chemical intercalation at 400 °C, was observed (XRD patterns are similar to equilibrium situation at 400 °C, Fig. SIII-21, SI).

The small OCV has also an important consequence on the kinetics. As we are thermodynamically close to equilibrium, the flux can be viewed as being proportional to the chemical potential gradient of Na whereby the chemical potential directly reflects the OCV. Thus, a sluggish kinetics is indeed expected. In fact, the qualitative parallelism between OCV and incorporation rate when the alkali metal is varied, may thus find a simple explanation (note that this presupposes that the ion conductivities and transfer coefficients are not too different).

Let us now consider the chemical diffusion coefficients (Fig. 2(d)). For the samples that do not show concentration profiles (Fig. SIII-6(d), SI), we can only give lower limits. The reason for giving lower limits is that at high temperatures the waiting times become too quick to be analyzed given the heating/cooling procedure, while at lower temperatures interfacial control has taken over. For 400 °C and 300 °C samples no concentration profiles remain after 1h, meaning that $D_{\mathrm{Na}}^{\delta}$ would be higher than $10^{-5}$ $cm^2/s$. These values are rather high and may not be exceedingly far away from Li and K diffusivities[29, 30]. In the cases where profiles could be observed, we may roughly assess upper limits. Such estimate is seconded by the fact that these values yield a reasonable activation energy (17 kJ/ mol, similar to values for Li and K cases[19]). At room temperature we thus may estimate a diffusion coefficient of $10^{-9}$ $cm^2/s$. This refers to electrochemical intercalation into coated HOPG. The higher value obtained for uncoated HOPG (see Fig. Fig. 2(d)) can be ascribed to co-intercalation, as mentioned earlier. We must however be aware of the fact that the diffusion scenario itself is complex and cannot be identified with a simple chemical diffusion process. In particular the diffusivity along the layers is highly non-linear. In the next section we can give a clearer account of this issue. Nonetheless we might conclude that for an insertion into HOPG to be battery-relevant, nanocrystalline samples are necessary for which then the necessary depression of the interfacial resistance becomes very severe.

The section after the next one primarily deals with the equilibrium values (summarized in Table SII-1, SI). The thermodynamic conclusions given there are not in contradiction with first principles calculations in the literature that report energetic instability of $NaC_6$ or $NaC_8$[21, 22]. The reasons are the neglect of entropic effects there as well as the fact that we here deal with low Na concentration values and an unexpected insertion pattern revealed by TEM which is going to be described in the following.

ELECTRON MICROSCOPY, XRD ANALYSIS AND CHEMICAL ANALYSIS: MICROSCOPIC PICTURE

For HT-equilibrated samples (cooled to 25 °C), TEM results show, contrary to expectations, predominant two-sided intercalation (Fig. 3 (a - d)), similar to alkali metal adsorption on both sides of a single graphene layer[31, 32]. These bilayers sandwiching single carbon layers are separated by 4 - 10 pure carbon-layers. (Do not confuse these "chemical Na bilayers" with electrostatic double layers, or with layers discussed for superdense sodiated graphite[15] obtained by high pressure synthesis where two Na layers are directly neighbored.) The spacing between the pure carbon layers is ~3.40 Å (Fig. SIII-22, SI), not perceptibly varying with distance from the Na-layer while the carbon-carbon layer spacing in the Na intercalated bilayers is 3.60 - 4.02 Å. Some bilayers show even varying spacings within the same bilayer (Fig. 3(c)) which we explain by varying under-occupancies (Ad 2, next section). In contrast to other alkali metals, we never observed monolayers in the TEM images. At HT we observed prevalent Na bilayers (more than 70 %, see TEM data, Fig. SIII-23, SI), only rare occurrences of trilayers, while at LT the vast majority of bilayers have disappeared and higher aggregates dominate (for additional TEM images on 25 °C chemical and electrochemical samples, see Figs. SIII-24, 25, SI).

XRD analyses and simulations (Part IV, SI) were performed for samples synthesized at HT, 25 °C, as well as electrochemical intercalation samples. According to these, the 400 °C chemically intercalated sample consists predominantly of Na bilayers (Fig. 3 (i), and Fig. SIV-3(a), Part IV, SI) what is consistent with the TEM results discussed above (with the exception that the XRD modelling suggests 20% monolayers, while TEM data do not show any.). For the chemically intercalated LT-samples (25 °C) a rather mixed structure with various kinds of intercalated layers is confirmed, even admixture with non-intercalated phases is possible (00l' peak, Fig. SIV-3(b), SI); for electrochemical intercalation at 25 °C the results are similar, but only if coating is used (Fig. SIV-3(c, d), SI). Otherwise, solvent co-intercalation occurs significantly changing the diffraction pattern.

Comparison of the microstructure with chemical analysis (Table SII-1, Part II-1, SI) suggests at an under-occupation of the Na bilayers. This would also explain the distribution of the unequal carbon-carbon distances of the Na bilayers (Fig. 3(c)). (Should there be remainders of adsorbed Na left or Na stored at higher dimensional defects, then the calculated occupation would be even less.) It is relevant to mention that for the other alkali metals the reduction of staging degree up to the $AC_6$ or $AC_8$ (A= alkali metals) composition is reported to be accompanied with increasing occupation within the layers[24, 33, 34].

As entropy contributions caused by under-occupation of that magnitude are small (Part V, SI) the following hypothetical picture suggests itself: Instead of occupying a layer fully ("fully" refers to the maximum occupancy of the interlayer space in a $AC_6$ or $AC_8$ structure) these Na ions are dispersed over the two layers more or less orderly so that the Na ions are not directly above each other. The energetic benefit of this configuration is explained below. The expected objective that there is no atomic mobility through the graphene layers, is not relevant, as the distribution can be established by in-diffusion along the layers. (On a more specific note, higher-dimensional defects could provide leakage in the perpendicular direction.) The perceptible electronic mobility should be sufficient to allow for extended delocalization also out-of-plane[35].

A clearer picture of the in-diffusion kinetics is provided by TEM at an early stage at 25 °C (Fig. 3(e-h)). For samples chemically sodiated at 25 °C, we are definitely referring to the regime controlled by (interfacial) kinetics. Accordingly, the Na content is small. TEM still reveals - apart from dominant two-sided intercalation (bilayers) - to a significant degree also trilayers, quadruple layers and some even more highly aggregated layers, separated by pure carbon-layers in between. The separation by pure carbon-layers is far beyond space charge repulsion (range of elastic fields is controversially discussed but it is less than 10 carbon layers in non-equilibrium, and very narrow in the ordered state[36, 37]) and indeed no ordering can be detected. Owing to our reasoning, ordering should start when the relevant interaction length scales are reached. How much repulsion - and hence which distance can be afforded - is a matter of Na activity ("chemical pressure"). The fact that after an initially expected random distribution, we see homogeneous bilayers right in the beginning and no diffusion profile (see Fig. SIII-3(d), SI) within the intercalation layers, means that the diffusivity is highly non-linear, more precisely self-amplified. It seems that a critical presence of the alkali element - probably because of the local expansion - accelerates diffusivity and thus favors further incorporation in these layers. (A similar acceleration is known for $FeCl_3$ intercalation[38, 39]). At a first glance it may be difficult to figure that the neighboring layer is favored also in the transient state in view of local expansion. But note that we deal with a two-dimensional situation where this problem does not critically arise and the above energetic preference predominates. For very fast diffusion within the layer, only filled or partially but homogeneously filled layers ought to be observed, and the progress of insertion to consist in realizing more of the occupied layers being initially isolated and eventually aggregated. The fact that we observe a homogeneous bulk situation (in the direction of Na flux) is a consequence of the interfacial reaction control. On a large scale and at times on the order of the diffusion time constant, one should of course also observe non-percolating layers. As will be shown in the next section, aggregation is energetically favored at LT, and the small aggregates hence must be seen as kinetic intermediates towards more massive aggregates. In the diffusion-controlled regime, "growth" of such layers should be observable (parallel to the increase of bilayer density) giving rise to the profiles already described above.

THERMODYNAMIC PICTURE

The thermodynamic analysis shall conclude our comprehensive study. For this, we primarily concentrate on the samples for which we can assume saturation equilibrium. In Fig. 4(a), the solubility values are logarithmically plotted versus $1/T$. At HT the graph is roughly linear while on lowering the temperature it becomes flatter with the clear tendency to increase again. Comparing the Na values with our values for Li and K we find much lower equilibrium content for Na than for the other alkali metals (~$NaC_{100}$). The OCV values follow the sequence (Na)<(Li)<(K)<(Rb)<(Cs) (Fig.4 (b)). This is in accordance with literature[22, 40] where even significantly lower if not zero values have been reported for Na. The OCV values (corresponding to the negative free enthalpies of reaction) are slightly above zero (This is consistent with ref.[8] claiming 2 mV at 300 °C.), in other words a high activity is required for obtaining this rather modest Na content.

When evaluating the $1/T$-dependence of the logarithms of the HT equilibrium contents we derive a small positive enthalpy of intercalation (unlike Li[41], K[42], Rb[43], Cs[44]) and a negative entropy value (like Li, K, Rb, Cs): $\Delta H_{\mathrm{ins}}^{0} = 8.5\ \mathrm{kJ/mol}$; $\Delta S_{\mathrm{ins}}^{0} = -26.6\ \mathrm{J/mol\ K}$. Given these small $\Delta H$-values, this is not in contradiction to the small but positive OCV data for the quenched HT samples as the OCV contains also the configurational contribution and relates to Na incorporation into Na-containing carbon-layer.

One should also not forget that the van't Hoff relation presupposes a random distribution. Thus, in view of the complex intercalation pattern these values are only effective values and are discussed again below. Moreover, if the $Na^+$ and $e^-$ were completely independent, the measured slope would be half of the incorporation enthalpy (Part SV-5, SI); but since we do not treat charge carriers separately, this is only a question of notation. For the HT samples TEM and XRD reveal overwhelmingly two-sided intercalation whereby these intercalated regions are (under-occupied) bilayers and separated by typically 4 - 10 pure carbon-layers. At LT, however, higher aggregates are observed even if equilibrium has not yet been reached.

There are several points to be explained: (1) What is the reason for the storage capacity minimum in the free energy for Na along the series of alkali metals? (2) Why do we observe two-sided intercalation and under-occupation? (3) Why are these layers typically separated by about 4 - 10 pure carbon-layers? (4) Why do we observe higher aggregates in the LT samples? Let us address these points step by step.

Ad 1): Because electrolyte and solvation influences disappear in the cell reaction $\mathrm{A} + x\mathrm{C} \rightarrow \mathrm{AC}_x$(with A being short for the alkali metal element) and because entropic differences for the formation free energy may be assumed to be rather small (see Part V-2, SI), the answer why Na shows a lower capacity than the other alkali metals must lie in the accommodation energetics. We have observed a positive intercalation enthalpy at HT (Fig. 4(a)), lowering its value with the clear tendency to change sign when the temperature is decreased. Unfortunately, the stability of the coating limits the time window of electrochemical intercalation. At any rate, the measured absolute values are small in contrast to the distinctly negative values in Li and K cases. Nobuhara et al.[21] and Liu et al.[22] could reproduce the OCV sequence (Na) < (Li) < (K) < (Rb) < (Cs) as far as the energy is concerned using first principles calculations for $AC_6$ and $AC_8$. Liu et al. decomposed the insertion process into atomization of A, straining (i.e. stretching and expanding) graphite to the state that is finally achieved, ionization of A plus transferring the electron to the carbon-substrate, and then coupling of the cation to the negatively charged graphite (involving electrostatic and quantum-chemical bonding effects). They found all individual contributions to be monotonic from Li to Cs. Coupling and straining were calculated to be costly when compared to Li (monotonically increasing from Li to Cs), while ionization and decohesion are beneficial when compared to Li (monotonically decreasing from Li to Cs). The overall process then shows a maximum in energy for Na insertion in agreement with the experiments. As on one hand the coupling dominates the costly contributions and the ionization the beneficial ones, Liu et al. could conclude that it is essentially their counteraction which leads to the non-monotonicity. (We interpret the distinctly lower coupling energy for the Li-ion to the negativated carbon-matrix simply as expression of the higher covalency of the Li-C bond.) Importantly, the overall formation energy is found to be positive for Na in the fully occupied $AC_6$ and $AC_8$ structures. This is not in contradiction to our findings that the slope in Fig. 4(a) changes sign (dashed region) or that the OCV is slightly positive, as (i) the inclusion of entropy could easily bring down the free energy to negative values as the absolute value of the energy is small; (ii) we deal with much smaller Na contents than considered by Liu et al. which are expected to substantially reduce the energy, and (iii) OCV and intercalation do not exactly refer to the same reaction (part V-7, SI) even though the difference may be small for very small Na contents.

Ad 2) The following scenario still needs to be explained: Li, K, Rb, and Cs form monolayers which become progressively less separated (decreasing staging number) as activity increases, and eventually form full aggregates (stage 1)[6, 40, 45, 46]. Na however prefers under-occupied bilayers with increasing ordering and aggregation on temperature reduction. Full aggregation is not seen even at 25 °C for $a_{\mathrm{Na}}$=1, very probably due to the sluggish kinetics.

Let us now discuss the benefit of the two-sided intercalation. As entropy advantages should be small (see Part V-2, SI), we have to seek the explanation in the gain of enthalpy. Incomplete filling (allegedly 50 % when compared to the $AC_6$ or $AC_8$ structure) as discussed above with the aggregation pattern from TEM as outlined in SI (Part II, 1-2) would indeed be consistent with a favorable energetic effect as then the Na ions can gain Madelung energy, in particular if the Na ions of the two layers are displaced with regard to each other in consistency with an under-occupation of about 50 %. (Interestingly, an under-occupation of 50 % would also be entropically favored if the energetic effect is negligible (see Part V-4, SI).) The fact that the two spacings within bilayers can be different (Fig. 3(c)), indicates varying under-occupancies. Unsurprisingly, literature reports under-occupancy for the other alkali elements too if the staging number is large[24, 33, 34].

To reiterate: Two-sided intercalation is not inconsistent with a zero mobility of Na through the layers, as the fast in-diffusion occurs along the layers. How the mechanism for bilayer filling looks like in detail, has to be clarified in the future. Certainly higher-dimensional defects, if present, would significantly facilitate the kinetics by allowing for perpendicular transport.

Strikingly, for the other alkali metals, besides fully aggregated (stage 1) states only monolayers (higher stage numbers) have been reported. In these cases obviously, and unlike Na, monolayers are tolerable and aggregation to fully occupied aggregates is, whenever possible, more favorable than bilayer formation. The cause is most likely to be sought in the counteraction of ionicity and size. Ionicity increases (covalency decreases) from Li to Cs and favors distribution over more than one layer involving under-occupation. Note that the lower Li ionicity is not directly determined by the only slightly higher ionization potential in the gas phase (compared to Na[47]), but by the value in the carbon environment which should be substantially varied due to the covalency of the Li-C bond (cf. coupling term in ref.[22]; the measured dipole moments have been reported to be markedly less than the value expected from a completely ionized bond (6 instead of 9.5 Debye[48]). SI (Part V-6) gives simple arguments that not only from a Madelung perspective but also from the perspective of delocalization, the energy gain from monolayers to bilayers is greater than from bilayers to higher aggregates making a preference for the Na-bilayers in terms of free energy plausible. This holds if the ions are ionized and do not induce strain. Obviously, Li is more covalent and the higher alkali metals too big to favor bilayers within the carbon matrix (Note that the energy connected with misfit stress depends quadratically on the misfit assuming that this effect is not compensated by the lower occupancy). Thus, Na appears to offer the best parameter compromise to favor bilayers. Certainly, all this offers an interesting task for ab initio calculations which might also clarify the relevance of the fact how important the fitting of bilayer stacking and matrix stacking is for the intercalation.

Ad 3): Let us now refer to the thermodynamic stability of staging compounds in general. In order for intercalated layers to be separated by non-intercalated layers rather than aggregating, repulsion between them must predominate over attraction. (Note again that the entropic effect on rearranging the layers is negligible.) Elastic interactions are either negligible (as for Li[49], or Na) and/or of low range. So we concentrate on the electric effect. In an adapted screening model[37, 50], it was shown that screening lengths can reach 5 - 8 Å, i.e. values higher than assumed by Dresselhaus et al. ("one atomic layer"[39]). A critical distance may be assessed to be 4 times this value[51] corresponding to about 10 layers nicely agreeing with the TEM results. This is nevertheless very approximate, also in view of the fact that the interaction of the alkali metal with the carbon layers can extend up to a few carbon layers[52]. For Li the higher covalency is an additional factor (in the case of zero ionization the electric double layer effect would be zero).

So far we have neglected electric double layers formed by the electron cloud screening the alkali ion arrays. The thermodynamics of a diffuse electric double layer interaction (do not confuse this with the Na bilayer intercalation) stemming from truncating the semi-infinite extent (very large distance, no interaction) of the electric potential to the finite profile with the value in the middle being larger than the corresponding semi-infinite values, has been treated by Overbeek in ref.[53]. As pointed out by Overbeek this confinement of the charge carriers reduces the configurational entropy which in most examples exceeds the energy effect connected with the approach of the diffuse space charge zones. Surprisingly, such entropic effects have never been discussed for staging compounds though being popular in colloid chemistry. Naturally the number of the separating carbon-layers and thus the degree of tolerated double layer repulsion is a function of the alkali metal activity (cf. voltage). In Na compounds the intercalation energy is not very favorable such that the intercalation layer separation is larger than for the other alkali metals for which a smaller if not vanishing distance (corresponding to an increasingly lower staging degree) is possible for the same condition. Fig. 5 gives a sketch of how we figure the formation of Na insertion thermodynamically if the Na activity and hence the Na content is increased. It can also be interpreted as sketch of the kinetic intermediates (cf. LT insertion path). In the very beginning, a random distribution is expected. Soon, bilayers are favored whose distances become smaller and smaller, eventually favoring aggregates such as $NaC_6$ or $NaC_8$ for extreme and possibly unrealistic waiting times. SI (Part V-5) calculates the energetic and entropic variations. An analogous feature can be applied to Li, K, Rb, Cs but by replacing bilayers with monolayers.

Ad 4): It remains to be explained why in the case of Na higher aggregates seem to be stable at LT (25 °C), while at HT (400 °C) almost exclusively chemical bilayers are observed. A tempting explanation would again be mass action considerations of the various aggregation degrees; but as already mentioned the configurational entropy for rearranging layers is much too small (as an example one may look at the annihilation of surfaces or grain boundaries: these energies are on the order of 0.1 or 1 $J/m^2$ or more, exceeding configuration free enthalpy terms by many orders of magnitude (Part V-4, SI). So one may seek the reason in the standard entropy of the aggregation reaction (i.e. the “local” entropy value).

Phononic effects do not appear sufficient to explain the behavior (the effective Debye temperatures for graphite and $LiC_6$ or $KC_6$ are similar[41, 54] and would only give rise to negligible entropy contributions.) Yet space charge layers can. Note that on aggregating two chemical bilayers to one quadruple layer - to select this example - one loses two space charge layers (per bilayer and hence for one Na per formula unit). As discussed above related entropy effects can and actually do exceed in many cases the related energy effects. Then the standard free enthalpy referring to a layer or a layer aggregate would be given by $\Delta H^0_{\mathrm{agg.}} - T\Delta S^0_{\mathrm{agg.}}$ where the first term is dominated by a chemical interaction term (outrunning the small electric double layer value) while the second stems essentially from space charge entropy (outrunning phononic contributions). Note that this may include elastic interactions. Should they be important, then the pure chemical bonding value would be even more negative. As shown in SI (Part V-4, 5), from the fact that the last term is dominant at HT, whilst overcompensated by the enthalpy term at LT, we can assess their values. The situation is sketched in Fig. 6 (More details are found in Part V, SI). Fig. 4(a) shows the van’t Hoff plot of the solubilities versus temperature. The HT behavior refers to intercalation in bilayers and the LT behavior is closer to the intercalation into massive aggregates. According to the rough estimate derived in SI (Part V-4) we can give $\Delta H^0_{\mathrm{agg.}} = (-20 \pm 4)\ \mathrm{kJ/mol}$, $\Delta S^0_{\mathrm{agg.}} = -34\ \mathrm{J/mol\ K}$, explaining the different slopes in Fig. 4(a) along with the different nanostructures. Fig.4(a) shows in addition to the HT equilibrium Na concentrations, the extrapolated LT values. According to the experiments, the values show an upward deviation from a van’t Hoff behavior. The insertion enthalpy corresponding to LT van’t Hoff

slope - which is to be understood as effective mean value - can be estimated to be $(-2.5 \pm 4)$ kJ/mol (Part V-5, SI). As this value is an average effective value, the equilibrium solubilities below 100 °C may be substantially higher than those obtained by the linear $\ln[\mathrm{Na}]$ vs. $1/T$ approach and even values comparable to $NaC_6$ or $NaC_8$ may be thermodynamically possible though kinetically hardly realizable. The thermodynamic cycle depicted by Fig. 6 shows the consistency of the thermodynamic arguments in terms of the respective enthalpies (For simplicity, in Fig. 6 the same Na-content was taken for HT and LT). In the SI (Refinements of the above $\Delta H$-cycle, Part V-5 and Fig. SV-6) it is shown that a different content ($y$) for HT and LT would not change the result as long as $y \ll 1$.

## SUMMARY AND CONCLUSIONS

This work investigates the significant chemical and electrochemical intercalation of Na into HOPG. On increasing temperature from 25 °C to 500 °C the storage behavior changes from interfacial control to diffusion control and finally to thermodynamic control. The Na content of these equilibrated samples is found to be on the order of 1 mol %. Thermodynamically higher Na concentrations are predicted for LT, but are kinetically out of reach (given practically relevant waiting times). The OCV is small but positive indicating thermodynamically stable insertion at $a_{\mathrm{Na}} = 1$ (i.e. equilibrated with Na). The small OCV values imply small driving forces and can be held responsible for the sluggish kinetics. The structure of these compositions is - at 400 °C - revealed by TEM to consist of carbon-layers that are embedded by Na layers which are under-occupied when compared to $AC_6$ or $AC_8$. The further evolution of the intercalation pattern on increased Na content can be understood by ordering of the bilayer sequence once the screening length is reached (the critical distance is typically on the order of 10 carbon-layers). Further aggregation to denser bilayer arrangements or higher layer aggregates (lower staging number) requires lower temperatures. In principle, an analogous qualitative picture can be assumed for the other alkali metals with the exception that here monolayers play the decisive role before full aggregates are realized on subsequently reducing the number of pure carbon-layers (reduction of staging number). The anomalous role of Na (within the column of alkali metals) regarding structure (bilayers), storage capacity (extremely small), OCV (close to zero) and kinetics (very sluggish) can be phenomenologically explained by the interplay of energy and entropy including space charges, and atomistically by the interplay of covalency, ionicity and size.

Interpreting the results in terms of battery research, it can be stated that for HOPG, in order to reach the necessary storage in practically relevant times at room temperature, coated nanocrystalline samples are needed. In view of the small OCV-values which are not safer than the values of hard carbon, the usefulness of graphite is not obvious. The most promising path for optimizing graphite electrodes should lie in chemically modified carbon enabling better thermodynamics, in the preparation of kinetically stable carbon structures with a high content of higher-dimensional defects (this includes high surface carbon) or in the preparation of high surface carbon whose surfaces are chemically modified.

## METHODS

### 1) Materials Synthesis

Na intercalated graphite compounds ($Na_yC$) were synthesized by chemical and electrochemical intercalation of Na (99.9%, Sigma-Aldrich) into uncoated and $TiO_2$ (or amorphous carbon, coating method is similar to ref. [55]) coated HOPG. Sample preparation for both chemical and electrochemical intercalation and post-treatment were in Ar-filled glove box ($O_2 \leq 0.1$ ppm, $H_2O \leq 0.1$ ppm). As a rule, every piece of HOPG used was around 5×5×0.25 $mm^3$ by cutting commercial HOPG pellets (Mosaic angle 0.8±0.2°, 1800±200 w/mK in basal plane, 8±2 w/mK perpendicular to basal plane, 10×10×1 $mm^3$, Thermo Fisher Scientific) without further treatment. $TiO_2$ coated samples were prepared by atomic layer deposition (ALD) method in clean room. Tetrakis-dimethylamido-titanium (TDMAT) and $H_2O$ were used as the precursors ($Ti(NMe_2)_4$ (g) + $2H_2O$ (g) → $TiO_2$ (s)+ $4HNMe_2$ (g)). Uniform $TiO_2$ coatings of different thicknesses were obtained by varying the number of cycles. The optimized deposition conditions and rates were given in ref. [56].

Chemical intercalation of Na (direct contact of HOPG with Na) into uncoated and coated HOPG (or uncoated polycrystalline graphite powder) was conducted at low/high temperatures (25-500 °C) at given time (Fig. SI-1(a, c), SI). HT-synthesized samples were cooled to room temperature under an argon atmosphere. Electrochemical intercalation of Na into uncoated and coated HOPG (electrolyte 1 M $NaPF_6$-EC/DMC (vol. 1:1) if without specific notice) was conducted at 25 and 40 °C by constant current discharge from OCV to 0 V (vs Na metal), followed by a hold at 0 V vs Na (voltage fluctuated between 0 and 1 mV, Arbin system BT 2000) for the specified duration. Li (electrolyte 1 M $LiPF_6$-EC/DMC (vol. 1:1), Aldrich) and K (electrolyte 1 M $KPF_6$-EC/DMC (vol. 1:1)) cases were similarly conducted. Polycrystalline graphite powder ($D_{50}$=11 µm, 99.95%, NGS Naturgraphit GmbH) was assembled as battery as usual, followed by investigating as similar to HOPG case, but negative voltages were also explored.

2) Kinetic and thermodynamic investigations by CV, OCV and XRD

For evaluating the kinetics of the electrochemical intercalation of Na into HOPG, cyclic voltammetry (CV) measurements of Na intercalation into (deintercalation from) uncoated and $TiO_2$ coated HOPG (10×10×1 $mm^3$) were performed using a Universal Pulse Dynamic-EIS Voltammetry (Voltalab, PGZ402) at 25 °C (see Fig. 1 (c, d, e)). Sweep rates ranging from 10 µV/s to about 10 V/s were applied with a potential window from 0 to 2.5 V in this study.

Potential (voltage) recordings were carried out as a function of time to evaluate kinetics of Na deintercalation and formation free energy of $Na_yC$. For $Na_yC$ chemically obtained at high temperatures, after cooled down to 25 °C, followed by cleaning their surface in an Ar-filled glovebox, then samples were assembled as coin cell batteries to record voltages vs. Na; For $Na_yC$ electrochemically obtained at 25 °C voltages were directly recorded without reassembling batteries.

Apparent diffusion coefficients ($D_{\mathrm{Na}}^{\delta}$) of (electro)chemical intercalation of Na into HOPG at high/low temperatures were estimated by XRD measurements on inner and outer parts (Fig. SI-1(d), SI) of the chemically and electrochemically synthesized samples $Na_yC$. $D_{\mathrm{Na}}^{\delta}$ can be estimated by $L^2 \simeq 2D_{\mathrm{Na}}^{\delta}t$ (where $L$ is diffusion length and $t$ diffusion time).

3) Materials Characterization

3a) X-Ray Diffraction (XRD)

Measurements were conducted under Bragg-Brentano configuration using programmable divergence slits, anti-scatter slits and a PIXcel 3D detector (PANalytical Empyrean, CuKα radiation). Samples were sealed in a polycarbonate domed holder before measurement. For data processing and analysis (diffraction), PANalytical software packages HighScore Plus were used.

3b) Na concentration measurements - Chemical analyses

After synthesis, samples were cleaned by scratching residual Na and other impurities on the surface. Acid-base titration routinely as well as inductively coupled plasma optical emission spectroscopy (ICP-OES, not routinely) was used to determine the overall Na concentration. For both techniques we used aqueous extractions of one sample, transforming the Na into NaOH. The sample (typically 5-20 mg) was placed in a Schlenk flask with 80 ml of freshly prepared bi-distilled water, and heated for approx. 20 h to 90 °C (under Ar atmosphere). Under these conditions the sodium excorporates from the graphite and reacts according to $Na + H_2O \rightarrow Na+ + OH^- + 0.5\ H_2$. The formed $OH^-$ was titrated with 0.01n HCl (Alfa Aesar) using a Metrohm Titrino 877 automatic titrator (sample solution still under Ar atmosphere to avoid reaction with $CO_2$); the amount of $Na^+$ directly corresponds to the titrated $OH^-$ amount. For samples with very low Na content, this method was not applicable because after heating the sample with bi-distilled water the pH was higher than for the pure water but still below 7 (bi-distilled $H_2O$ made from water coming from a cation exchanger is still slightly acidic). In these cases the pH of the pure bi-distilled water and of the sample solution was measured, and the amount of $Na^+$ calculated from the decrease of $OH^-$ concentration (and sample solution volume).

3c) Scanning transmission electron microscopy (STEM) and electron energy loss spectroscopy (EELS)

Experimental details on STEM sample preparation is shown in ref. [57]. STEM studies were conducted using a spherical aberration-corrected STEM (JEM-ARM200F, JEOL Co. Ltd.) equipped with a cold-field emission gun and a DCOR probe Cs-corrector (CEOS GmbH) operated at 200 kV. The high-angle annular dark-field (HAADF-STEM) images were obtained by an ADF detector with a convergent semi-angle of 20.4 mrad and collection semi-angles of 70−300 mrad. In order to make precise measurements of lattice constants, six serial frames were acquired with a short dwell time (2 µs/pixel), aligned, and added afterward to improve the signal-to-noise ratio (SNR) and to minimize the image distortion of STEM images. EELS acquisition has been performed with a Gatan GIF Quantum ERS imaging filter equipped with a Gatan K2 Summit camera and a CCD camera with a convergent semi-angle of 20.4 mrad and a collection semi-angle of 111 mrad. An image processing pipeline was developed to identify intercalated areas to enable statistical analysis of Na intercalation in HOPG.

3d) X-Ray photoelectron spectroscopy (XPS)

XPS measurements were performed on a Kratos Axis Ultra system equipped with a monochromatized Al $K_\alpha$ X-Ray source. Samples were prepared and transferred to the XPS chamber in an airtight transfer tool in an Ar-filled glove box. After measurement of sample surface, for depth profiling, a scanning MiniBeam III Argon ion sputter gun (from Kratos) with a beam energy of 4.0 keV and an emission current of 20 mA was used (1, 3, 8, 16, and 30 min). An immediate measurement was performed after each sputtering.

3e) Scanning Electron Microscopy (SEM) and Energy Dispersive X-ray Spectroscopy (EDX)

The SEM micrographs were obtained with a Zeiss Merlin SEM operating at a voltage of 4 kV. EDX mapping was done with an Oxford Instruments detector (Ultim Extreme).

3f) Electrochemical Impedance Spectroscopy (EIS) measurement

Electrochemical Impedance Spectroscopy (EIS) was employed to characterize the interfacial stability at room temperature. Electrochemcially intercalated samples were synthesized as described in Section 1). Measurements were performed using a Novocontrol Alpha-A analyzer (Novocontrol Techologies, Germany, 2-wire measurement, $10^6$ to $10^{-3}$ Hz, amplitude 0.01 or 0.05 V).

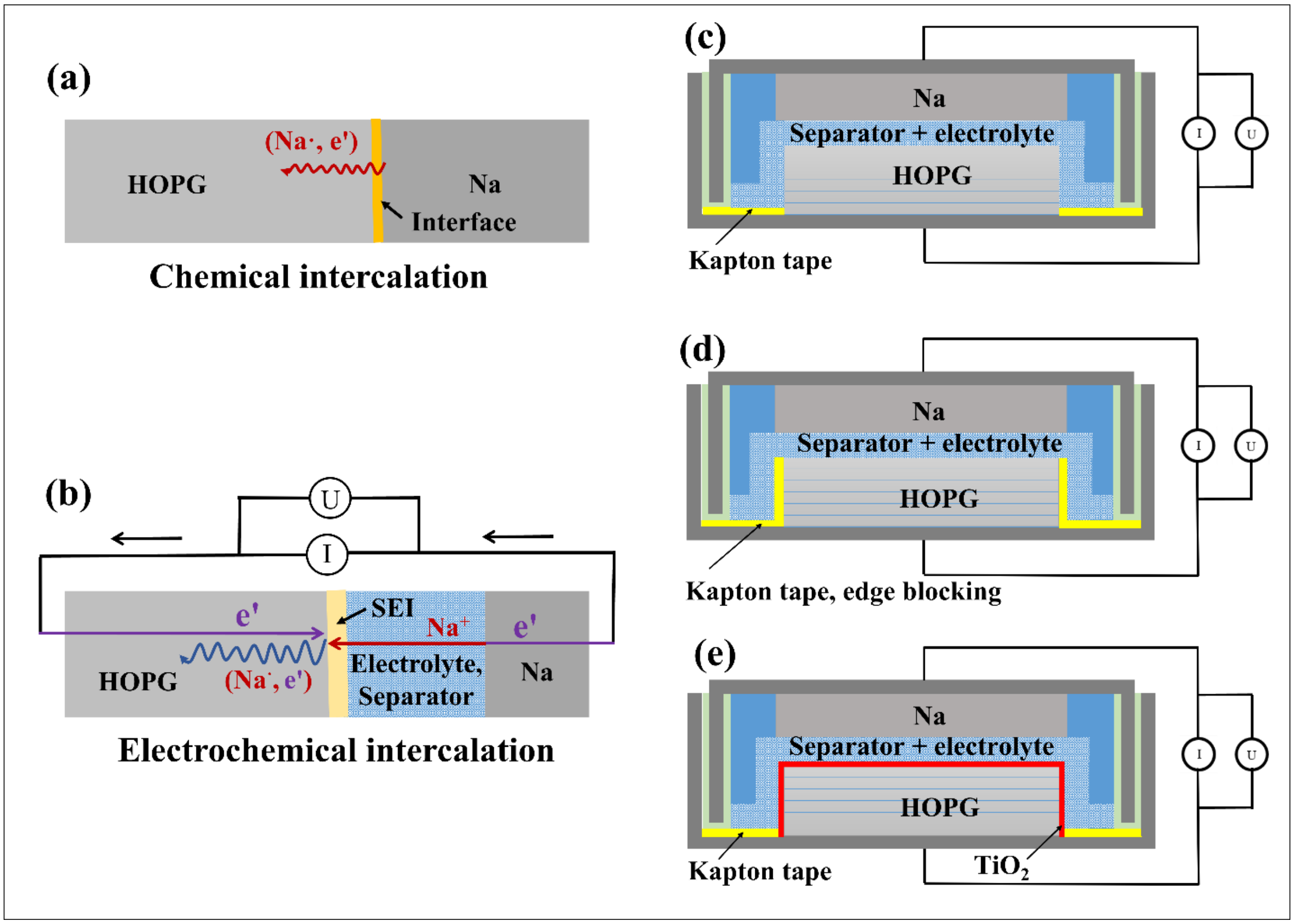

Fig. 1 Sketches of the cells for Na intercalation into HOPG via chemical method (a) and electrochemical method (b), schematic set-ups for CV measurements of pure HOPG (c), of side surface blocked HOPG (d) and of $TiO_2$ coated HOPG (e).

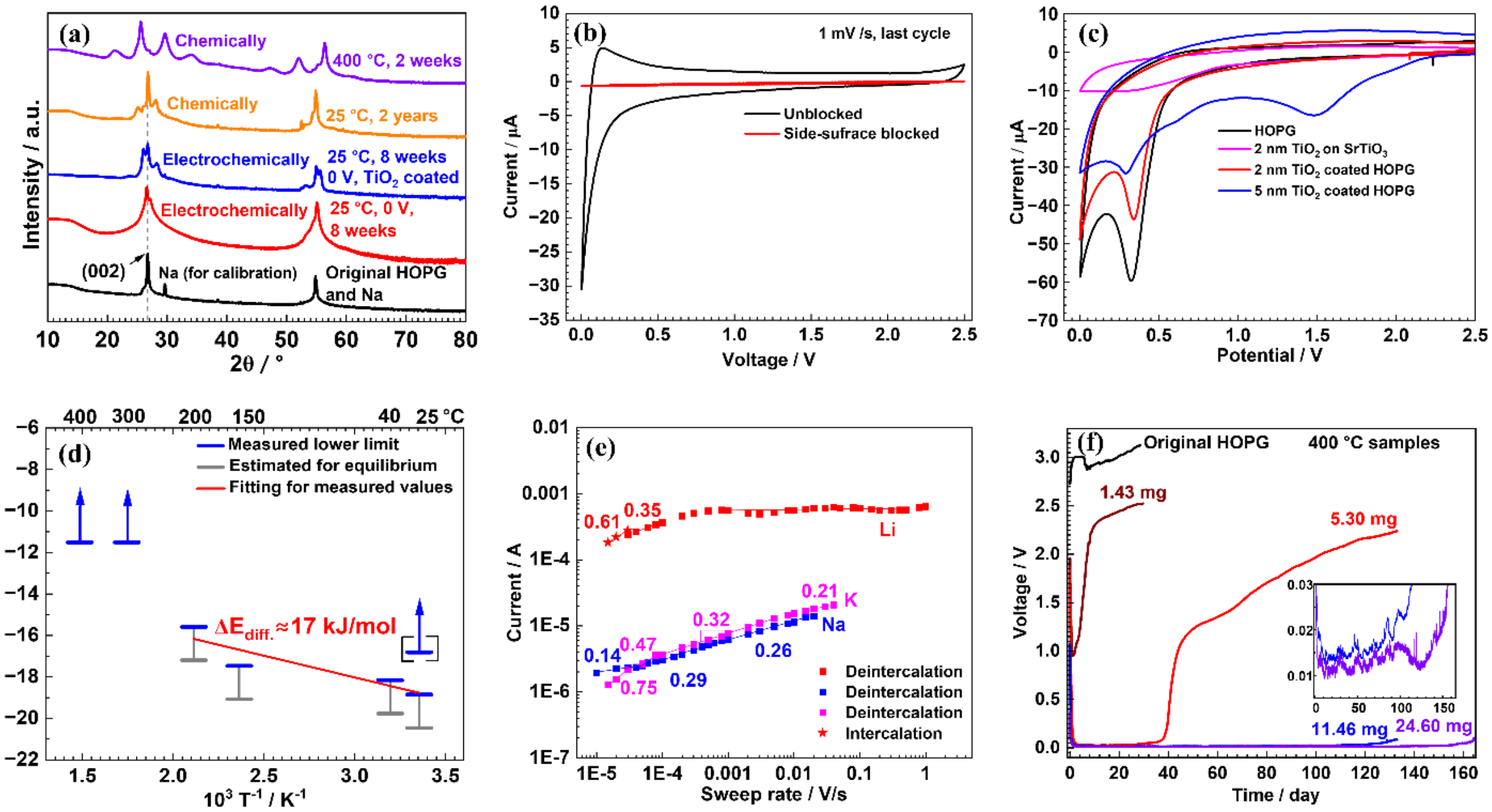

Fig. 2 (a) XRD patterns for chemical and electrochemical intercalation of Na into HOPG ($TiO_2$ coated-5 nm $TiO_2$ coating on HOPG). CV measurements for Na (de)intercalation into / from (b) pure and side-surface blocked and (c) $TiO_2$ coated HOPG at 25 °C. (d) Apparent chemical diffusivities as a function of temperature. Solid-arrow data points indicate lower limits under thermodynamic control or interfacial control. The other data refer to HOPG for which profiles are observed and lower and upper limits could be estimated (profiles, see Fig. SⅢ-6(d), and 150 °C-2 weeks case in Table SII-1, SI). Apart from the 25 and 40 °C experiments (which are from electrochemical intercalation into 2 and 5 nm $TiO_2$ coated HOPG, respectively), the data stem from chemical intercalation of uncoated samples. The black bracketed 25 °C value (electrochemical intercalation in uncoated HOPG) suffers from co-intercalation which seems to accelerate the process. The apparent activation energy for the roughly linear LT part is ~17 kJ/mol (see red fitting line). (e) Current vs potential sweep rate 25 °C. (f) OCV for Na intercalated HOPG at 25 °C (inset is zooming into the plateau range; for more details, see Fig. SIII-18, SI).

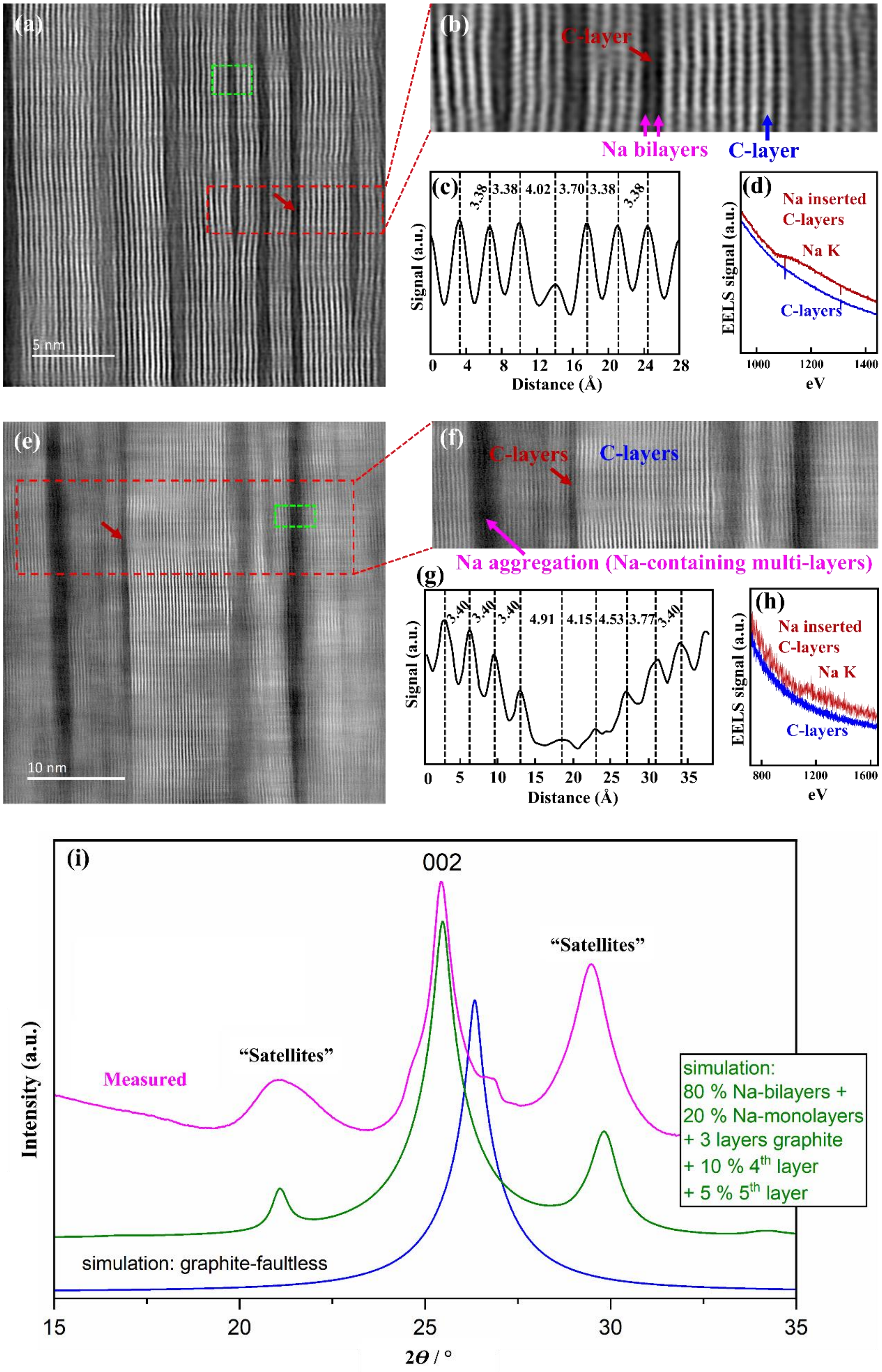

Fig. 3 Na intercalated HOPG chemically synthesized at 400 °C for 2 weeks: (a, b) HAADF-STEM images, (c) interlayer distances for the green dashed square region and (d) Na-K edge EELS spectra for the red arrow marked region in (a). Na intercalated HOPG chemically synthesized at 25 °C for 10 months: (e, f) HAADF-STEM images, (g) interlayer distances for the green dashed square and (h) Na-K EELS spectra for the red arrow marked in (e). The TEM images combined with the interlayer distances and the EELS data can only be explained by dominance of Na bilayers. (i) Na intercalated HOPG chemically synthesized at 400 °C for 2 weeks: Comparison of the measured (magenta) with the simulated patterns (green and blue).

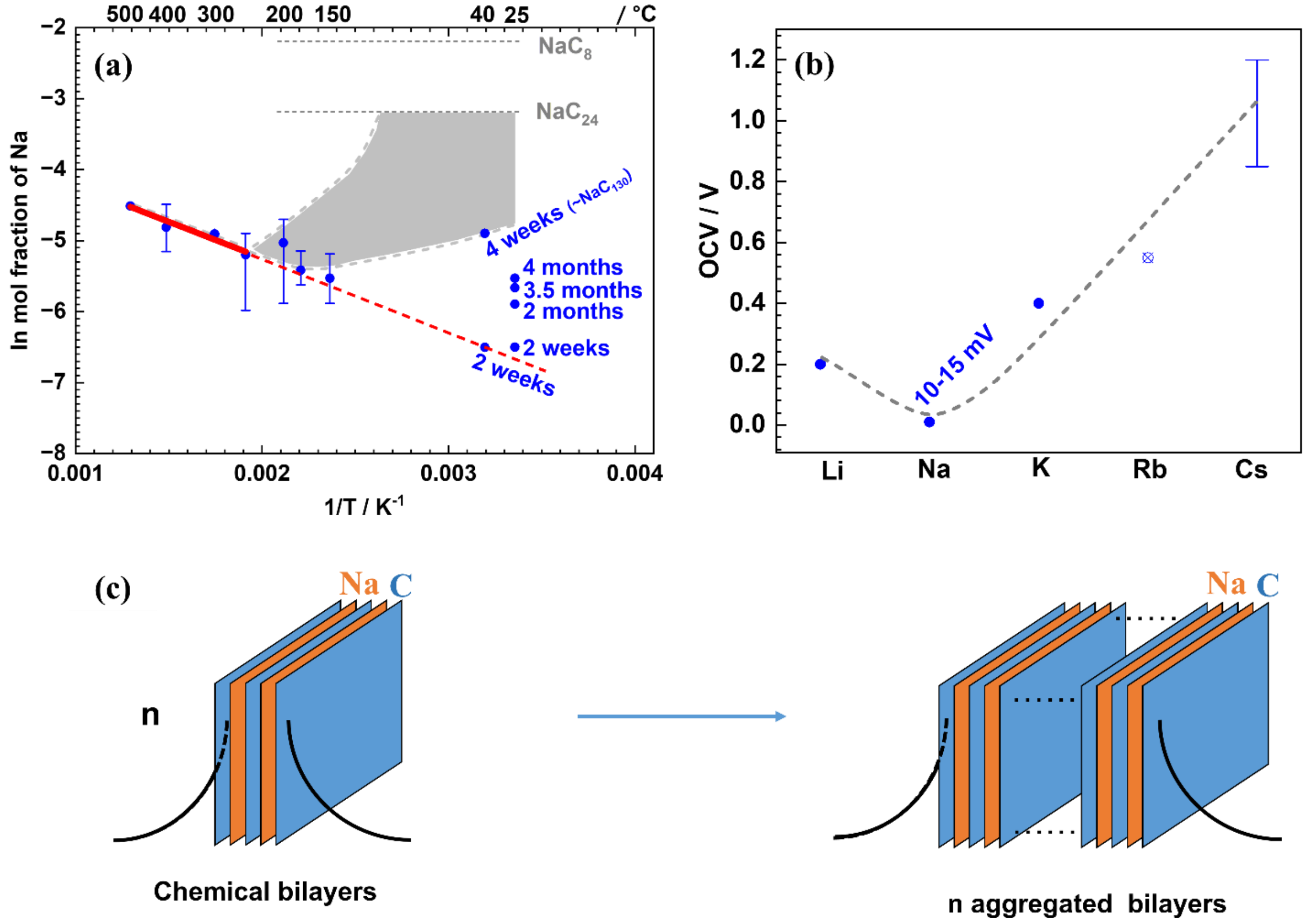

Fig. 4 (a) Na concentration vs temperature plot (grey coloured area indicates the range of expected equilibrium values; bold-red solid line indicates regime of thermodynamic control (equilibrium-established). Data at 25 and 40 °C were obtained from electrochemically intercalated samples ($TiO_2$ coated). The bars include not only the error of analysis, but also the distribution within a given batch. For 2- and 8-weeks intercalation experiments (25 °C), 2 nm $TiO_2$ coated HOPGs were used, and 5 nm coated samples were utilized for the longer time (longer than 8 weeks) or higher temperature (40 °C) experiments. Data at other temperatures (above 40 °C) were acquired from chemically intercalated samples. For detailed concentrations, Table SII-1, SI). (b) OCV at 25 °C (OCV for $NaC_{\sim100}$ synthesized at 400 °C, and $AC_{40\text{-}100}$ for the other alkali metal intercalated graphite compounds. The Rb value is from room temperature electrochemical data of ref.[40], while the Cs value is estimated from 600 °C chemical data of ref.[44]. Dashed curve is a guide to eye.). (c) Sketch for chemical bilayer aggregation to higher aggregates, resulting in loss of space charge zones. The electrons compensating for the $Na^+$ are partly in the space charge layers and partly in-between the Na-layers.

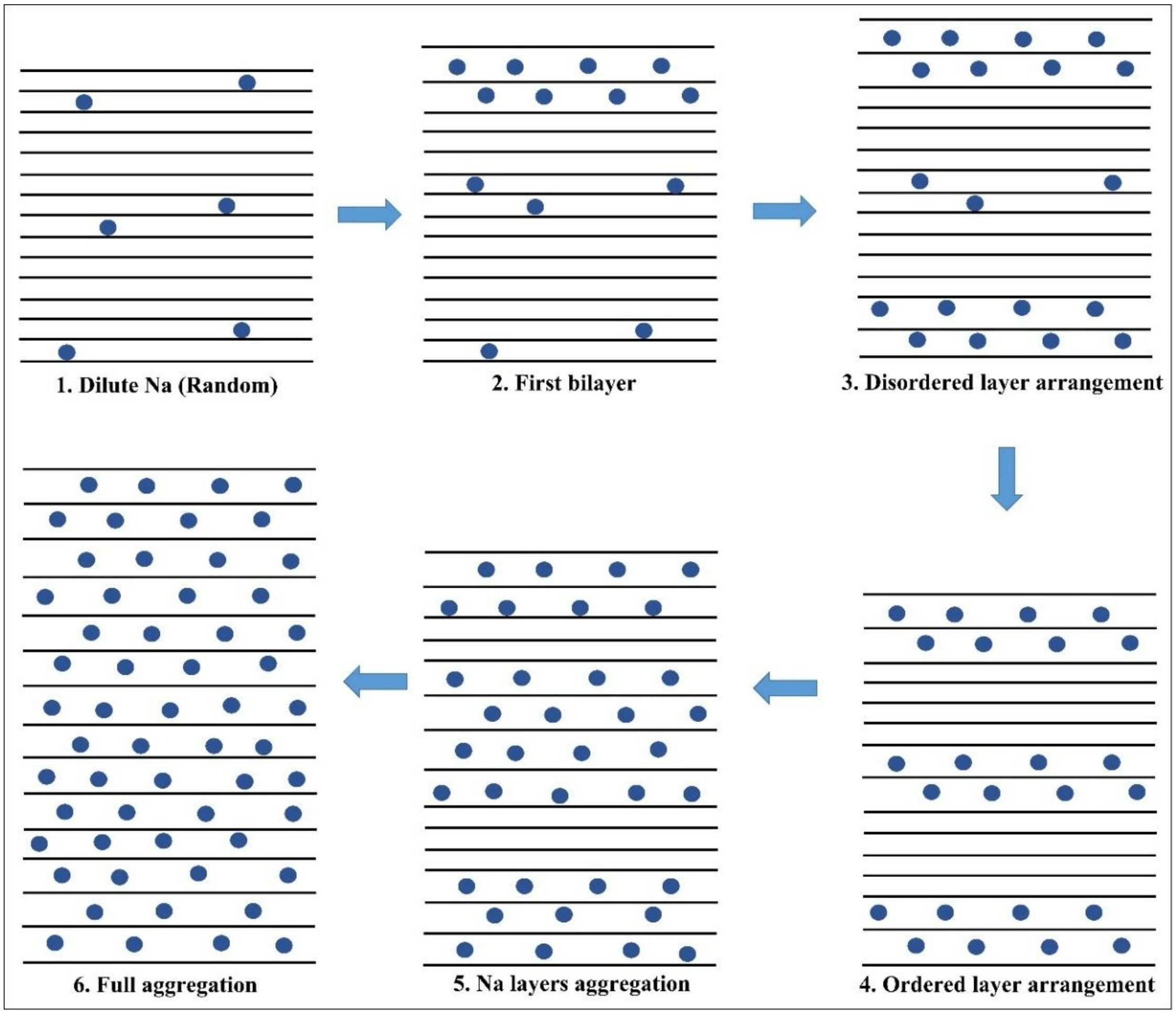

Fig. 5 Sketch of the equilibrium intercalation pattern at increasing activity of Na (without defects consideration). The figures can also be viewed as transient cases towards full aggregation at given high Na activity.

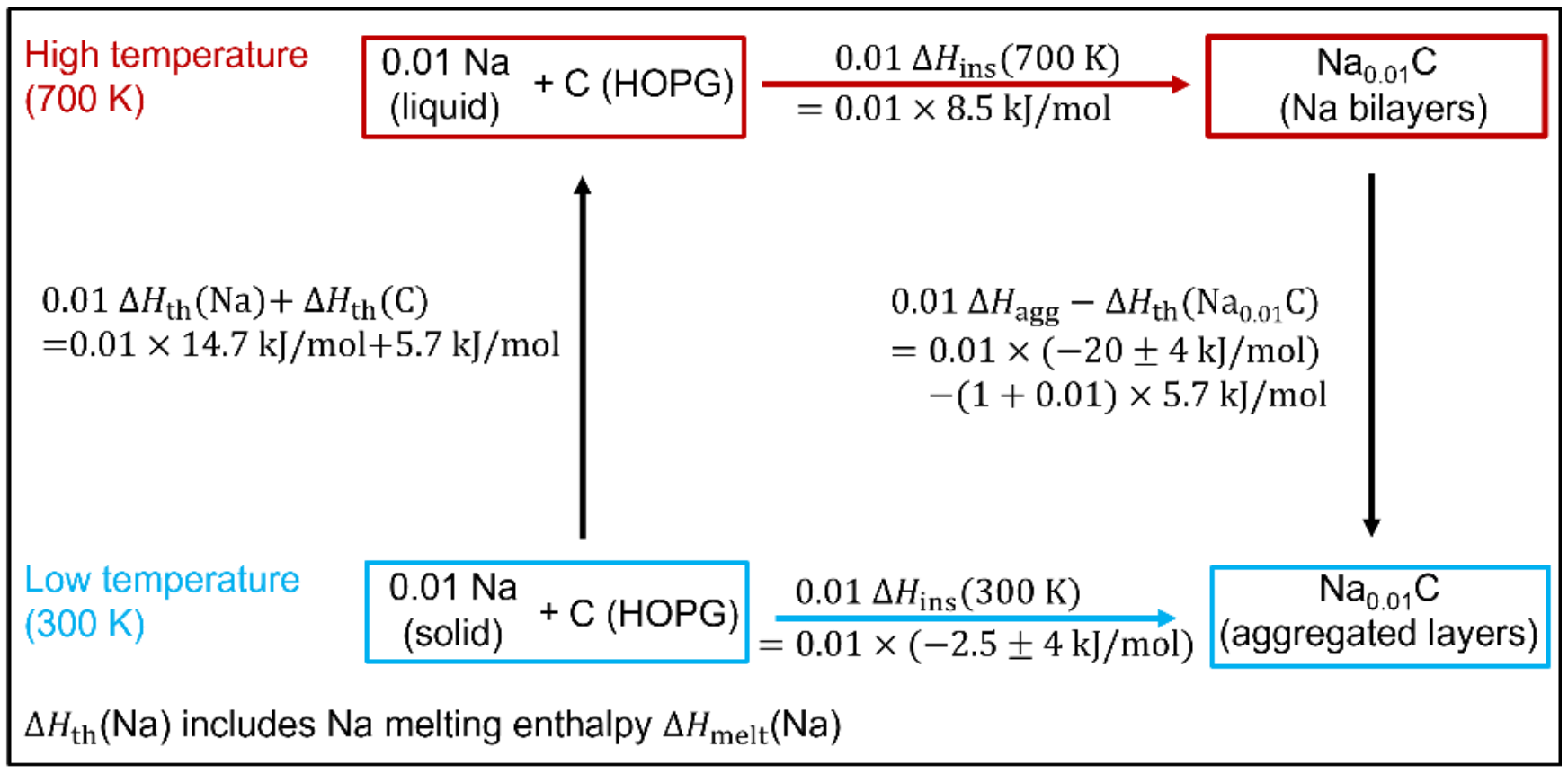

Fig. 6 Thermodynamic cycle for the enthalpy changes as discussed in the text. The figure refers to the same $y$ for HT and LT. Part IV-5, SI gives a more general figure and shows that a difference in $y\ (700\ K)$ and $y\ (300\ K)$ does not perceptibly vary the results as long as $y \ll 1$.

**ACKNOWLEDGEMENTS**

We are grateful to the following persons: Dr. H. Hoier and A. Sorg (XRD measurement), Dr. K. Küster (XPS), B. Fenk (EDX), S. Hammoud (ICP), F. Kaiser, U. Klock, U. Traub, and A. Fuchs (technical support), Dr. J. Deuschle (TEM sample preparation), Dr. Y. Zhu, Y. Zhang, Dr. J. Popovic-Neuber, Dr. I. Moudrakovski, and Dr. J. Smet (discussions), Dr. E. Kotomin (discussion and internal review). Support from MPG (all authors) and from CSC-DAAD (C. G.) is acknowledged.

**AUTHOR CONTRIBUTIONS**

C. G. and C. X. contributed equally to the work.

C. G. performed the storage experiments, C. X. performed electrochemical experiments, H. W. and P. v. A. are responsible for the TEM measurements, R. M. did the chemical analysis, S. B. did the XRD simulations, B. V. L. and J. M. supervised the experiments and conceived the project. J. M. is responsible for the thermodynamics analysis. All the authors discussed the results together and assisted in writing the paper.

**SUPPLEMENTARY INFORMATION**

# Anomalous Sodium Insertion in Highly Oriented Graphite: Thermodynamics, Kinetics and Evidence for Two-Sided Intercalation

Chuanhai Gan[*1], Chuanlian Xiao[1], Hongguang Wang, Peter A. van Aken, Rotraut Merkle, Sebastian Bette, Bettina V. Lotsch, Joachim Maier*

Max Planck Institute for Solid State Research, 70569 Stuttgart, Germany

[1]These authors contributed equally.

*Corresponding authors, c.gan@fkf.mpg.de; office-maier@fkf.mpg.de

**Table of Contents**

## Part Ⅰ Experimental Schemes for Chemical Intercalation

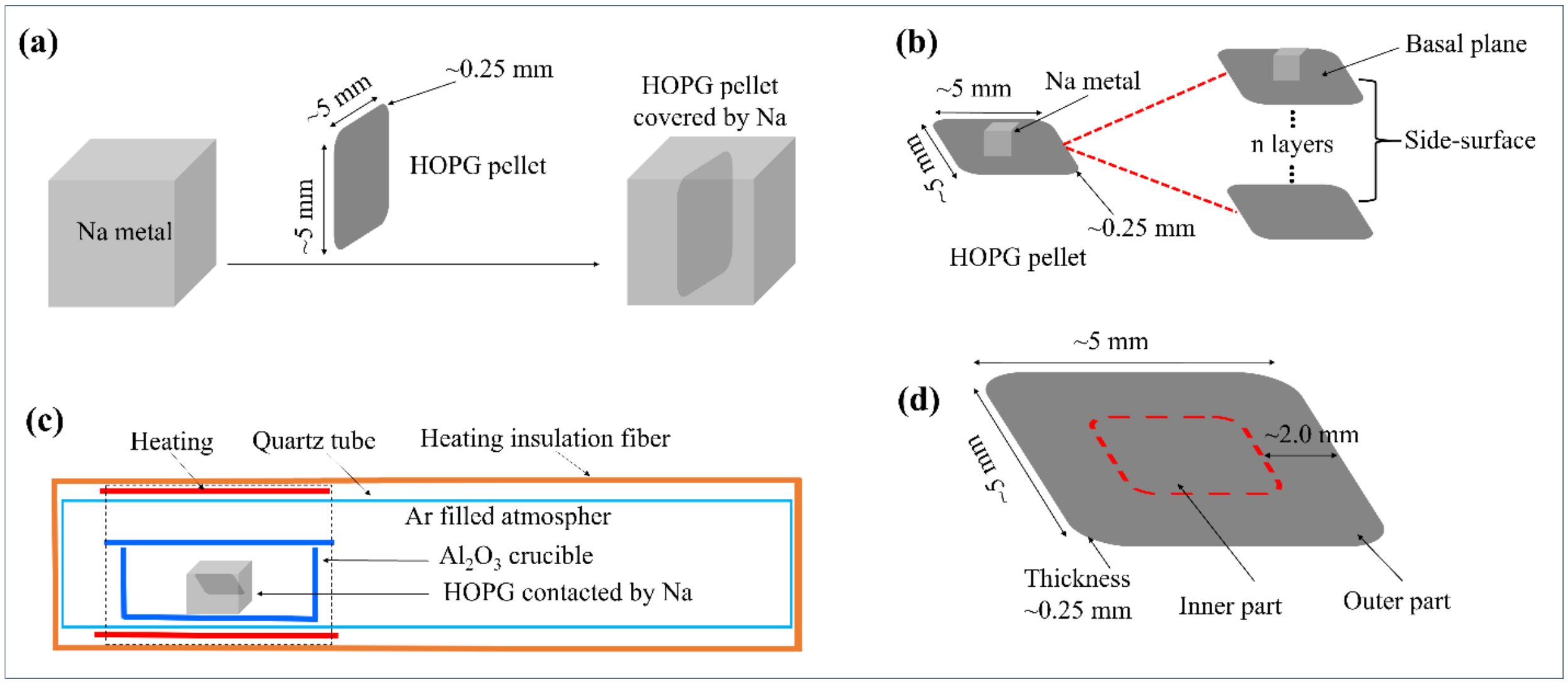

Fig. SI-1. Experimental schemes of 25 °C chemical intercalation (a) by fully covering HOPG pellet using Na metal, (b) by attaching Na metal to the basal plane of HOPG pellet. (c) Schematic set-up for high temperature chemical intercalation by fully covering HOPG pellet using Na metal. (d) Experimental schemes for separating inner and outer parts of synthesized sample.

## Part Ⅱ Na Concentration, Interface Stability, and Kinetic Analysis of CV Measurements

### 1-1) Na-compositions of samples for different storage procedures

Table SⅡ-1 Na intercalated HOPG composition ($NaC_x$) by titration after hydrolysis [a, b]

| Temperature (°C) \ Time | 2 days | 1 week | 2 weeks | 3 weeks | 4 weeks | 8 weeks | > 8 weeks | Average |
|---|---|---|---|---|---|---|---|---|
| 500 (chemically) | | | $NaC_{90}$ | | | | | $NaC_{90}$ |
| 400 (chemically) | $NaC_{91}$<br>$NaC_{88}$<br>$NaC_{148}$<br>$NaC_{172}$<br>$NaC_{119}$<br>$NaC_{115}$ [c] | | $NaC_{106}$<br>$NaC_{115}$<br>$NaC_{102}$<br>$NaC_{96}$ (outer)<br>$NaC_{153}$ (inner) | | $NaC_{109}$ | | | $NaC_{122}$ |
| 300 (chemically) | | | $NaC_{133}$<br>$NaC_{135}$<br>$NaC_{135}$ | | | | | $NaC_{134}$ |
| 250 (chemically) | $NaC_{216}$<br>$NaC_{419}$ | $NaC_{163}$<br>$NaC_{156}$ | $NaC_{133}$<br>$NaC_{135}$ | $NaC_{209}$ (outer)<br>$NaC_{395}$ (inner) | | | $NaC_{138}$ (outer)<br>$NaC_{158}$ (inner)<br>(2.5 months)<br>$NaC_{193}$<br>(5 nm $TiO_2$ coated, 2.5 months) | $NaC_{180}$ |
| 200 (chemically) | | | $NaC_{358}$<br>$NaC_{194}$<br>$NaC_{194}$<br>$NaC_{185}$<br>$NaC_{132}$ | $NaC_{290}$<br>$NaC_{315}$<br>(5 nm $TiO_2$ coated) | | | $NaC_{122}$ (outer)<br>$NaC_{191}$ (inner)<br>(2.5 months)<br>$NaC_{109}$ (outer)<br>$NaC_{139}$ (inner)<br>(2.5 months) | $NaC_{152}$ |
| 180 (chemically) | | | $NaC_{276}$<br>$NaC_{171}$<br>(5 nm $TiO_2$ coated) | | | | | $NaC_{224}$ |
| 150 (chemically) | | | $NaC_{239}$ (outer)<br>$NaC_{14660}$ (inner) | | | $NaC_{359}$ (outer)<br>$NaC_{374}$ (inner) | | $NaC_{324}$<br>(14660 and 16225 not in) |
| | | | $NaC_{16225}$ [d]<br>(5 nm $TiO_2$ coated) | | | $NaC_{178}$<br>(5 nm $TiO_2$ coated) | | |
| 100 (chemically) | | | $NaC_{12266}$ [d]<br>$NaC_{10269}$ [d] | | | | | |
| | | | $NaC_{14263}$ [d] (outer)<br>$NaC_{17823}$ [d] (inner)<br>(5nm $TiO_2$ coated) | | | | | |
| 25 (chemically) | | | $NaC_{15200}$ [d]<br>$NaC_{11850}$ [d] | | | | $NaC_{13861}$ [d] (1 year) | |
| | | | $NaC_{13700}$ [d]<br>(2 nm $TiO_2$ coated) | | | | $NaC_{18315}$ [d] (1 year)<br>(2 nm $TiO_2$ coated) | |
| 40 (electrochemically, 5 nm $TiO_2$ coated) | | | $NaC_{666}$ | | $NaC_{133}$<br>(outer)<br>$NaC_{1115}$ [d]<br>(inner) | | | |
| 25 [e] (electrochemically) | | $NaC_{10650}$ [d]<br>(outer)<br>$NaC_{7740}$ [d]<br>(inner) | $NaC_{539}$ | | | $NaC_{276}$ | $NaC_{19320}$ [d]<br>(3 months) | |
| 25 (electrochemically, $TiO_2$ coated) | | | $NaC_{664}$ (2 nm $TiO_2$) | | | $NaC_{362}$<br>(2 nm $TiO_2$) | $NaC_{573}$ [f] (2 nm $TiO_2$)<br>$NaC_{904}$ [f] (2 nm $TiO_2$)<br>(3 months) | |
| | | | $NaC_{13040}$ [d] (5 nm $TiO_2$) | | | $NaC_{466}$<br>(5 nm $TiO_2$) | $NaC_{287}$ (5 nm $TiO_2$, 3.5 months)<br>$NaC_{251}$ (5 nm $TiO_2$, 4 months) | |

a) Estimated error bar: ±10%; Original HOPG: $NaC_{100000}$.

b) Note that the time elongation of intercalation may also be influenced by corrosion. This may be the reason for non-monotonicity of concentration, e.g. 200 °C cases.

c) Sample was tested by ICP-OES. In ICP-OES analysis, the dissolved sample was introduced into the inductively generated argon plasma via an atomizer system and excited. The emitted radiation was transferred to the ICP spectrometer where it was split into the individual wavelengths and evaluated. The intensities of the spectral lines were measured with CCD semiconductor detectors. Calibration was carried out with Na solution mixed from standard solutions.

d) Calculation based on pH for x>2000. It has to be noted that these cases come with a larger uncertainty of the Na content. It has to be noted that this calculation assumes that the base formed as reaction product of Na with water if fully dissociated; if this does not hold then the Na content determined for this analysis is underestimated.

e) Solvent co-intercalation probably occurs due to unstable interfaces (see Part Ⅱ-2).

f) 2 nm $TiO_2$ coated samples under prolonged time (eg. >8 weeks) would result in Na concentration decrease, which is attributed to unstable interfaces.

**1-2)** Na under-occupancies of chemically 400 °C sample

As theoretical stoichiometry is, for Na intercalated graphite, $NaC_8$ or $NaC_6$ are discussed in the literature. According to TEM data (see Fig. 3 (a) of the main text and Fig. SⅢ-23), more than 70 % of the intercalates are in Na bilayers, 0 % in monolayers, the rest is in trilayers or higher layers aggregates. Empty carbon layers in-between intercalated Na layers are typically 4-10 layers in average.

Let us assume full occupancy, i.e. Na intercalated carbon layers are in the form of $NaC_8$ (or $NaC_6$), with n empty carbon layers in between (Fig. SⅡ-1).

a) If we assume n is 7, empty carbon layers in between intercalated Na layers in an ideal unit cell, then we end up with stoichiometry of $NaC_{36}$, which is distinctly less than the value from analysis, corresponding to ~$NaC_{100}$ (Table SⅡ−1 above). A composition ~$NaC_{100}$ would imply having 20 carbon layers in between which exceeds the average value by about a fact of 2.

b) If we also take account of the fact that we have higher aggregates or higher-dimensional defects where Na adsorbed (cavities, triple-phase junctions, etc.), the Na-under occupation in bilayers would be even more significant.

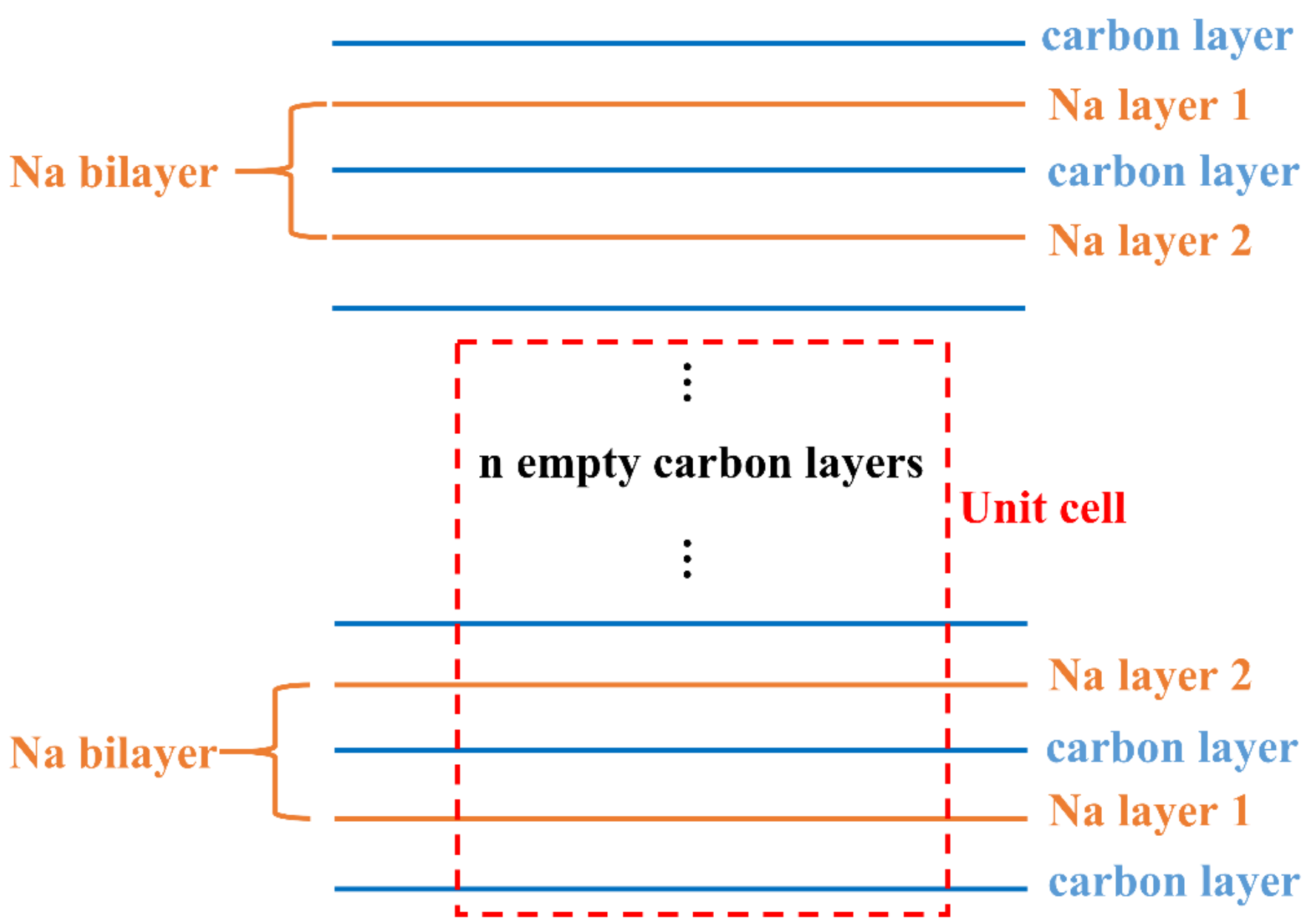

Fig. SⅡ-1 Sketch of Na intercalated HOPG

**2) Interface stability and $TiO_2$ coating for electrochemical storage**

Since the interface of Na-containing HOPG-electrolyte is unstable, $TiO_2$ coating is used to improve the interface stability of the HOPG pellet. The following consideration shows that $TiO_2$ would not react with Na-containing HOPG. It suffices to show that it is stable even against Na.

$\mathrm{TiO_2} + 4\mathrm{Na} \rightleftharpoons \mathrm{Ti} + 2\mathrm{Na_2O}$

Using

$\Delta G^{\circ}_{\mathrm{TiO_2}} = -889\ \mathrm{kJ/mol}$[1]

$2\Delta G^{\circ}_{\mathrm{Na_2O}} = 2 \cdot (-378)\ \mathrm{kJ/mol} = -756\ \mathrm{kJ/mol}$[1]

we see that

$2\Delta G^{\circ}_{\mathrm{Na_2O}} > \Delta G^{\circ}_{\mathrm{TiO_2}}$

and $\Delta G^{\circ} (\simeq \Delta G)$ of the redox reaction is positive.

**3) Kinetic analysis of CV measurements**

Cyclic voltammetry (CV) is a popular technique for obtaining information about complex electrode reactions over a wide potential range in a fairly rapid fashion. Usually, the potential is linearly varied with time, i.e., is characterized by a constant sweep rate. After having reached the maximum potential, the process is symmetrically reversed.

Useful thermodynamic and kinetic information can then be obtained by analyzing the measured current-voltage relation. The mathematical analysis of the sweep experiments was first solved by Randles[2] and Sevcik[3]. The obtained Randles-Sevcik equation is often used to predict the peak current as a function of sweep rate for a reversible reaction in a semi-infinite situation.

$$I_p = 0.4463(\frac{F^3}{RT})^{1/2} n^{3/2} A {D_0}^{1/2} {C_O}^* v^{1/2} = KnFA{C_O}^*(\frac{nFvD_0}{RT})^{1/2}$$

where $I_p$ is the peak current, $K$ is a constant, $n$ is the number of electrons transferred in the redox reaction, $A$ is the area of electrode, $F$ is Faraday constant, $D_0$ is diffusion coefficient, ${C_O}^*$ is the concentration and $v$ is the sweep rate.

The dependence of the peak current on sweep rate is often investigated in order to estimate kinetic parameters. When the kinetics are limited by diffusion, there is consensus that a power law relation should be valid: $I_p \propto v^b$, with $b = 0.5$.

For a sample of finite size (the sample size is very small or the time-scale of a measurement is long enough), the diffusion profile reaches the other side of the sample and the semi-infinite linear diffusion model is no longer valid. Then the theory of linear sweep voltammetry with finite diffusion space should be considered following the work of Aoki et al.[4]. An approximate expression of peak current as a function of sweep rate using the boundary conditions in the system of finite space can be obtained as:

$$I_p = 0.4463\left(\frac{{C_O}^* D_0 nF}{d}\right) w^{1/2} \tanh\left(0.56\, w^{1/2} + 0.05\, w\right)$$

where $d$ is the thickness of diffusion space and $w = (nF/RT)vd^2/D$. The additional term $\tanh\left(0.56\, w^{1/2} + 0.05\, w\right)$ is the only difference in comparison to semi-infinite diffusion. When the films are thick and sweep rate is large, $\tanh\left(0.56\, w^{1/2} + 0.05\, w\right)$ approaches 1 and the equation becomes identical to the semi-infinite diffusion case.

The exponent in $I_p \propto v^b$ is expected to change from $b = 0.5$ to 1.0 as sweep rate decreases.

The situation changes for surface reaction limitation will be shown elsewhere (C. Xiao, R. Usiskin, J. Maier, in preparation). The typical exponent will then be distinctly lower.

## Part Ⅲ Supplementary Figures

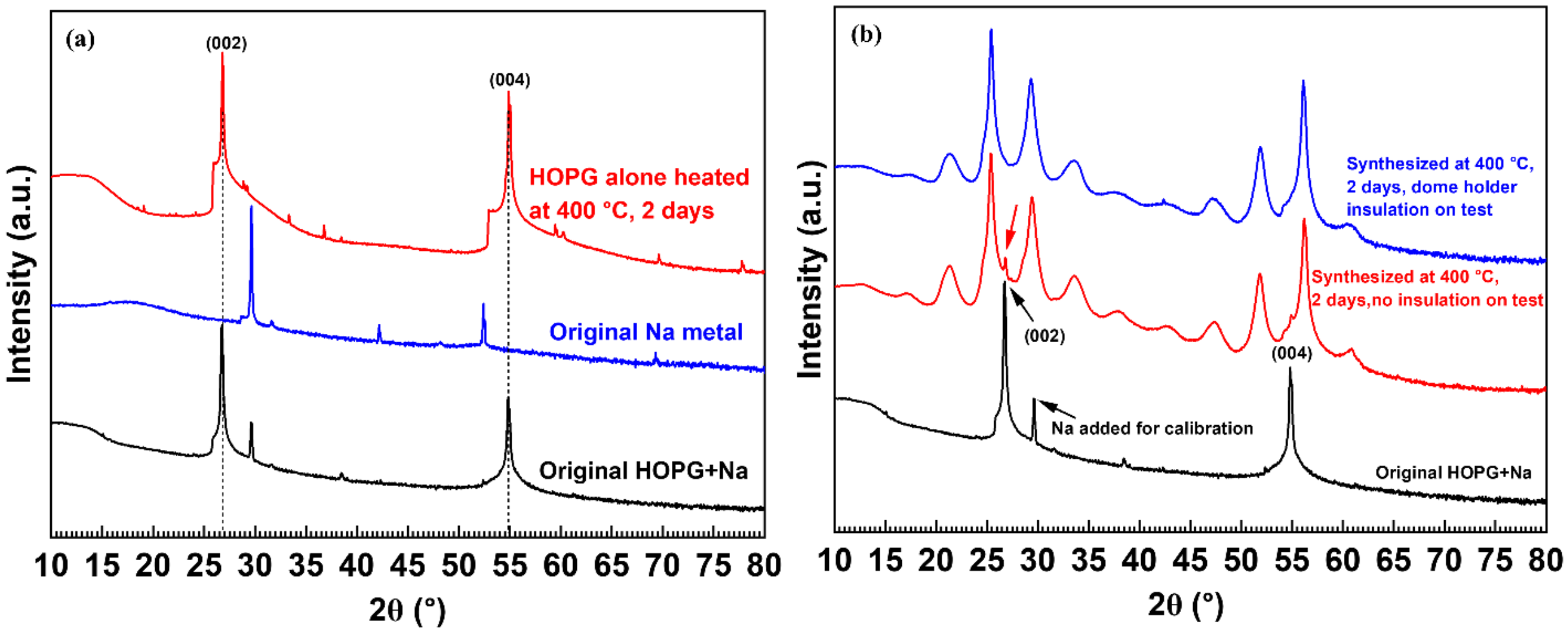

Fig. SⅢ-1 XRD patterns for (a) original Na metal, original HOPG plus Na metal, and pure HOPG after heating treatment at 400 °C of 2 weeks, (b) Na intercalated HOPG samples exposed to/insulated from air during test (Without dome holder insulation during test, there was a weak graphite signal of (002) as marked by red arrow).

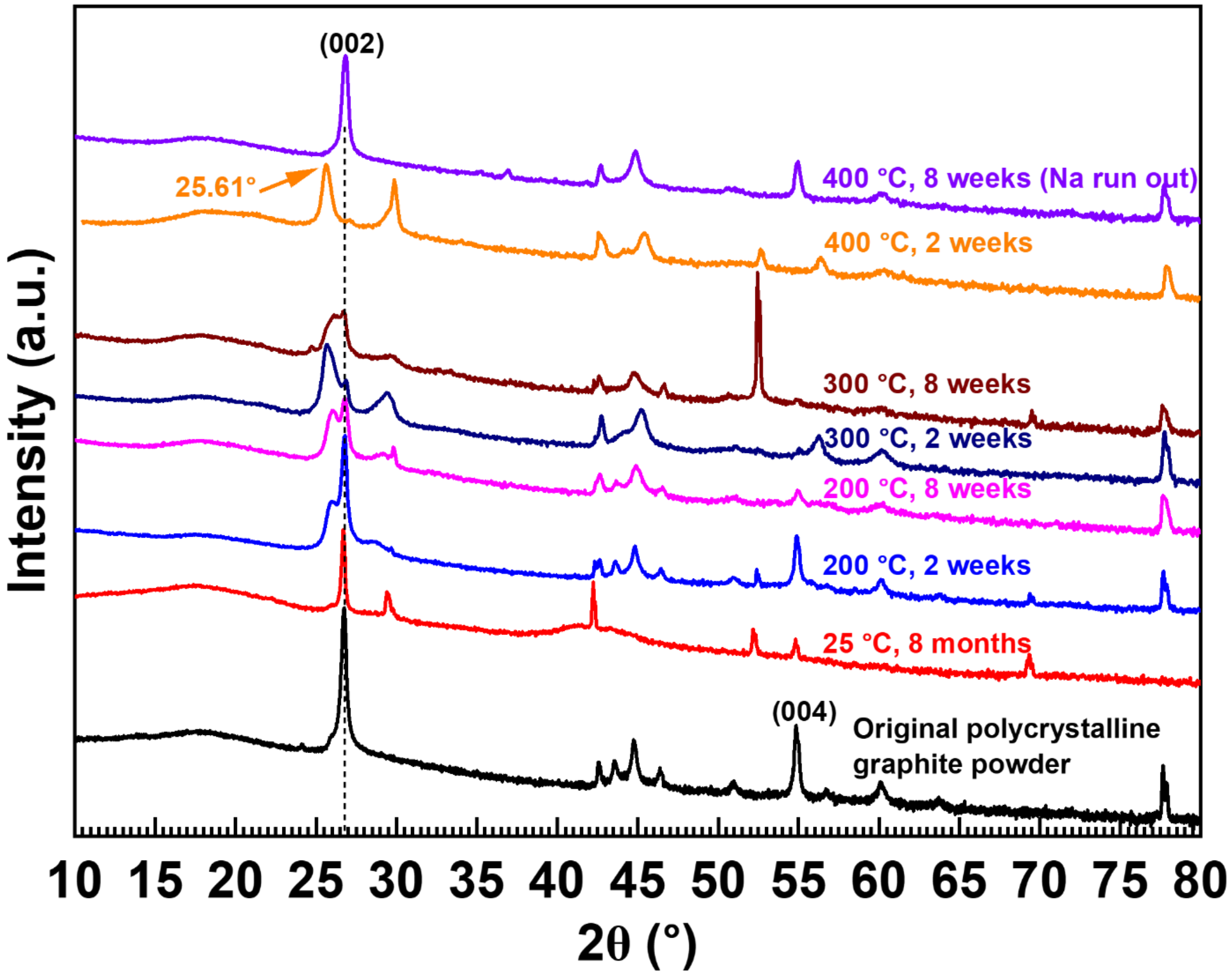

Fig. SⅢ-2 XRD patterns for chemical intercalation of Na into polycrystalline graphite powder (Na metal was added intentionally for the peak position calibration).

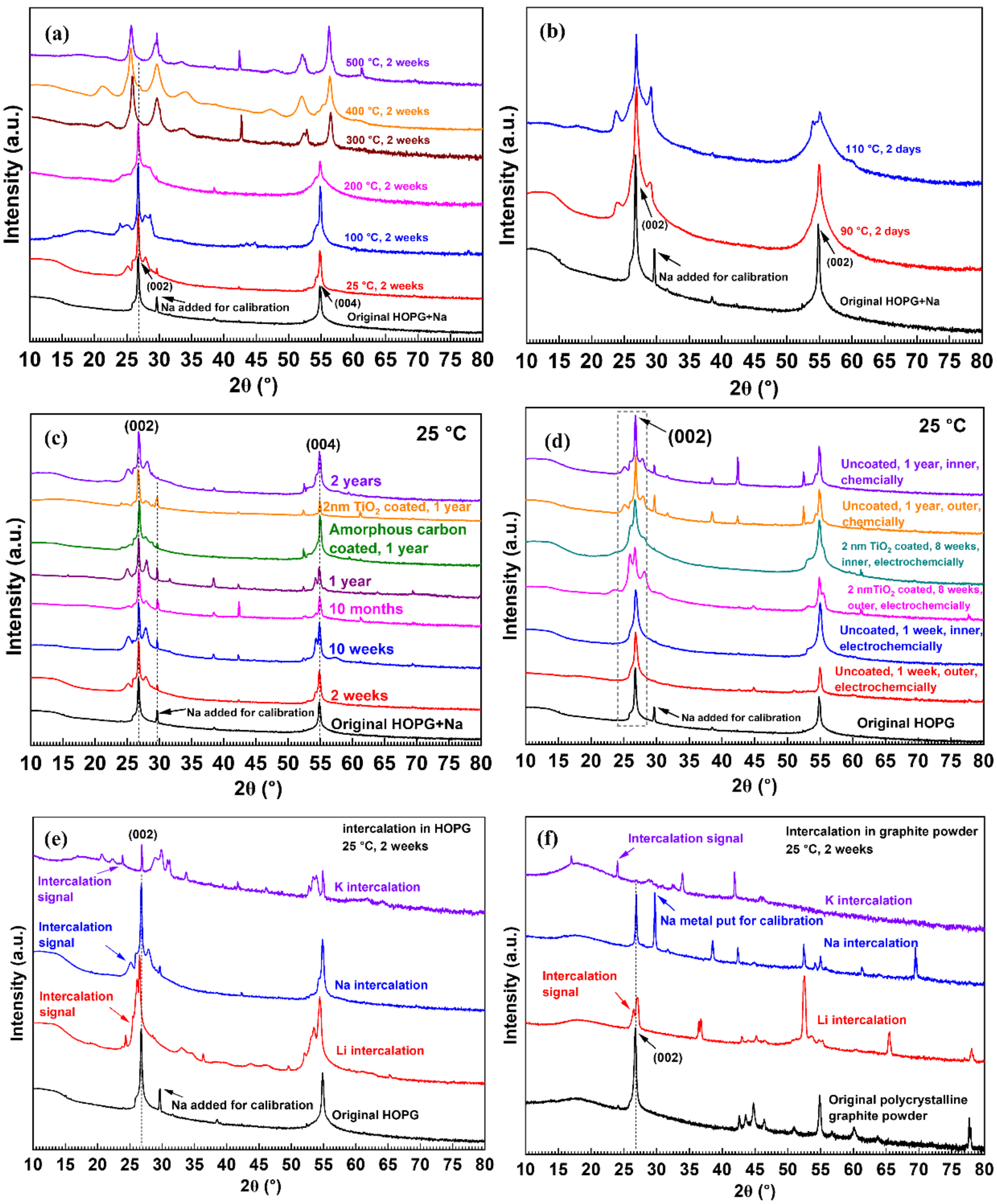

Fig. SⅢ-3. XRD patterns for chemical intercalation of Na into (a, b) HOPG at various temperatures with given time, (c) uncoated and coated ($TiO_2$ or amorphous carbon) HOPG at 25 °C, (d) chemically and electrochemically Na intercalated compounds for comparing the inner part and outer part of sample at 25 °C (dashed grey box shows intercalation signals at lower and higher angles compared to pure (002) peak of HOPG), chemically Li, Na and K intercalation into HOPG (e) and polycrystalline graphite powder (f) for 2 weeks at 25 °C. Na metal was added for the peak position calibration.

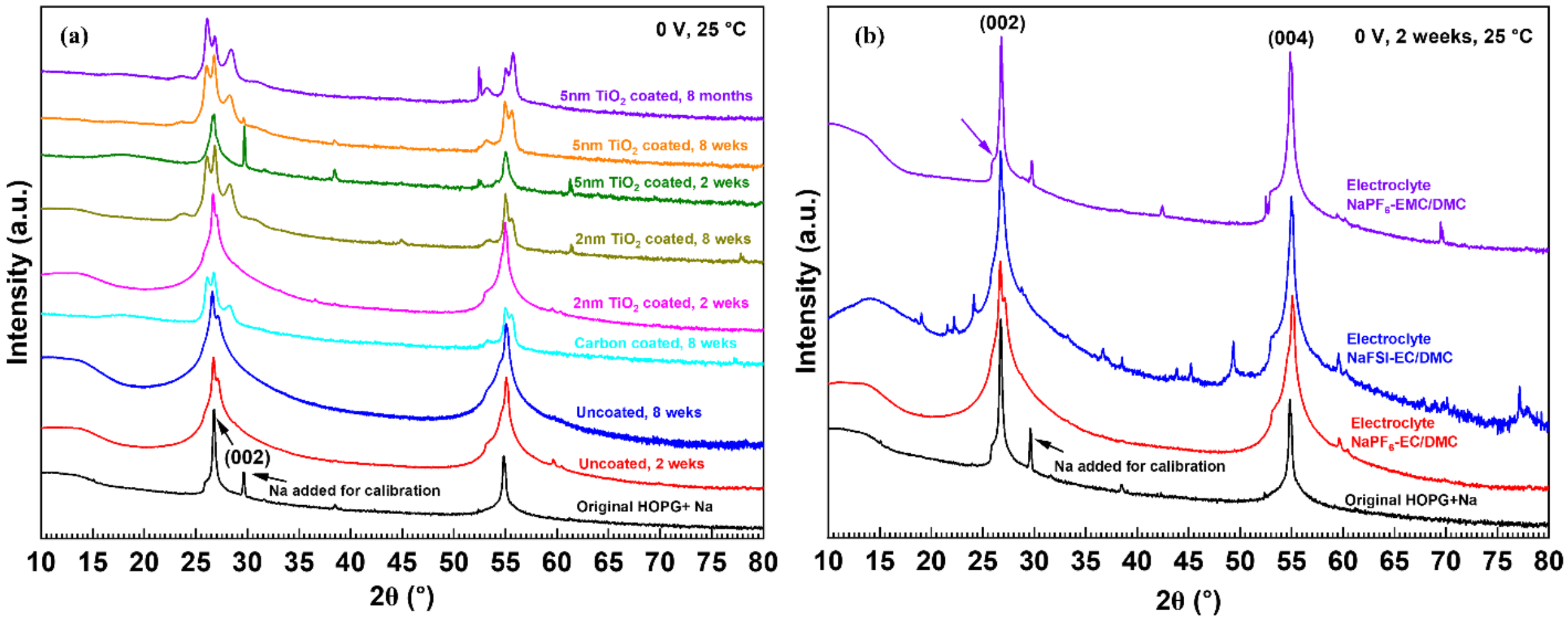

Fig. SⅢ-4 XRD patterns for electrochemical intercalation of Na into (a) $TiO_2$ or amorphous carbon coated HOPG at 0 V vs Na metal at 25 °C, (b) HOPG with different electrolytes for 2 weeks. Note that Na metal was added for the peak position calibration; the violet pattern (using $NaPF_6$-EMC/DEC electrolyte) indicates absence of observable intercalation.

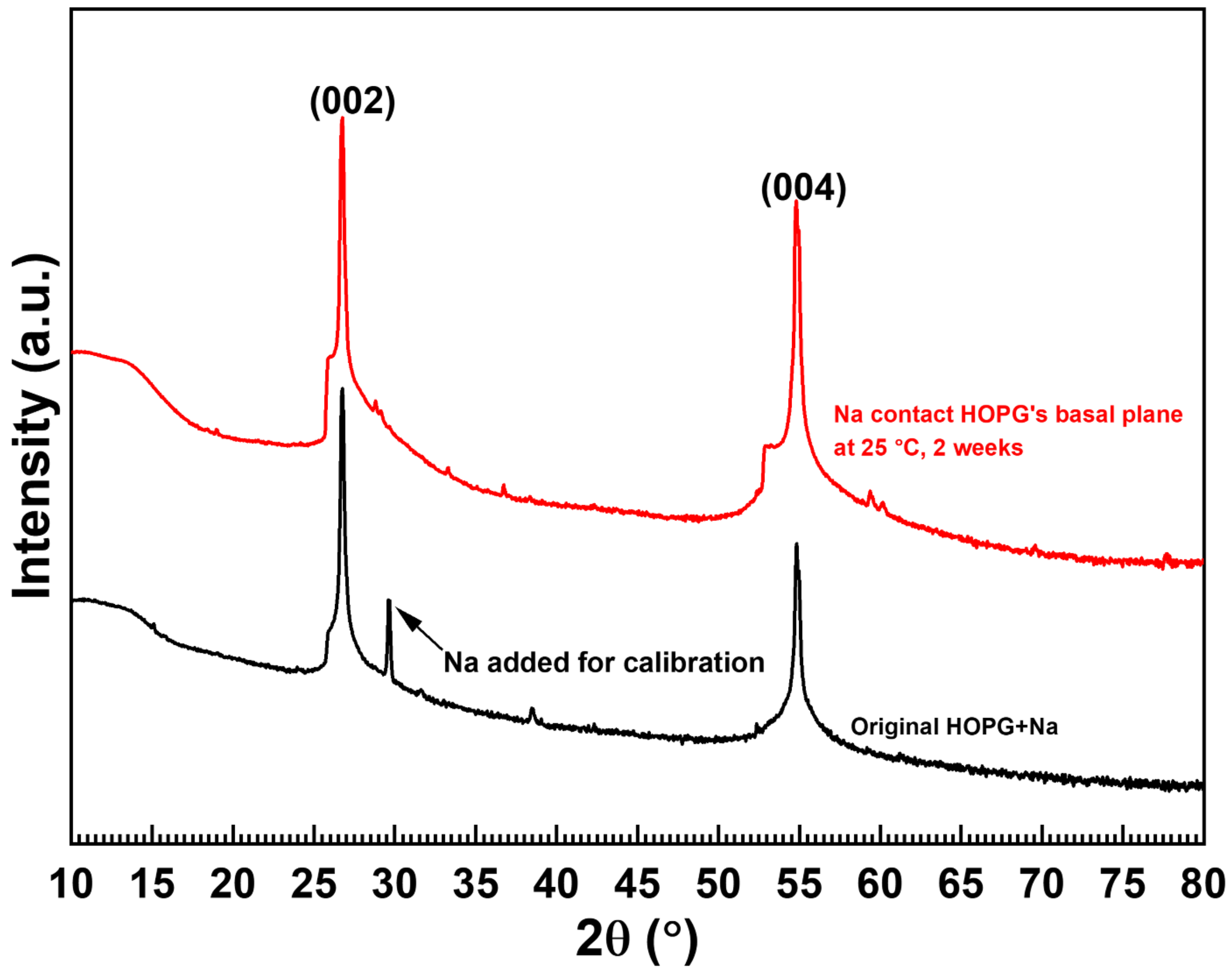

Fig. SⅢ-5. XRD patterns of direct contact of Na on the basal plane of HOPG at 25 °C for 2 weeks. Na metal was added for the peak position calibration.

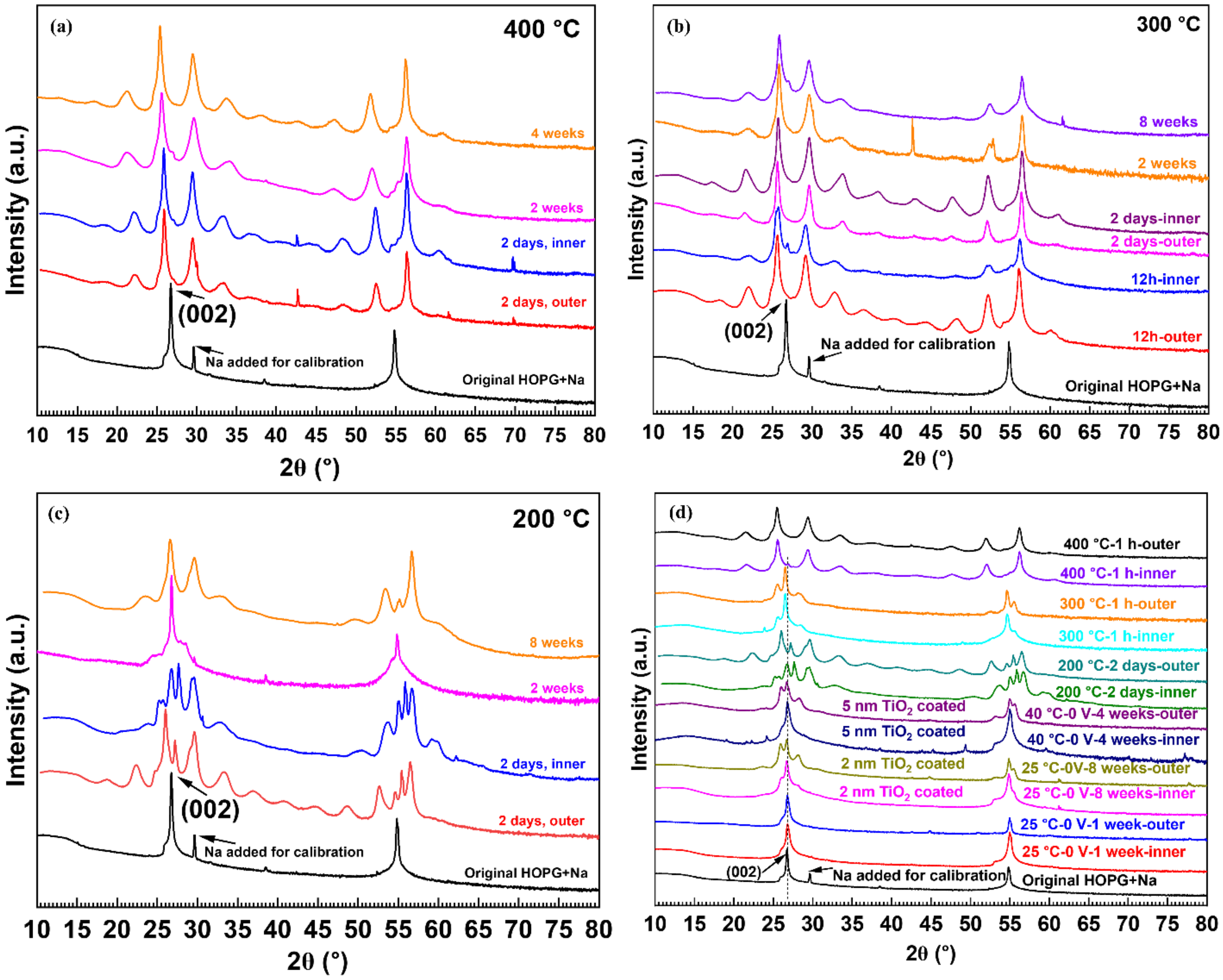

Fig. SⅢ-6. XRD patterns for chemical intercalation of Na into HOPG at (a) 400 °C, (b) 300 °C, (c) 200 °C and (d) much shorter holding time at various temperatures (25 and 40 °C samples were synthesized by electrochemical method. Inner and outer indicate inner and outer parts of sample were tested, respectively. Na metal was added for the peak position calibration).

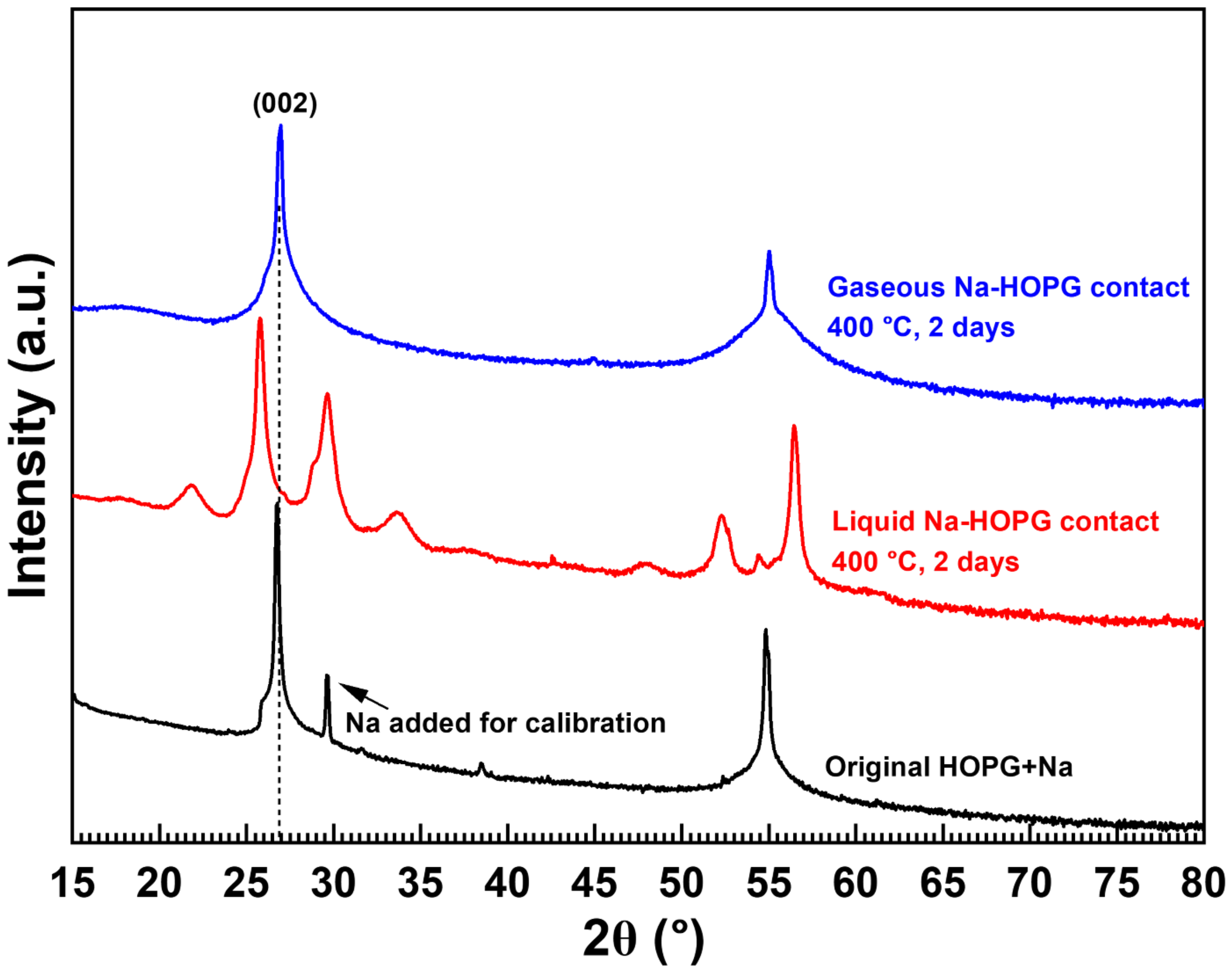

Fig. SⅢ-7. XRD patterns comparison for chemically gaseous and liquid Na intercalation into HOPG at 400 °C for 2 days. Na metal was added for the peak position calibration.

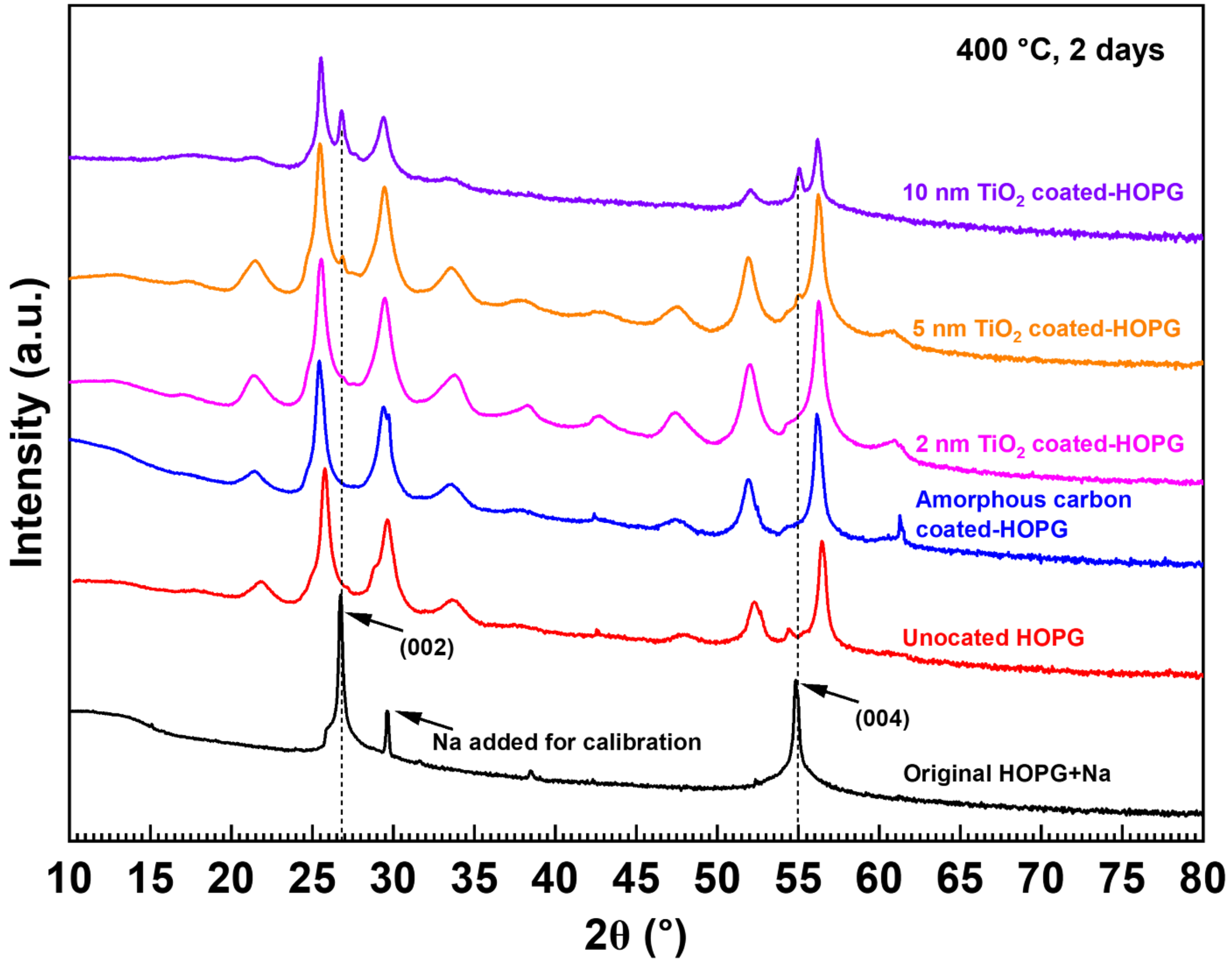

Fig. SⅢ-8. XRD patterns for chemical intercalation of Na into uncoated-, amorphous carbon coated-, and $TiO_2$ (of various thicknesses) coated- HOPG at 400 °C for 2 days. Noting that the (002) graphite peak still exists for the thicker $TiO_2$ coated sample (Na metal was added for the peak position calibration).

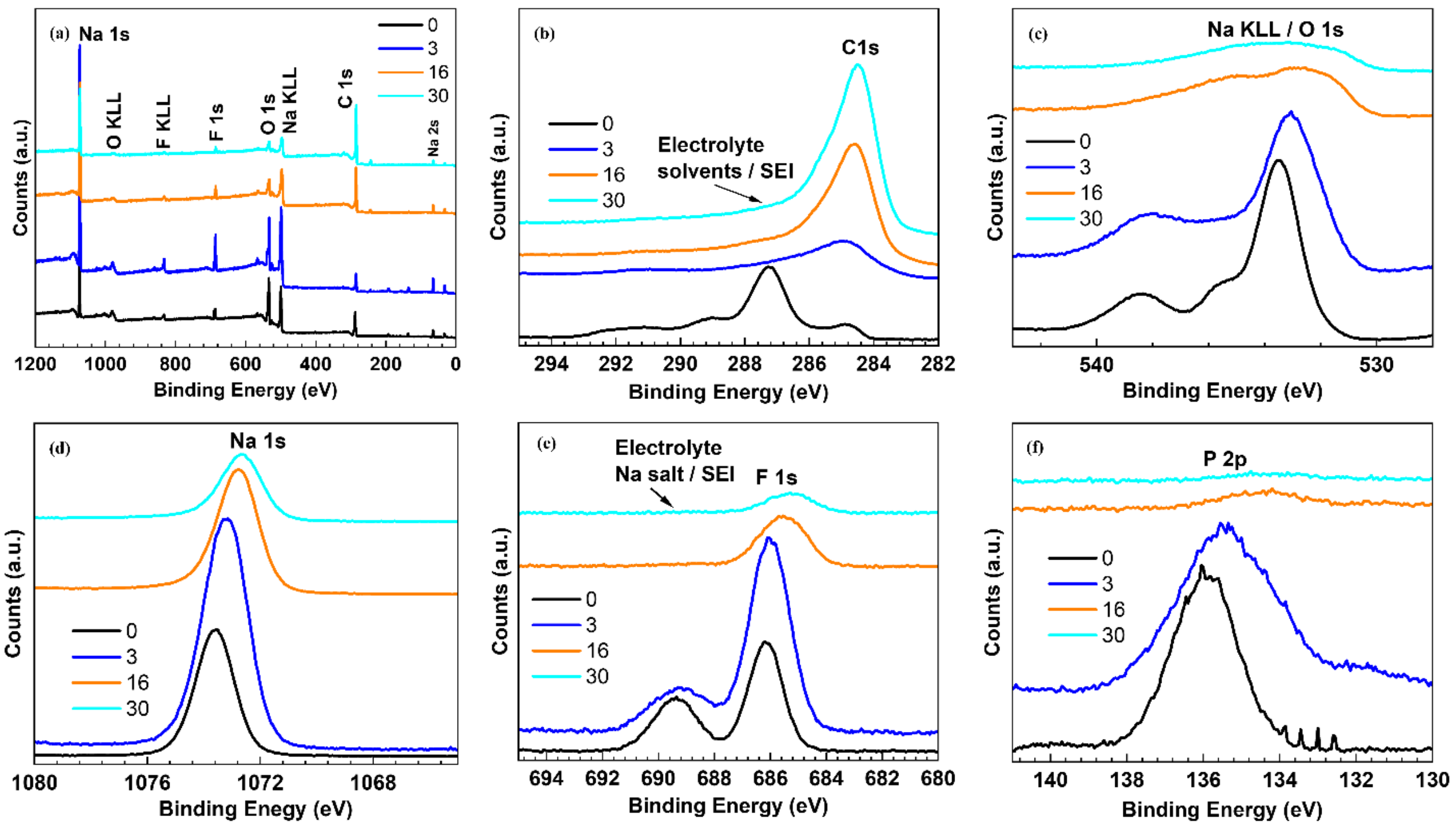

Fig. SⅢ-9 XPS spectrum for the electrochemical intercalation of Na into HOPG at 0 V vs Na with a waiting time of 3 days at 25 °C (The numbers 0, 3, 16 and 30 indicate $Ar^+$ sputtering time in minute, i.e. corresponding detected depth from sample surface).

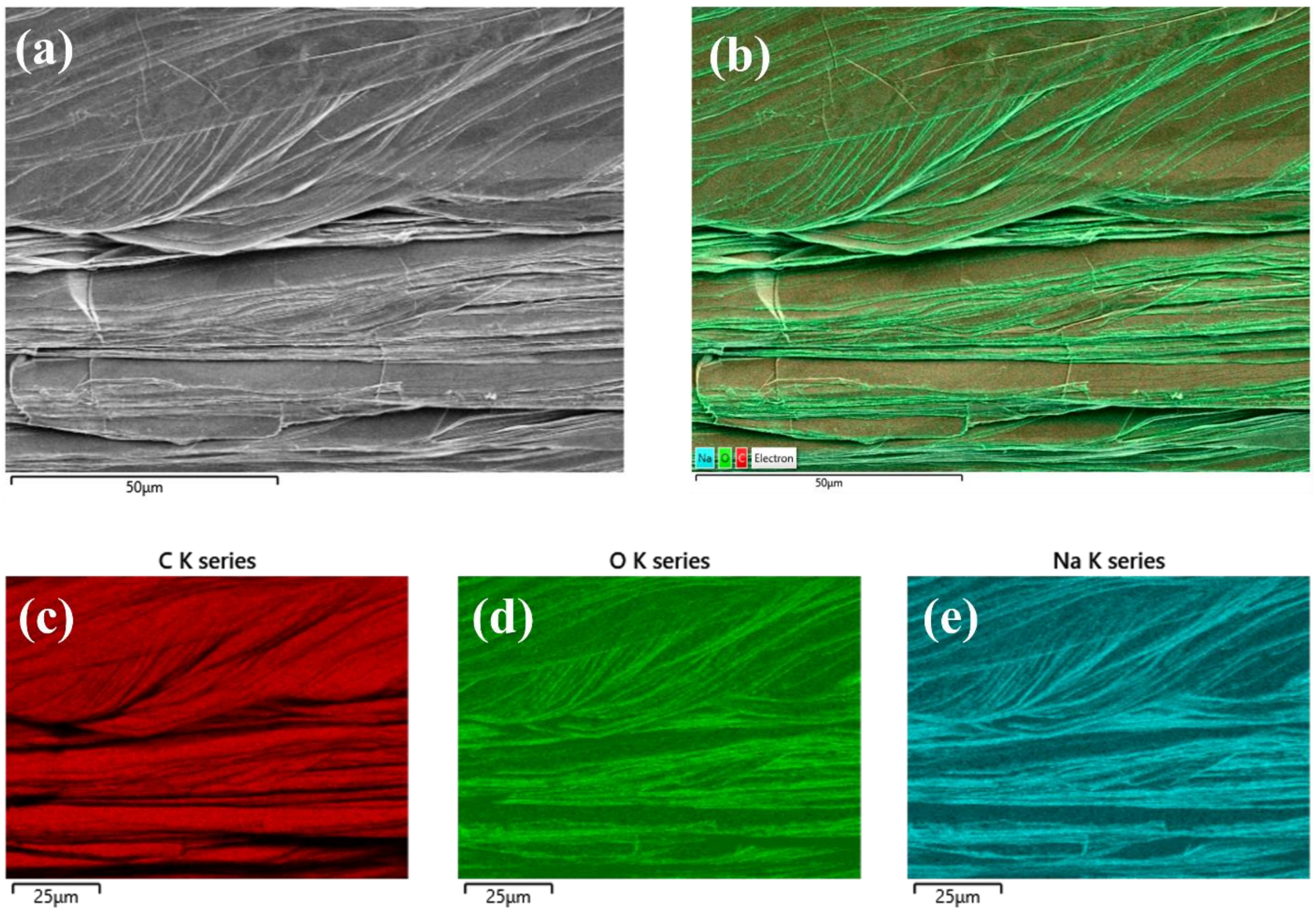

Fig. SⅢ-10. EDX elemental mapping of the cross-section (inner part) of electrochemically Na intercalated HOPG (Electrochemical test cell (coin cell) was assembled in an argon-filled glove box ($O_2 \leq 0.1$ ppm, $H_2O \leq 0.1$ ppm), with HOPG as working electrode, Na metal as the reference electrode. 1 M $NaPF_6$ solution in a 1:1 vol/vol mixture of ethylene carbonate (EC) and dimethyl carbonate (DMC) was utilized as the electrolyte and Whatman® glass microfiber (Grade GF/D) was used as separator. The cell was discharged to 0 V from OCV with a constant current and then held at 0 V for three weeks by using Arbin system (BT 2000). Then the cell was disassembled in an argon-filled glove box and the intercalated HOPG was washed using solvent DMC in order to remove the electrolyte residue. The electrochemically intercalated HOPG was cut into two pieces. The samples were mounted to a SEM holder in a glove box and transferred to SEM chamber in the protective environment using a vacuum transfer tool. The fresh cross-section of inner part was measured.).

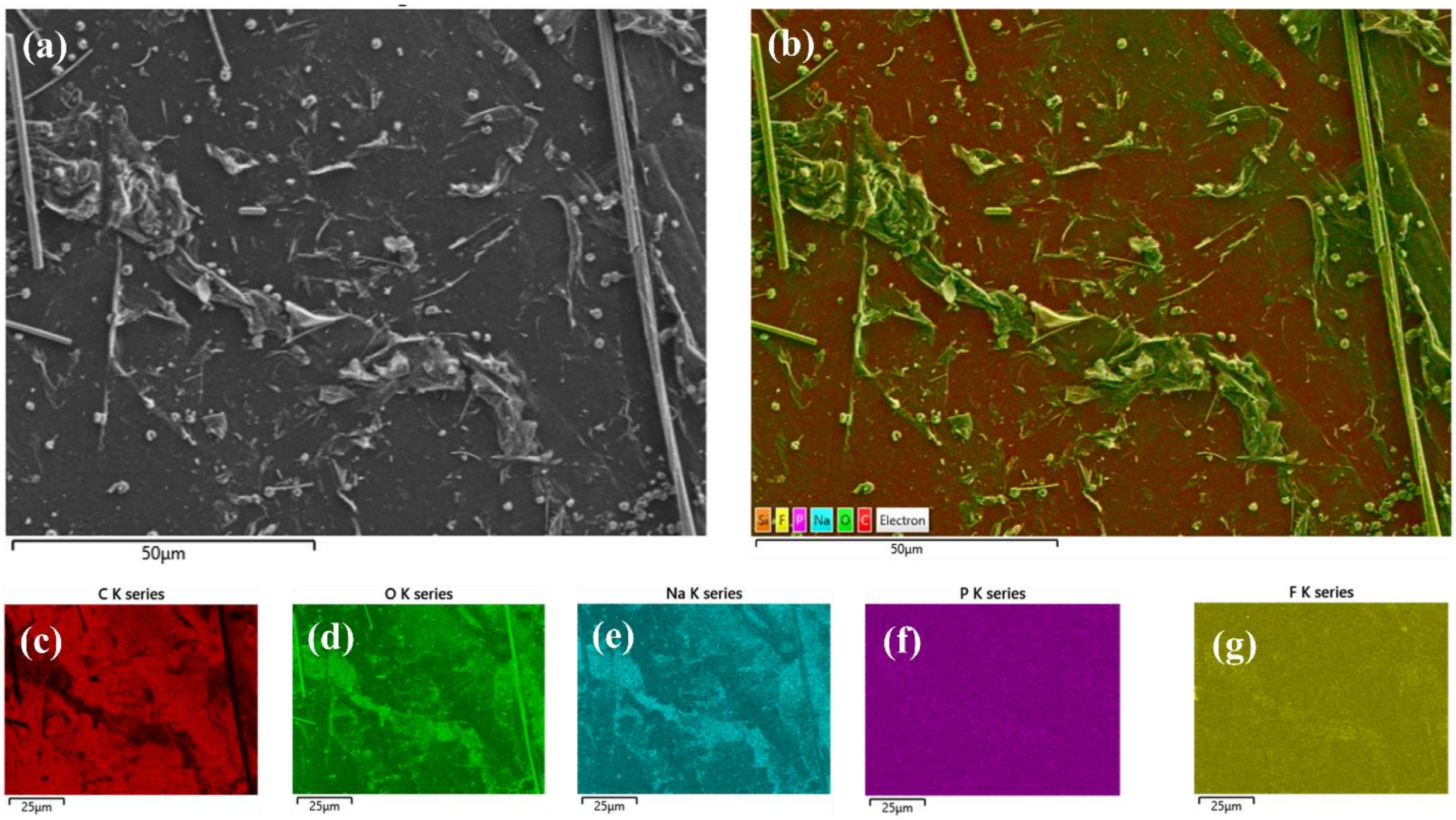

Fig. SⅢ-11. EDX elemental mapping of the top surface (basal plane) of electrochemically intercalated HOPG (for detailed sample preparation procedure, see Fig. SⅢ-10.).

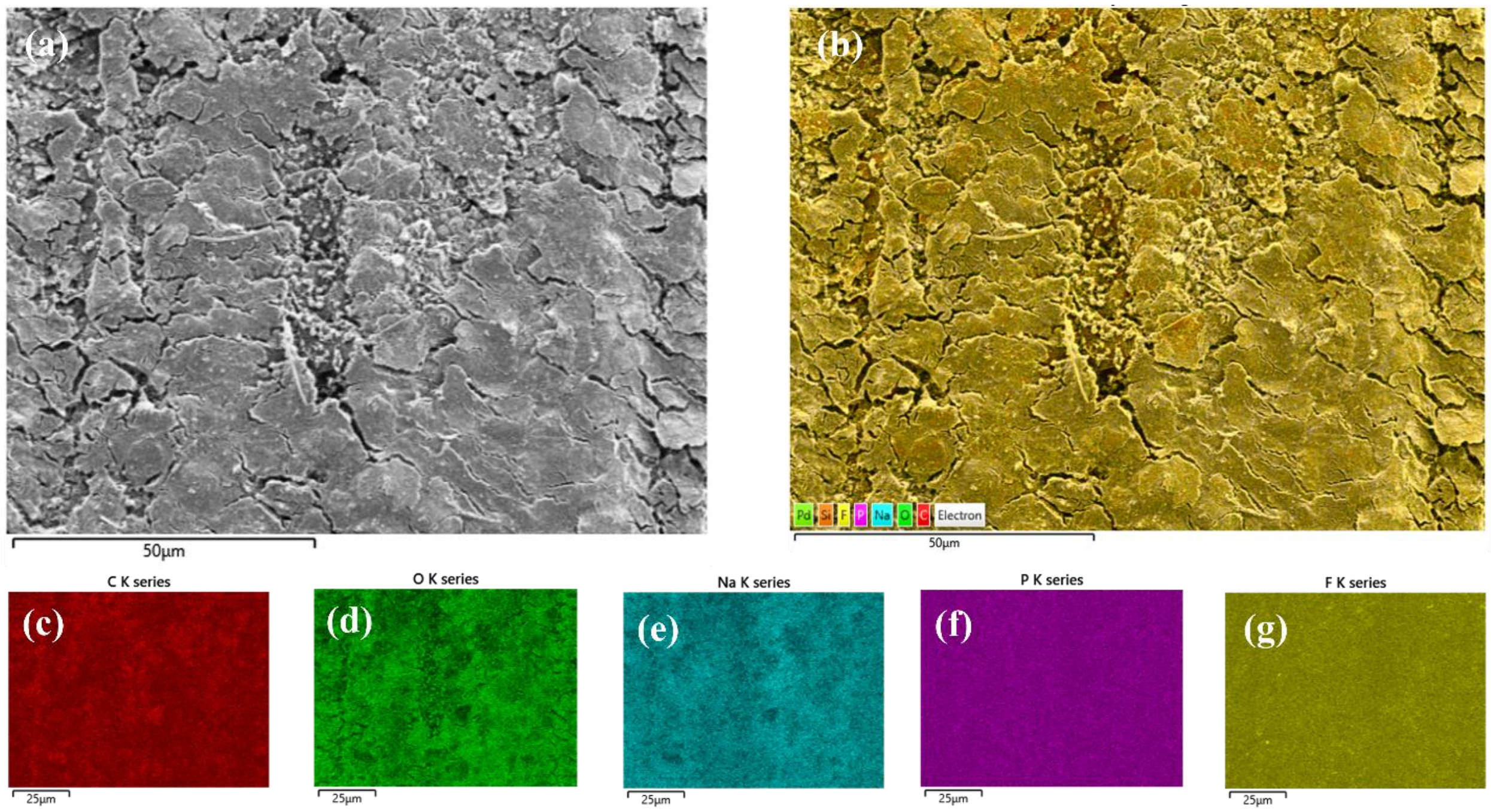

Fig. SⅢ-12. EDX elemental mapping of the cross-section (outer side-surface) of electrochemically intercalated HOPG (for detailed sample preparation procedure, see Fig. SⅢ-10.).

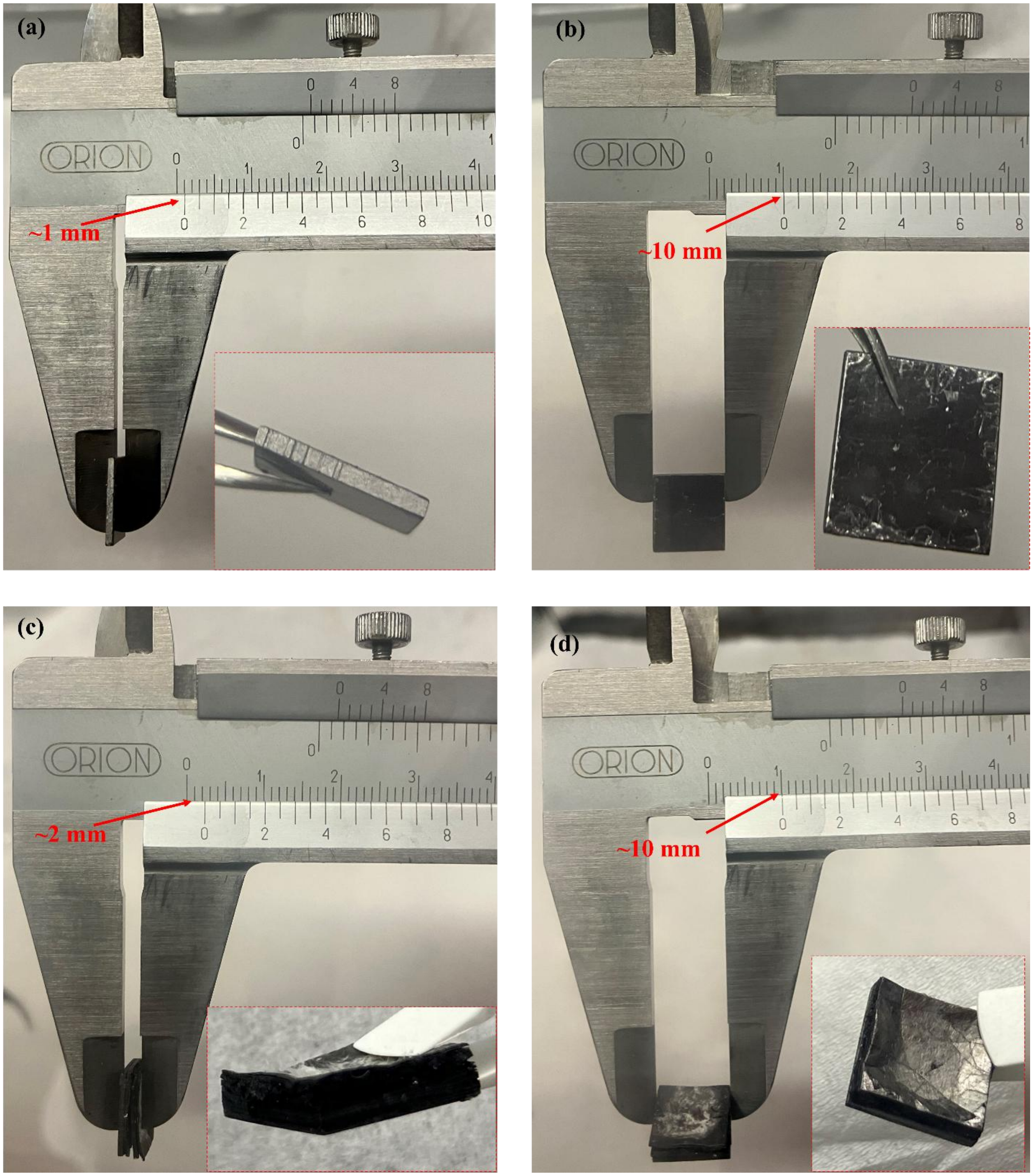

Fig. SⅢ-13. Photos of original HOPG (a, b), and treated HOPG (electrochemical intercalation of Na into HOPG at 0 V vs Na for 5 months at 25 °C) (c, d). The comparison with the scale indicates the expansion.

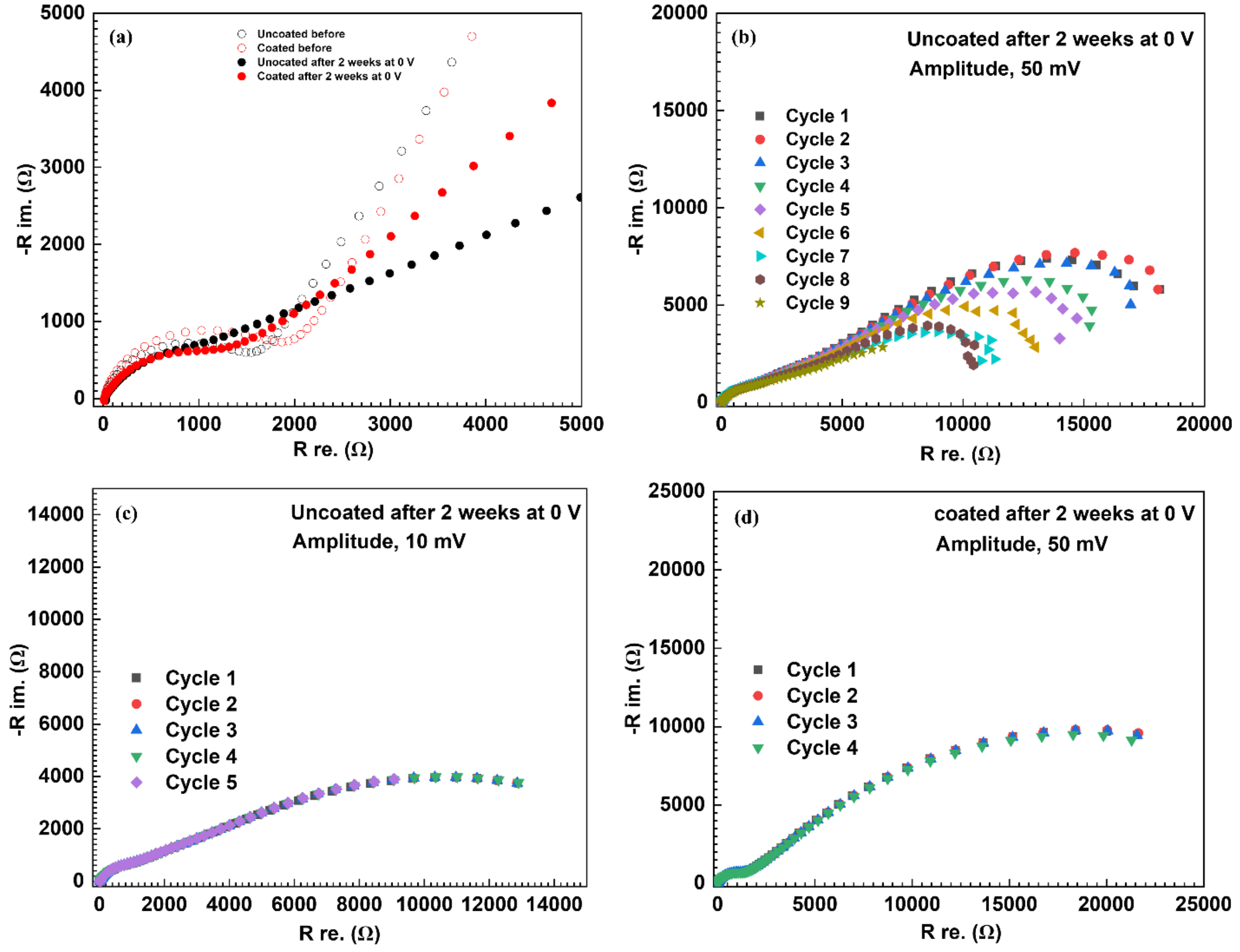

Fig. SⅢ-14. EIS spectra for electrochemical intercalation of Na at 0 V vs Na at 25 °C into uncoated- and 5 nm $TiO_2$ coated- HOPG. (a) Before and after Na intercalation applying 50, 10 mV (AC output amplitudes), respectively, (b) uncoated test sample and 50 mV, (c) uncoated sample and 10 mV, (d) coated sample and 50 mV.

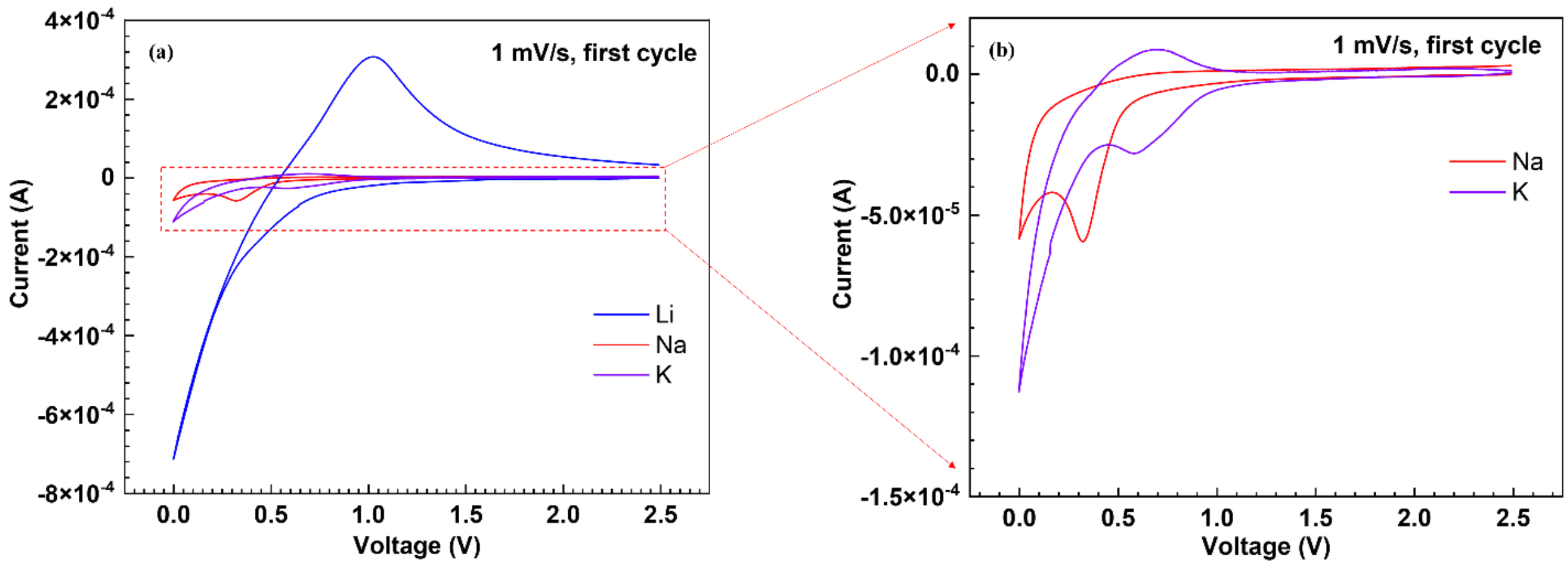

Fig. SⅢ-15. CV for Li-, Na- and K- intercalation into HOPG at 25 °C.

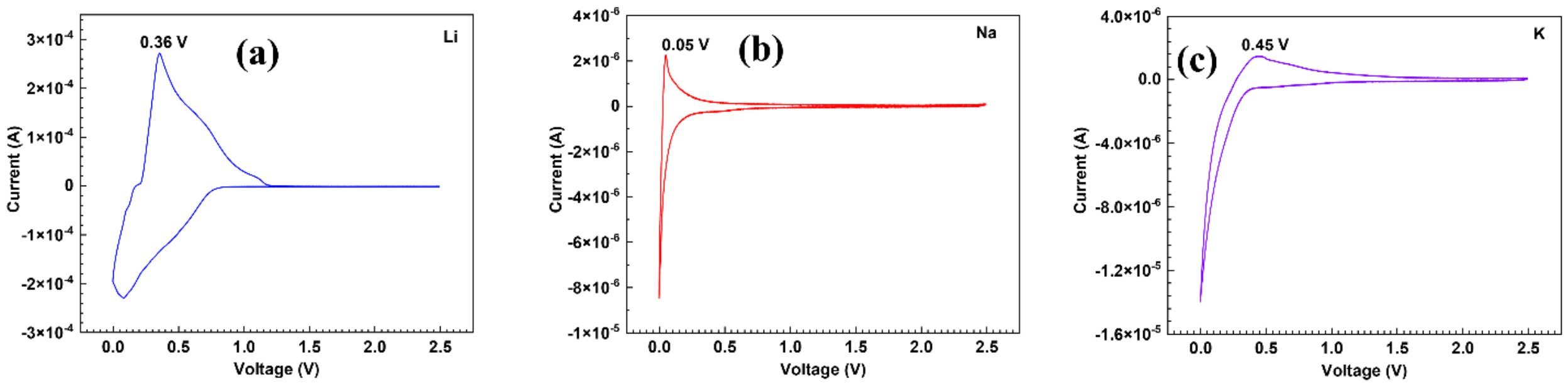

Fig. SⅢ-16. Cyclic voltammograms (at 25 °C) at a sweep rate of 20 μV/s for (a) Li- HOPG, (b) Na-HOPG, and (c) K- HOPG. The peak current potentials of de-intercalation are 360 mV, 50 mV, and 450 mV for Li-, Na- and K- HOPG, respectively. The true OCV values should be in between de-intercalation and intercalation (close to 0 V) potentials.

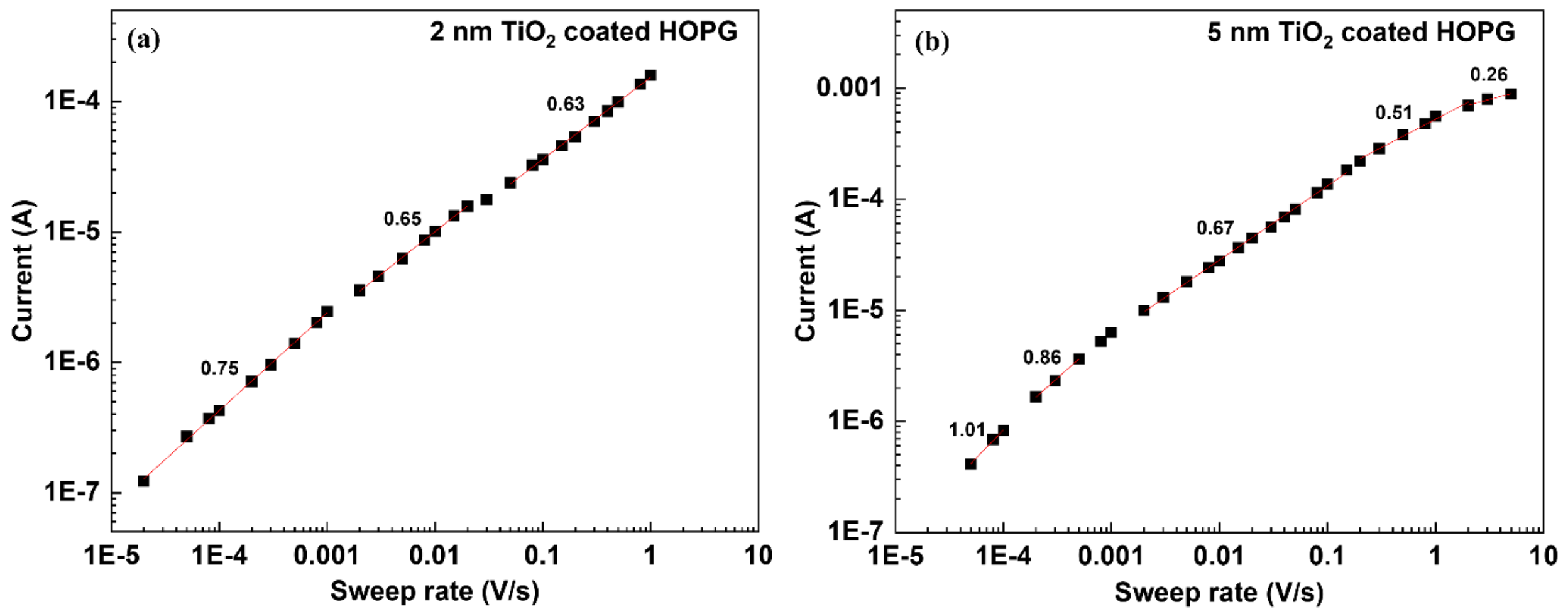

Fig. SⅢ-17. Peak current at various sweep rates ν of CV measurements for the 2 nm- and 5 nm-$TiO_2$ coated HOPG.

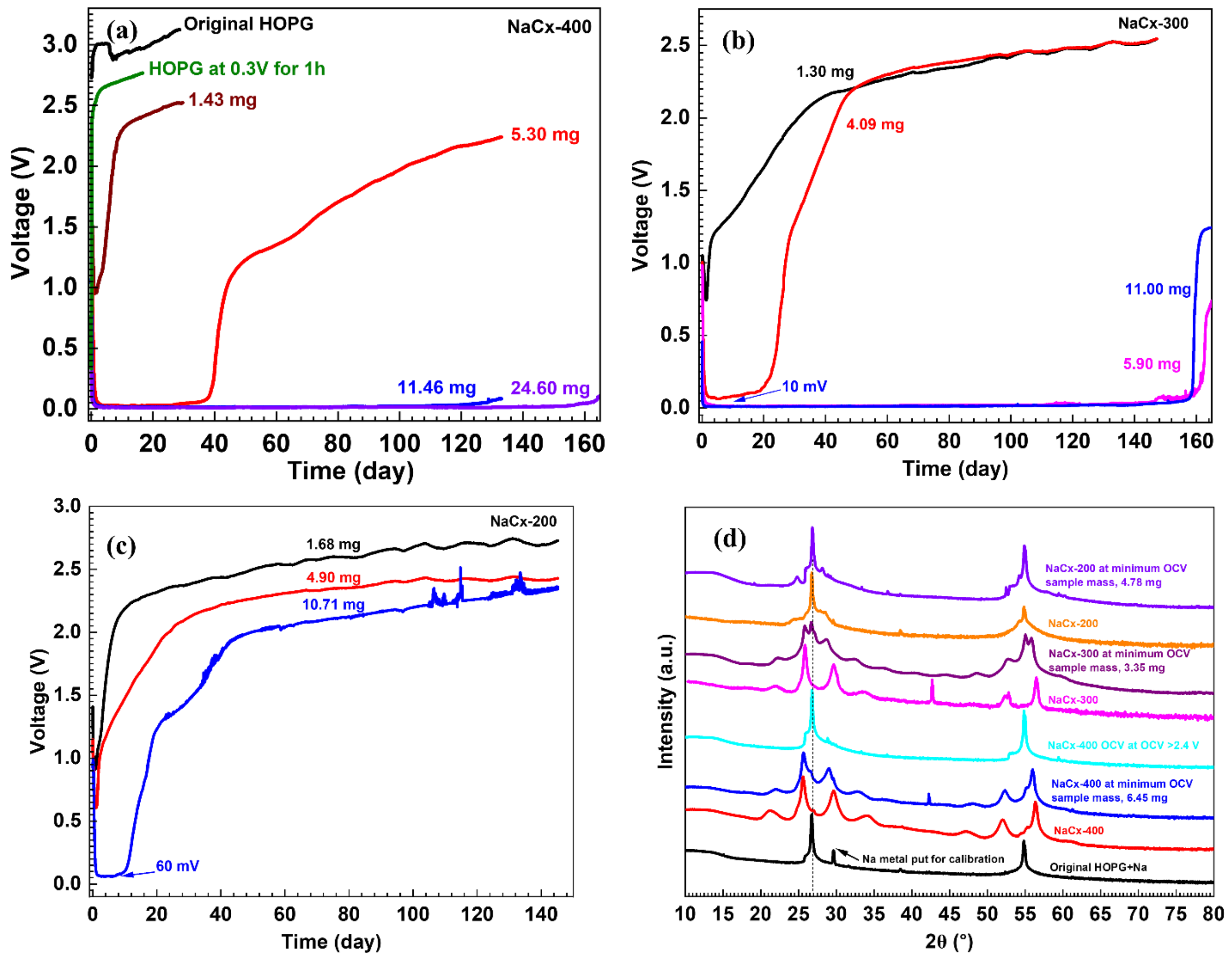

Fig. SⅢ-18. (a, b, c) Open circuit voltages (OCV) for chemical intercalation of Na into HOPG at 400 ($NaC_x$-400), 300 ($NaC_x$-300), and 200 °C ($NaC_x$-200), respectively. (d) XRD patterns for samples held at various states (minimum OCV denote stable OCV, e.g. 10 mV for $NaC_x$-300). Noting that both OCV and XRD measurements carried out at 25 °C.

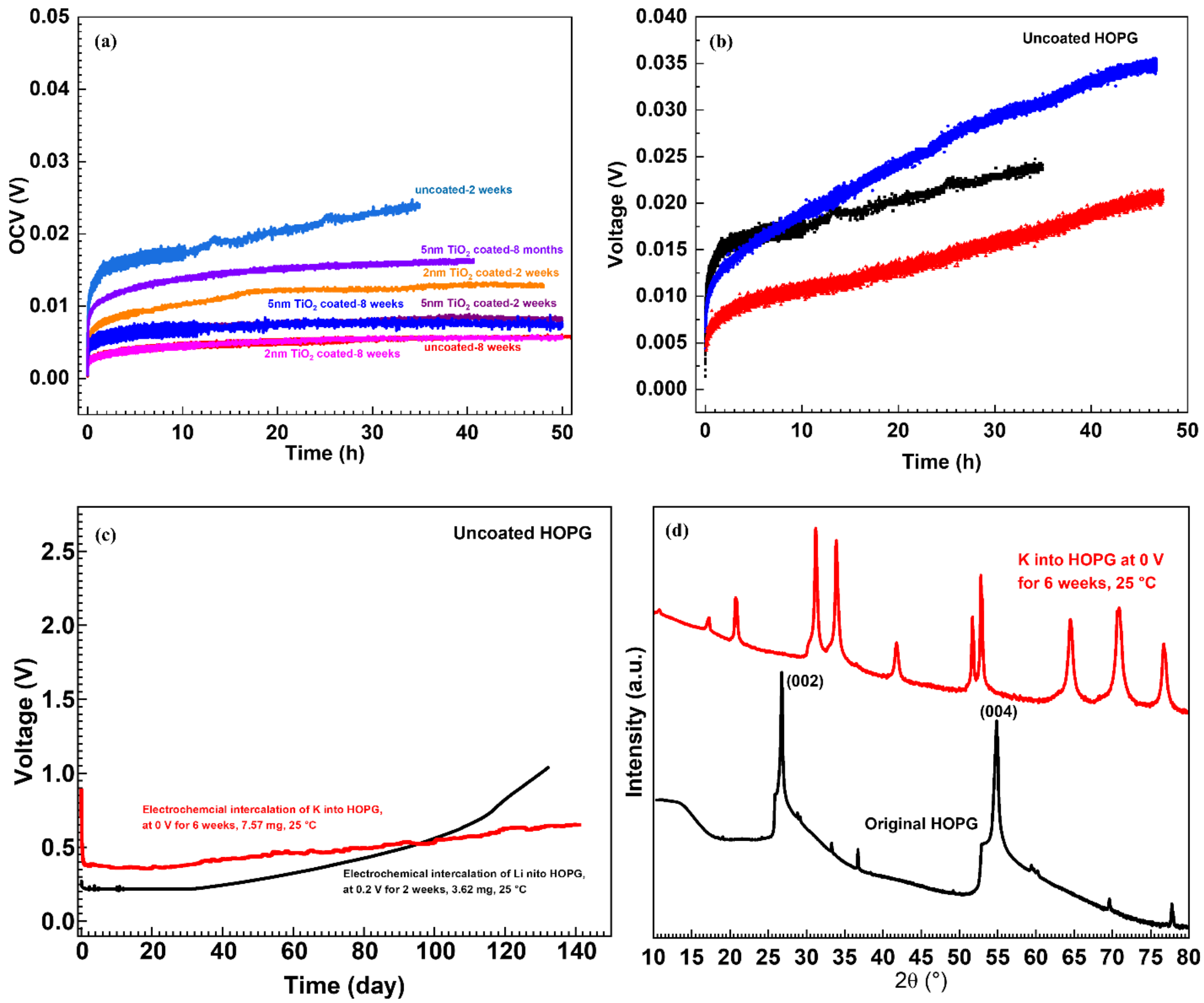

Fig. SⅢ-19. Open circuit voltages (OCV) for samples synthesized by electrochemical intercalation of Na (a, b) into uncoated and coated HOPG, (c) Li and K into HOPG, at 0 V at 25 °C. (d) XRD patterns corresponding to K intercalated HOPG of (c). The XRD pattern of Li intercalated HOPG is shown in Fig. SⅢ-20 below.

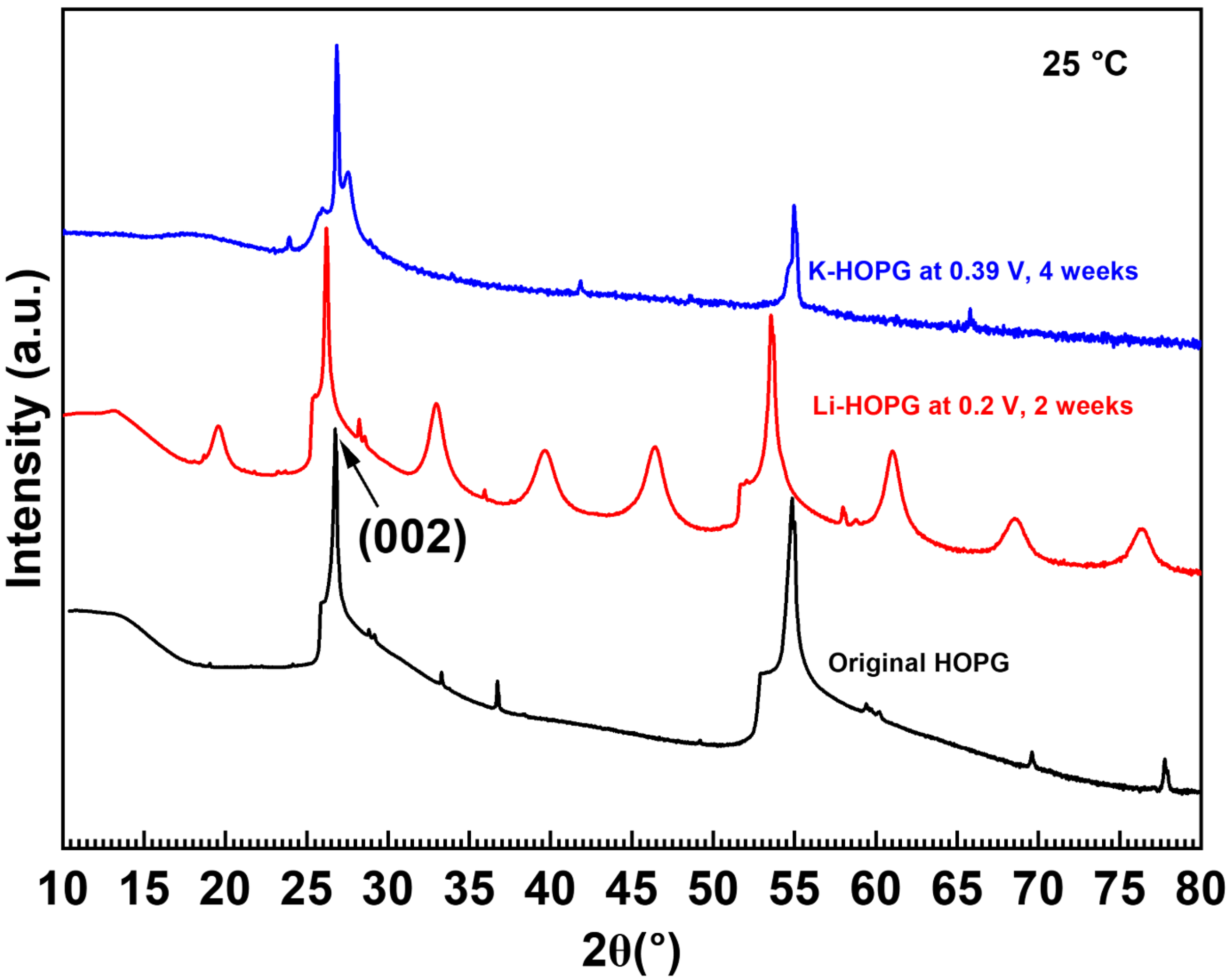

Fig. SⅢ-20. XRD patterns for electrochemical intercalation of Li and K into HOPG at 0.2 V and 0.39 V at 25 °C, respectively (note that (002) plane of graphite disappears after 2 weeks in Li case while not in K case, indicating K intercalation kinetics is poorer than Li intercalation).

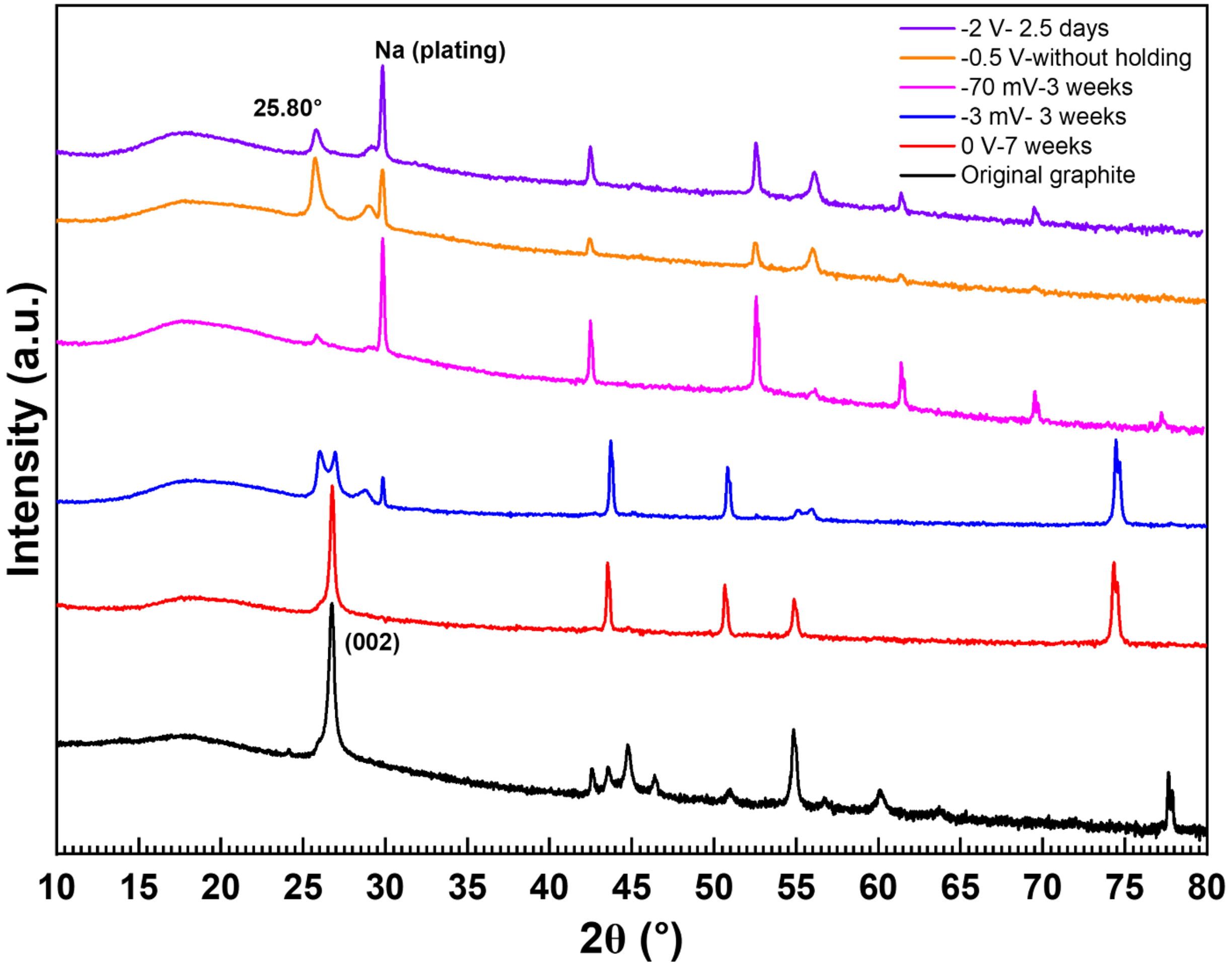

Fig. SⅢ-21. XRD patterns for electrochemical intercalation of Na into polycrystalline graphite powder at negative voltages at 25 °C (the intercalation was conducted by discharging the cell to 0 V vs Na with a constant current, followed by potential sweep control of -0.05 mV/s to certain voltages, then holding at the given voltages, e.g. -0.50 V).

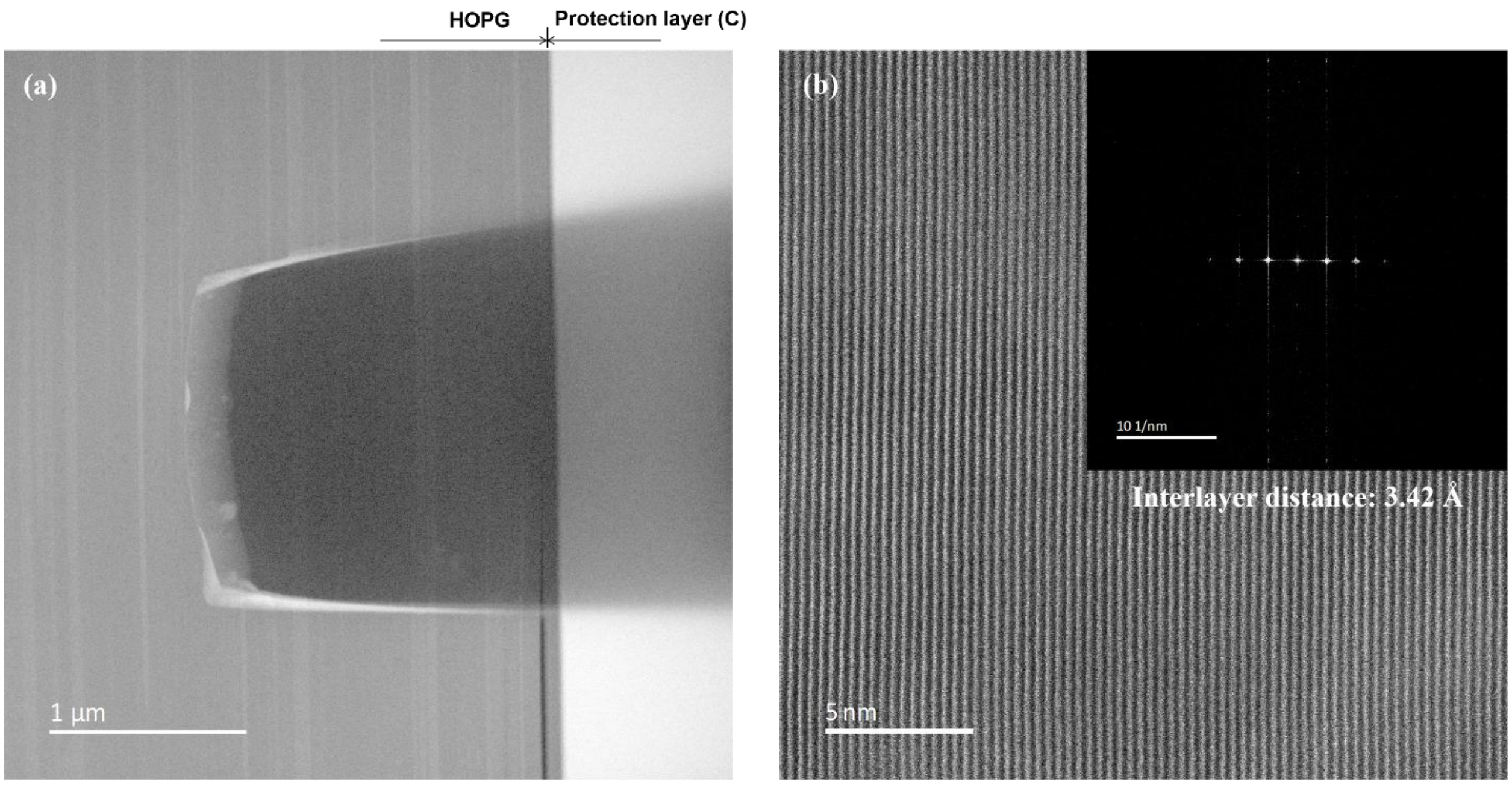

Fig. SⅢ-22. Original HOPG. (a) Low-magnification TEM image, (b) HAADF-STEM image (inset, Fast Fourier transform image).

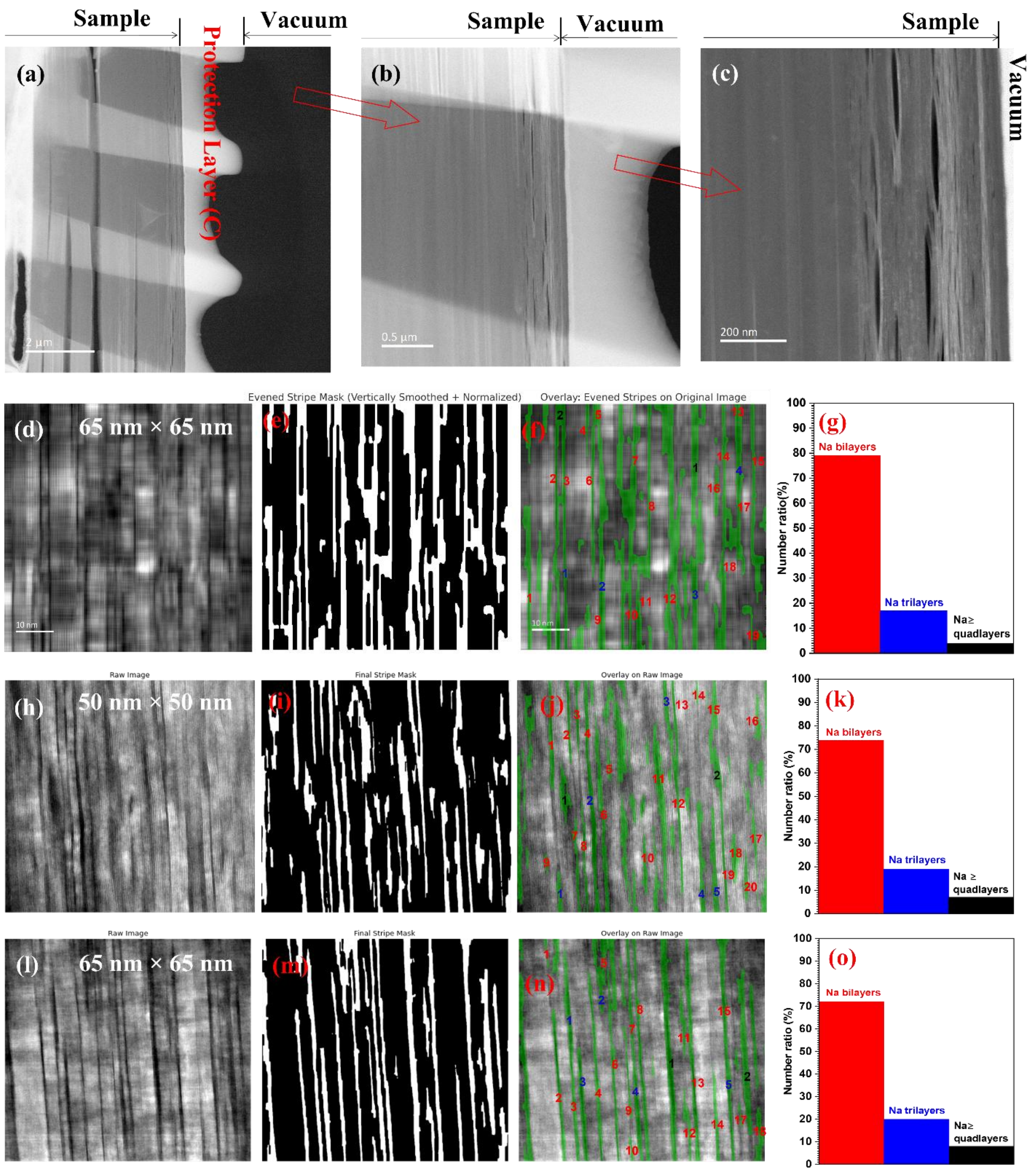

Fig. SⅢ-23. Na intercalated into HOPG compounds chemically synthesized at 400 °C with a waiting time of 2 weeks. (a-c) Low-magnification TEM image, (d, h, l) ADF-STEM images for representative regions, (e-g, i-k, m-o) statistics of number of Na bilayers, trilayers and ≥ quadlayers (note in images (f, j, n), red, blue and black numbers denote Na bilayers, trilayers and ≥ quadlayers counted, respectively).

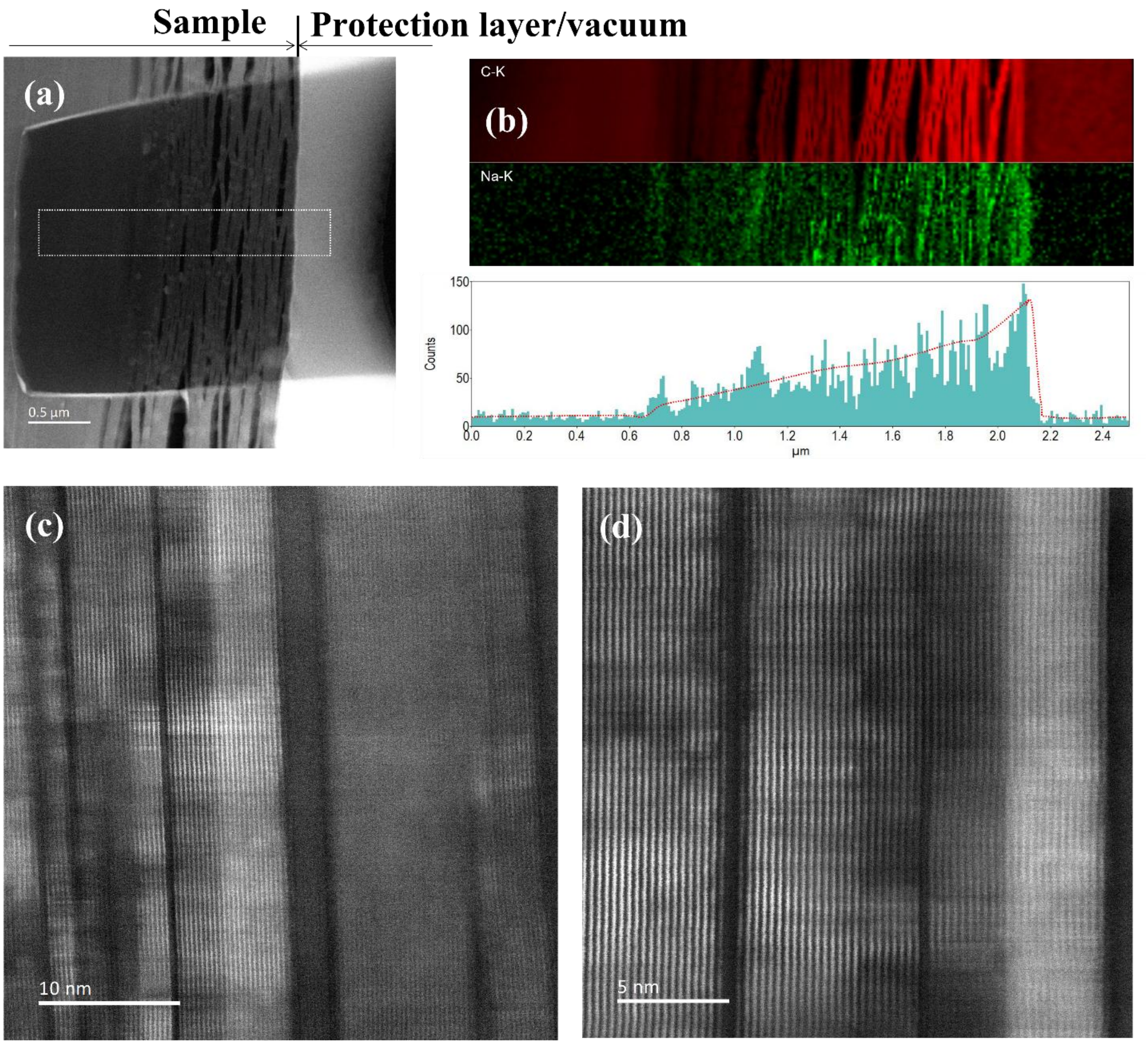

Fig. SⅢ-24. Na intercalated into HOPG compounds chemically synthesized at 25 °C with a waiting time of 10 months. (a) Low-magnification TEM image, (b) EELS mapping, (c, d) HAADF-STEM images.

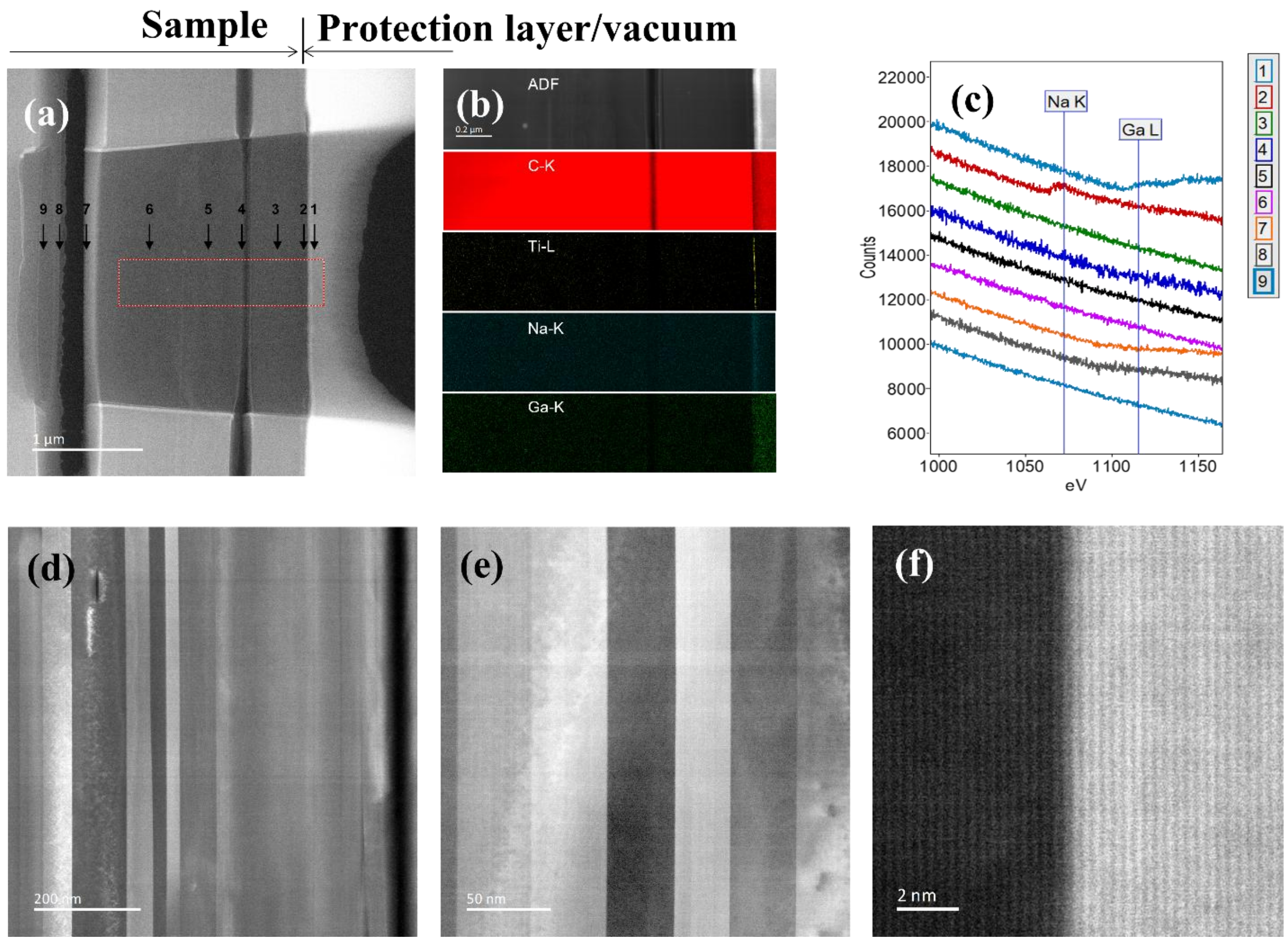

Fig. SⅢ-25. Na intercalated into 5 nm $TiO_2$ coated-HOPG compounds electrochemically synthesized at 25 °C, 0 V vs Na with a waiting time of 2 weeks. (a) Low-magnification TEM image, (b) EEELS mapping, (c) EELS spectra, (d, e, f) ADF-STEM images (note the bright and dark stripes arise from variations in the stacking configuration between layers, not intercalation signal. There is no confirmed Na intercalation signal based on EELS or distance of carbon-carbon interlayer which is attributed to extremely low Na concentration, but Na intercalation can be concluded by combing the cracks in image (a) and XRD data (Fig. SⅢ-4(a)), titration analysis (Table SⅡ-1) as well. Thus, it can be concluded that weakly Na intercalated in random distribution).

## Part Ⅳ XRD Simulations and Analyses

XRD pattern of the synthesized samples were recorded in Bragg Brentano geometry under inert atmosphere in order to avoid sample damage and degradation. The inert atmosphere sample holder did not allow for sample rotation. Hence, all XRD data suffered from preferred orientation. Moreover, it was not possible to adjust the z-position of the sample holder to the height of the film, this led to a sample height error, which artificially introduced a shift of the peak position. This can be clearly observed by the mismatch of the Na 011 peak in the XRD patterns of pure Na metal and Na metal admixed with original HOPG (Fig. SⅣ-1). Hence fully weighted Rietveld refinement cannot be performed on these samples and we resorted to a qualitative comparison with simulated XRD patterns. For the XRD simulations, we have to make assumptions: The intercalation of Na in-between graphite layers, leads to an increase in interlayer distance. However, there are no reliable data on this increase in interlayer distance available in literature. Hence, we use 4.02 Å, as observed by TEM measurement, as the interlayer distance of Na intercalated HOPG (see Fig. 3(c), Main Text). In addition, we assume that Na is situated above and below the centers of the $C_{6/3}$ rings and therefore, Na intercalation alters the ABAB stacking order of hexagonal graphite locally to an eclipsed AAAA stacking.

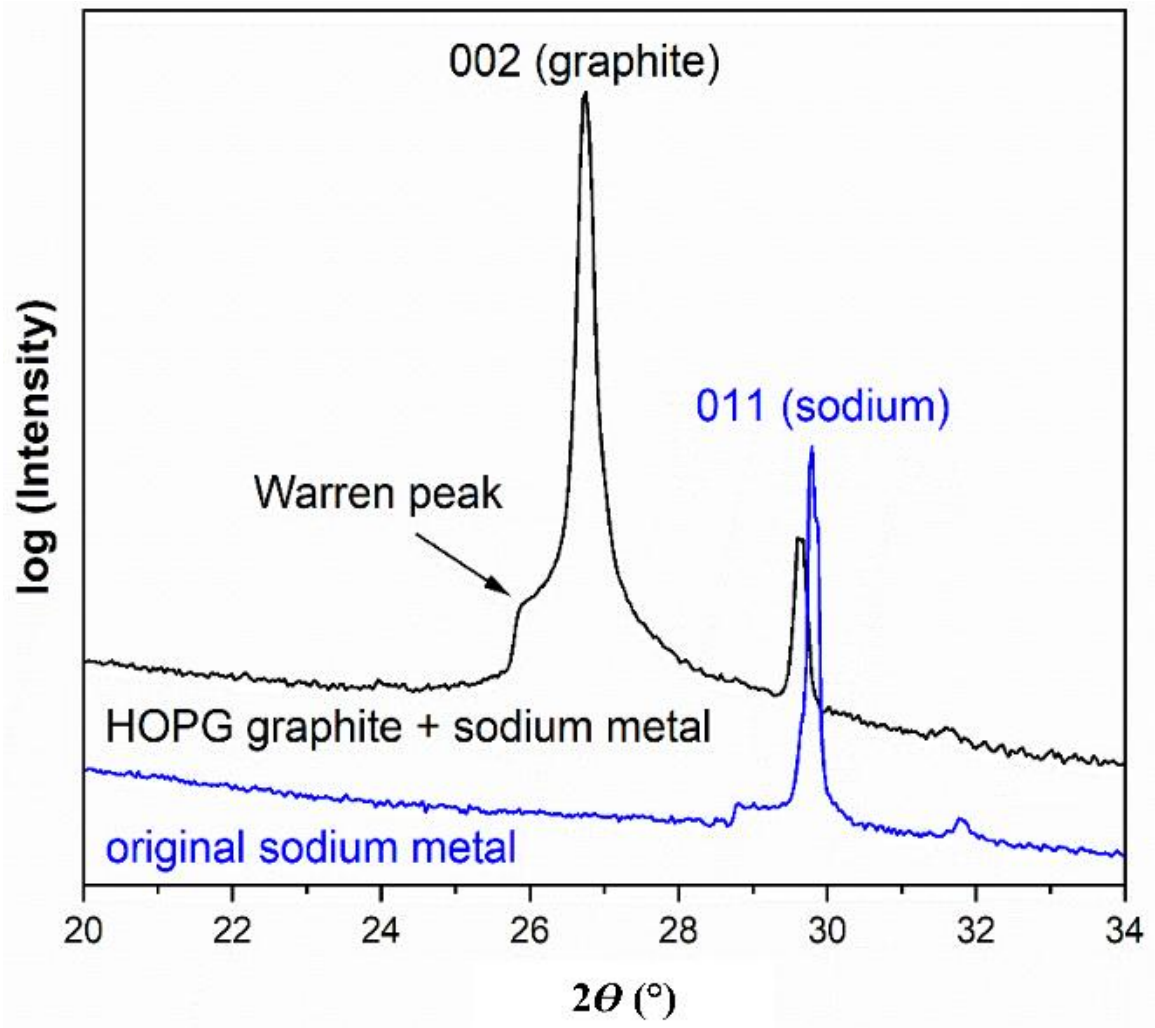

Fig. SⅣ-1. Comparison of the XRD pattern of pure Na metal and Na metal admixed with HOPG including selected reflection indices. The small, symmetry forbidden, trigonally shaped reflection in HOPG, denoted as “Warren peak”[5], indicates the presence of stacking disorder[6] in HOPG.

In the first series of simulations, it refers to random intercalation, i.e. no staging of ordered or disordered Na layers in-between graphite sheets, applying an intercalation probability of 25 %. Both scenarios lead to a considerable downshift (to lower diffraction angle) and broadening of the 002 reflection from 26.5 ° (2θ) to 25.3 ° (2θ) (Fig. SⅣ-2(a)). It should be noted, that the downshift of the 002 reflection depends on the increased interlayer distance of Na intercalated layer and on the intercalation probability. A higher intercalation probability and a larger increase in interlayer distance, would lead to more pronounced downshift of this peak. In case of ordered Na layers, an additional, triangular Warren-type peak can be observed at around 20.8 ° (2θ). In the second series of simulations, we tested the intercalation of randomized Na monolayers and bilayers (= two subsequent intercalation events, i.e. double Na) combined with three and four non-intercalated graphite layers. This microstructural motif is often referred to a staging[7]. Staging commonly implies that Na layers are filled first, before additional Na layers are intercalated in-between graphite layers. The occupation of Na within the intercalated layers, primarily effects the peak intensity, which is corrupted by pronounced preferred orientation. Hence, we will rather use the crystallographic term "Reichweite", describing the range of a fault, i.e. a Na intercalation event in the stacking of graphite, as we cannot make any quantitative statements on the population of the Na layers. Both the intercalation of Na monolayers and bilayers combined with a Reichweite of three or four, leads to a downshift of the 002 reflection but not to a considerable broadening of the peak (Fig. SⅣ-2(b)). In addition, satellite peaks appear both at lower and higher diffraction angles, with the satellites situated at higher diffraction angles having a significantly higher integral peak intensity than the ones at lower angles. The larger the Reichweite of the intercalation event, the lower are the satellites split apart from the 002 diffraction line. An intercalation of Na monolayers, also leads to less intense satellites than the intercalation of Na bilayers. In the next series of simulations, we compare the intercalation of an ordered vs. a randomized Na layers with a Reichweite of three layers (Fig. SⅣ-2(c)). Both simulated patterns of faulted intercalated-graphite are indistinguishable. This is attributed to the stacking faulted nature of graphite (Fig. SⅣ-1), as the position of an ordered Na, *n*, is randomized with respect to the preceding Na layer *n*-4 and the subsequent Na layer *n*+4. As a consequence of the strongly averaged information content of a powder pattern, the ordered Na positions appear to be smeared out and therefore disordered. There are, however, slight differences in the peak shape and the resulting peak maxima, in particular for non 00*l* reflections, but due to the pronounced preferred orientation of the samples, we cannot analyze the data in this depth. As a consequence, we cannot draw any conclusions of the atom arrangement of the intercalated Na layers from the XRD data. In the last series of simulations, we softened the criteria for monolayer and bilayer intercalation and for the Reichweite. At first, we simulated the single intercalation of Na applying a Reichweite of three (Fig. SⅣ-2(d), green line). As previously seen, this leads to a downshift of the position of the 002 reflection compared to faultless graphite (Fig. SⅣ-2(d), blue and green lines) and to the emergence of satellite peaks at higher and lower diffraction angles. In the next step, we allow for an intercalation of a Na bilayer with a probability of 10 % ($P_{double\ Na}$ = 0.1) and a Reichweite of three but with a 10 % probability for the existence of a 4$^{th}$ ($P_{4th}$ = 0.1) and a 5 % probability of a 5$^{th}$ ($P_{5th}$ = 0.05) non-intercalated graphite layer within the stack (Table SⅣ-1). This leads to a broadening of the 002 reflection, whereas the position of the peak maximum is hardly affected (Fig. SⅣ-2(d), green and magenta lines). The satellites, however are broadened to a higher degree and appear nor much broader than the 002 reflection. In addition, they are shifted more closer to the 002 peak. Finally, we simulated a diffraction pattern in which an intercalation of a Na layer occurs with a probability of 90 % ($P_{double\ Na}$ = 0.9) and a Reichweite of three but with a 10 % probability for the existence of a 4$^{th}$ ($P_{4th}$ = 0.1) and a 5 % probability of a 5$^{th}$ ($P_{5th}$ = 0.05) non-intercalated graphite layer within the stack. The position of the 002 reflection is now shifted towards significantly lower diffraction angles, as the mean interlayer distance is increased compared to the other simulated patterns (Fig. SⅣ-2(d), magenta and cyan lines). The satellite peaks appear less broadened but still broader than the 002 reflection. The downshift of the satellite at higher diffraction angles is considerably larger than the upshift of the satellite at lower angles.

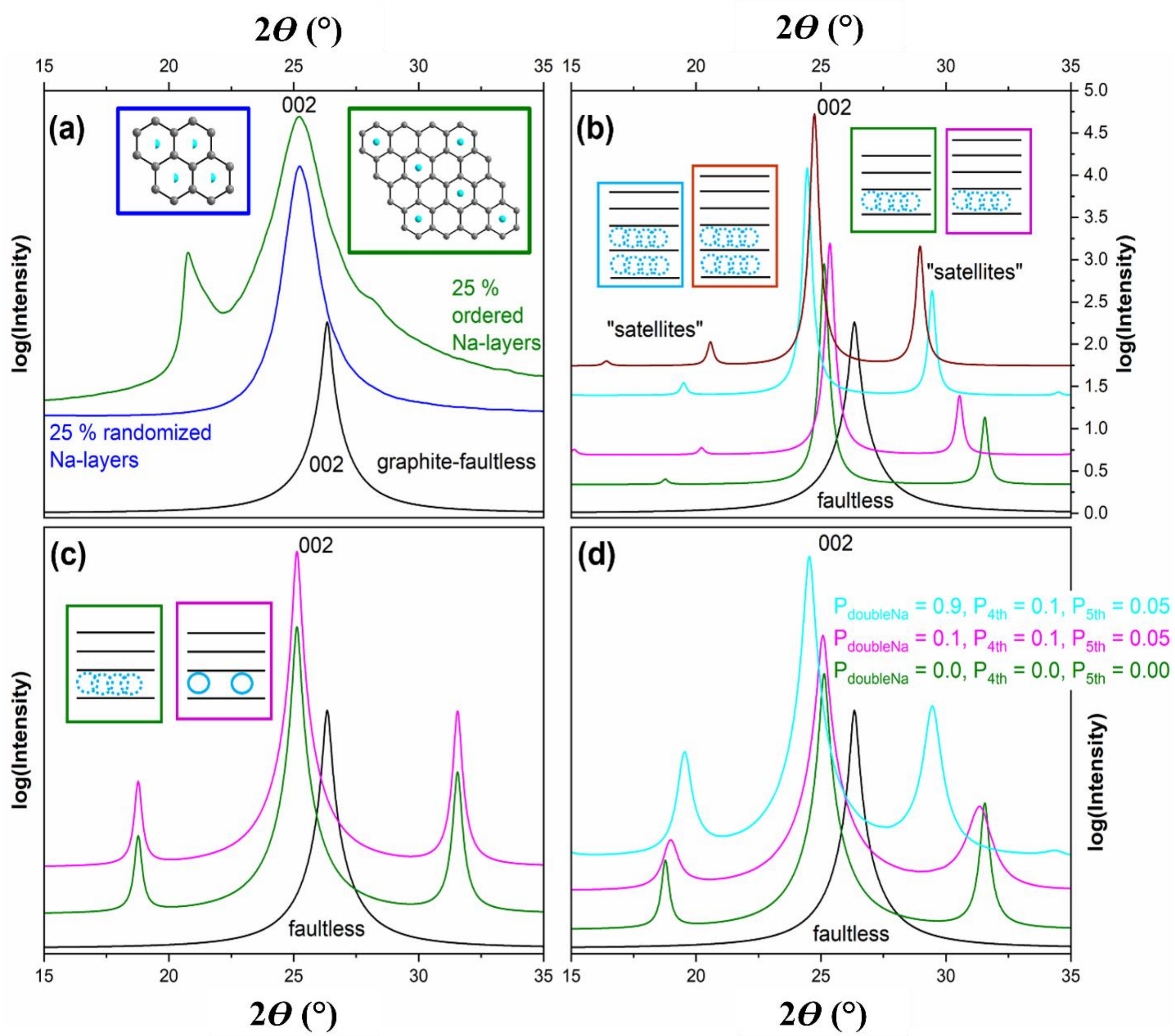

Fig. SⅣ-2. Simulations of Na intercalated hexagonal graphite using different faulting scenarios: (a) random intercalation of ordered vs. disordered Na layer (no staging), (b) intercalation of Na monolayers and bilayers followed by three and four non-intercalated graphite layers, (c) intercalation of Na monolayers, followed by three non-intercalated graphite layers, (d) intercalation of Na bilayers, with randomization of Reichweite of the intercalation (Table SⅣ-1).

The comparison of the measured XRD patterns of the starting materials, HOPG with Na-metal and Na-intercalated HOPG obtained by chemical intercalation at 400 °C after two weeks (Fig. SⅣ-3(a), blue and green lines) shows, that the 002 reflection is significantly downshifted by almost 3 ° (2θ) and slightly broadened. In addition, satellite peaks appear at lower and higher diffraction angles, which are significantly broader than 002 diffraction line. The overall shift on the 002 peak position is an indicator for a successful intercalation of Na in-between the graphite layer, which corresponds to an increase of the mean interlayer distance. The satellite peaks indicate a Reichweite of the intercalation event (Fig. SⅣ-2(b, c, d)) and broadening of both the 002 diffraction and of the satellites point to the co-existence of both Na monolayers, bilayers and perhaps even multi-layer Na intercalation. The measured peak positions do not exactly match the reflection positions in the simulated pattern, which corresponds to an intercalation of a Na layer occurs with a probability of 90 % ($P_{double\ Na}$ = 0.9) and a Reichweite of three but with a 10 % probability for the existence of a 4$^{th}$ ($P_{4th}$ = 0.1) and a 5 % probability of a 5$^{th}$ ($P_{5th}$ = 0.05) non-intercalated graphite layer within the stack. This can be assigned to multiple factors: differences in the overall intercalation probability, further modulation of the interlayer distance by partial occupation of the Na layers or a different or broader distribution of the Reichweite of the defect. Given to the limited data quality (preferred orientation, poor particle statistics, sample height error), the multi-dimensionality of the anticipated microstructural model and the fact that a modulation of the Reichweite of defects turns the dimension of the transition probability matrix into a parameter, which needs to be optimized[8], we refrain from any quantitative statements. The chemical intercalation of Na at room temperature (25 °C) seems to lead to an incomplete conversion, as neither the position nor the broadening of the 002 diffraction line changes (Fig. S Ⅳ-3(b)). After 10 months (magenta line), a small and broad satellite appears at higher diffraction angles. Eventually, after one year (cyan line), another broad reflection emerges at lower diffraction angle. Considering the peak position, it could be both a satellite or another 00l' diffraction line. The latter would indicate a two-phase character of the sample with a non-intercalated phase co-existing with a Na-intercalated HOPG phase, in which defects are associated with a considerable Reichweite, most likely exceeding quadlayers. Another explanation would be a low intercalation probability, which is connected with a high Reichweite, as an increase in the Reichweite, leads to smaller splitting of the peak position of the 002 diffraction line and its satellites (Fig. SⅣ-2(b)). The electrochemical intercalation of Na in HOPG seemingly leads to a splitting of the 002 diffraction line (Fig. SⅣ-3(c)) into a larger peak at higher and a smaller reflection at lower diffraction line. This however, would mean that the electrochemical intercalation of Na leads both to an increase and decrease of the interlayer distance compared to original HOPG. The latter fact seems to be very unlikely. Hence, we interpret the XRD pattern as a low Na intercalation, with large Reichweite. The satellite at lower diffraction angles might be broadened to a degree that it merges with the background. It should be noted, that this satellite is also considerably more broadened in the chemically intercalated samples (Fig. SⅣ-3(a, b)). When the electrochemical intercalation is extended from 2 weeks (Fig. SⅣ-3, green line) to 8 weeks (magenta line), the 00l reflection slightly shifts towards lower diffraction angles and the peak becomes broader, the satellite peak also broadens. This and the fact, that the shape of the 00l reflection becomes slightly more trigonal shaped leads to the conclusion, that the intercalation process proceeds over time, but the Na layers are distributed more randomly within the stack (Fig. SⅣ-2(a)). The electrochemical intercalation leads to different diffraction patterns, when HOPG is coated by 5 nm $TiO_2$ (Fig. SⅣ-3(d)). Here additional satellite peaks 011 appearing at higher diffraction angles, are reflections attributed to Na metal intentionally put for calibration. Overall, these patterns are comparable to the XRD data obtained from chemical intercalation at room temperature (Fig. SⅣ-3(b)). Hence, the diffraction effects could be explained by similar microstructural effects. As the satellite peak at, lower diffraction angles, which is most likely another 00*l*' line, we assign this to a two-phase character of the sample with a non-intercalated phase co-existing with a Na-intercalated HOPG phase, in which defects are associated with a considerable Reichweite, most likely exceeding quadlayers. The intercalation probability is considerably higher than for the chemically intercalated sample. Due to the broadening of the 00*l*' reflection and its triangular shape, we also conclude that the distribution of Na layers within the stack is partially randomized.

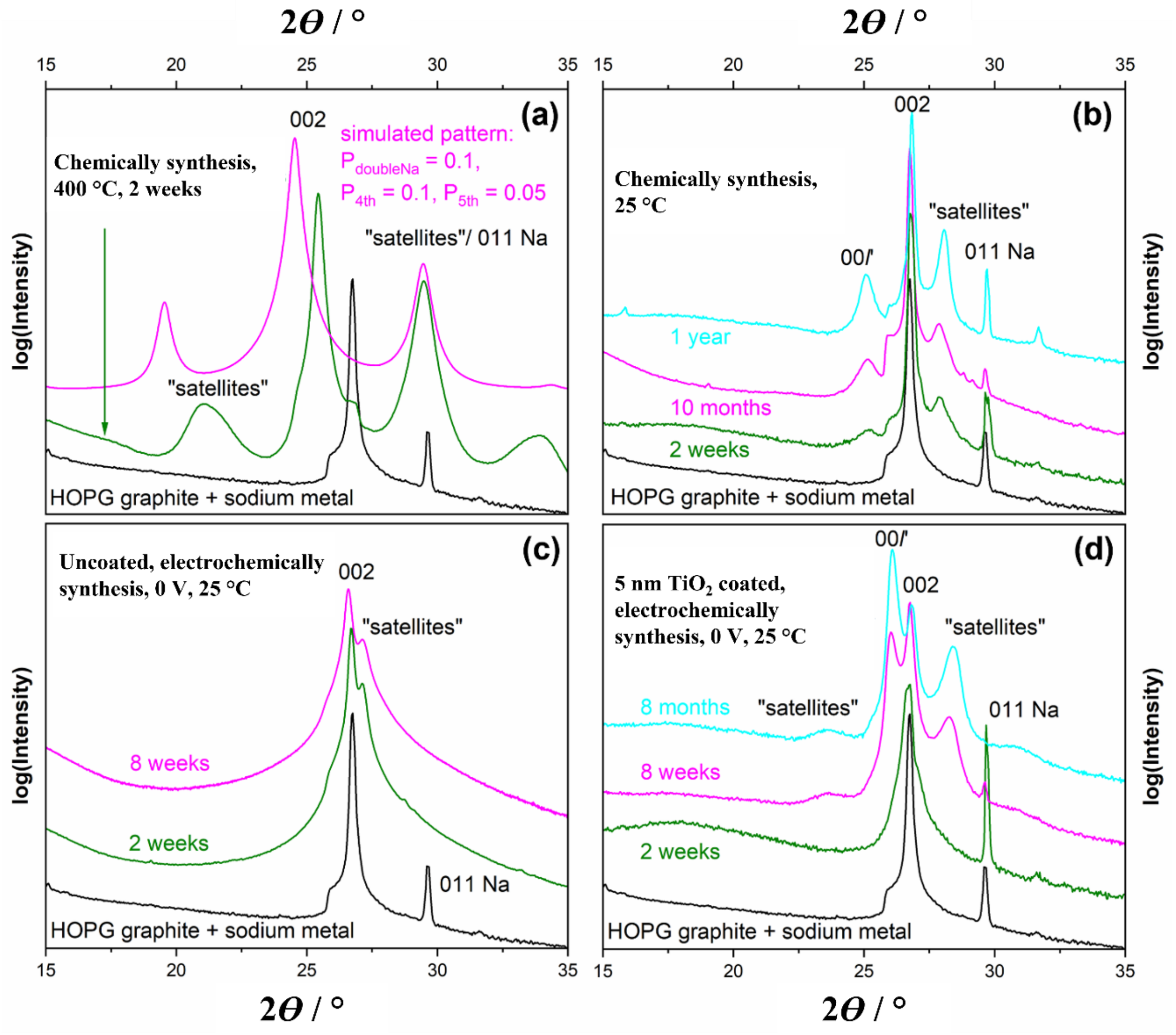

Fig. SⅣ-3. (a) Comparison of the measured XRD patterns of HOPG with Na-metal, the starting materials (blue line), Na-intercalated HOPG obtained after two weeks of chemical intercalation at 400 °C (green line) and the simulated pattern intercalation of Na bilayers, with randomization of Reichweite of the intercalation event (magenta line, Fig. SⅣ-2(d), cyan line), (b) comparison of the measured XRD patterns of HOPG with Na-metal, the starting materials (blue line) and Na-intercalated HOPG obtained after two weeks (green line), ten months (magenta line) and one year (cyan line) of chemical intercalation at 25 °C, (c) comparison of the measured XRD patterns of HOPG with Na-metal, the starting materials (blue line) and Na-intercalated HOPG obtained after 2 (green line) and 8 weeks (magenta line) of electrochemical synthesis at 0 V at 25 °C, (d) comparison of the measured XRD patterns of HOPG with Na-metal, the starting materials (blue

line) and Na intercalated-5nm $TiO_2$ coated HOPG after 2 (green line) and 8 weeks (magenta line) and 8 months (cyan line) of electrochemical synthesis at 0 V at 25 °C.

Table SⅣ-1. Transition probability matrix used for the faulting scenario of intercalation of Na bilayers, with randomization of Reichweite of the intercalation event (Fig. SⅣ-2(d))

| to→ / from↓ | Na+Graphite | Na+Graphite | Graphite | Graphite | Graphite | Graphite | Graphite | Graphite |
|---|---|---|---|---|---|---|---|---|
| Na+Graphite | 0 | $\mathbf{P}_{\textbf{double Na}}$ | 1-$\mathbf{P}_{\textbf{doubleNa}}$ | 0 | 0 | 0 | 0 | 0 |
| Na+Graphite | 0 | 0 | 1 | 0 | 0 | 0 | 0 | 0 |
| Graphite | 0 | 0 | 0 | 1 | 0 | 0 | 0 | 0 |
| Graphite | 0 | 0 | 0 | 0 | 1 | 0 | 0 | 0 |
| Graphite | 0 | 0 | 0 | 0 | 0 | $\mathbf{P}_{\textbf{4th}}$ | 0 | 1-$\mathbf{P}_{\textbf{4th}}$ |
| Graphite | 0 | 0 | 0 | 0 | 0 | 0 | $\mathbf{P}_{\textbf{5th}}$ | 1-$\mathbf{P}_{\textbf{5th}}$ |
| Graphite | 0 | 0 | 0 | 0 | 0 | 0 | 0 | 1 |

## Part Ⅴ Entropy and Enthalpy Problems

### 1) Configurational entropy of intercalation planes

At a first glance the higher degree of aggregated layers when the temperature is reduced, might be explained by the loss of configurational entropy ($S_{\mathrm{conf}}$) of layer arrangement and then by mass action laws for (dis-)aggregation. (Whether we consider monolayer or bilayer is not relevant) However, $S_{\mathrm{conf}}$ is much too small in comparison to the aggregation energy. It suffices to study the site exchange of $N_{\mathrm{I}}$ monolayers within the ensemble of $N_{\mathrm{I}} + N_{\mathrm{C}}$ layers (Fig. SⅤ-1).

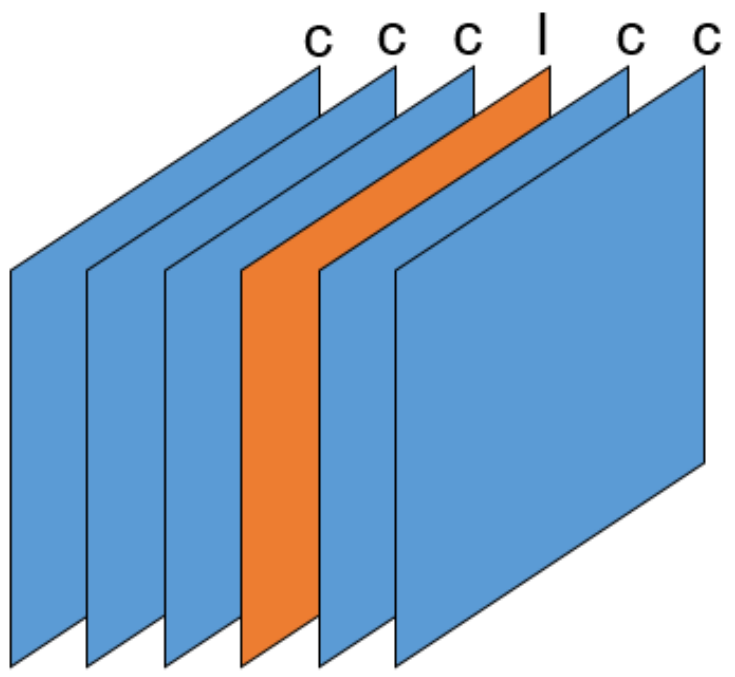

Fig. SⅤ-1. Statistics of a random layer arrangement.

Let us assume a random distribution, then

$$S_{\mathrm{conf}} = k \ln \binom{N_{\mathrm{I}} + N_{\mathrm{c}}}{N_{\mathrm{I}}}$$

and the chemical potential of a layer

$$\mu_{\mathrm{I}} = \mu_{\mathrm{I}}^{\circ} + kT \ln \left(\frac{N_{\mathrm{I}}}{N_{\mathrm{c}}}\right)$$

where $k$ is Boltzmann constant, $\mu_{\mathrm{I}}^{\circ}$ standard chemical potential, and $T$ temperature in kelvin.

For the equilibrium concentration we find

$$\frac{N_{\mathrm{I,eq}}}{N_{\mathrm{c}}} = \exp\left(-\frac{\Delta G_{\mathrm{I}}^{\circ}}{RT}\right)$$

where $\Delta G_{\mathrm{I}}^{\circ}$ is the standard free enthalpy of forming a single intercalated layer (I). The value refers to the entire plane and is massively larger than value for single atomic constituents (at).

Moreover, the reference quantity $N_\mathrm{c}$ is vastly smaller than the reference value for single atomic constituents ($N_\mathrm{c} \sim N_\mathrm{at}^{1/3}$). In total the equilibrium concentration is much smaller than unity, i.e. the configurational entropy of layer rearrangement is absolutely negligible (see page 119-120, ref.[9]).

**2) Configurational entropy of Na distribution within a layer**

Let us consider the configurational entropy of the rearrangement of atomic constituents (Na) within a given plane and again assume a random distribution (Fig. SⅤ-2).

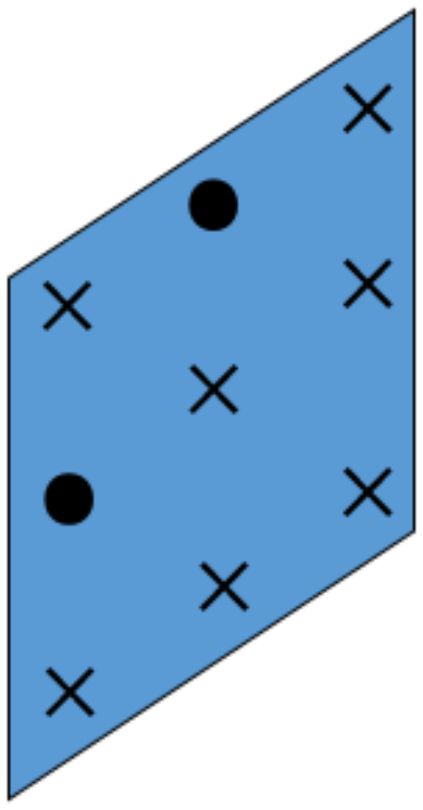

Fig. SⅤ-2. Statistics of random Na distribution for a given plane.

We can adopt the above equations when $N_\mathrm{I}$ is relabeled as $N_\mathrm{Na}$, and $N_\mathrm{c}$ is now to be understood as the number of rings not "occupied" by Na in say the $NaC_6$ structure (the rings that are not occupied even in the fully intercalated state are not included). The entropy value is much higher than in the previous example as we refer to point defects (but less than in the 3D case to be considered below; $N_\text{total sites in plane} \sim N^2_\text{total planes} \sim N^{2/3}_\text{total sites in total volume}$). Specifically, we will estimate the free energy change when Na atoms of a fully occupied monolayer (fully occupied or vacant are defined relative to a $NaC_6$ structure) are redistributed over two neighboring layers such that the chemical bilayer is formed. The free energy change is at minimum, if approximately

$$\mathrm{const} + kT \ln \frac{x}{1-x} = 0$$

where $x = \frac{N_\mathrm{Na}}{N_\mathrm{c}+N_\mathrm{Na}}$ and *const* refers to the change of the standard potential on this transfer.

When ordering effects are negligible, *const* is rather small, and we find $x \simeq 0.5$. Like in the previous section mass actions can be formulated, whose solution gives complete preference to the state of lowest free standard enthalpy (here the aggregated state).

The other extreme would be a strongly ordered situation. Here now the first term is significant and variable, while the configurational term is small. Then according to the main text, the equilibrium $x$ might be also 0.5.

Whether in the bilayers the above entropy argument holds for both layers separately or if double the sites are available, depends on the exchangeability between the layers which appears to be only relevant if there are imperfections that allow for perpendicular transport.

**3) Entropy and enthalpy of charged particles in an ideal space charge layer**

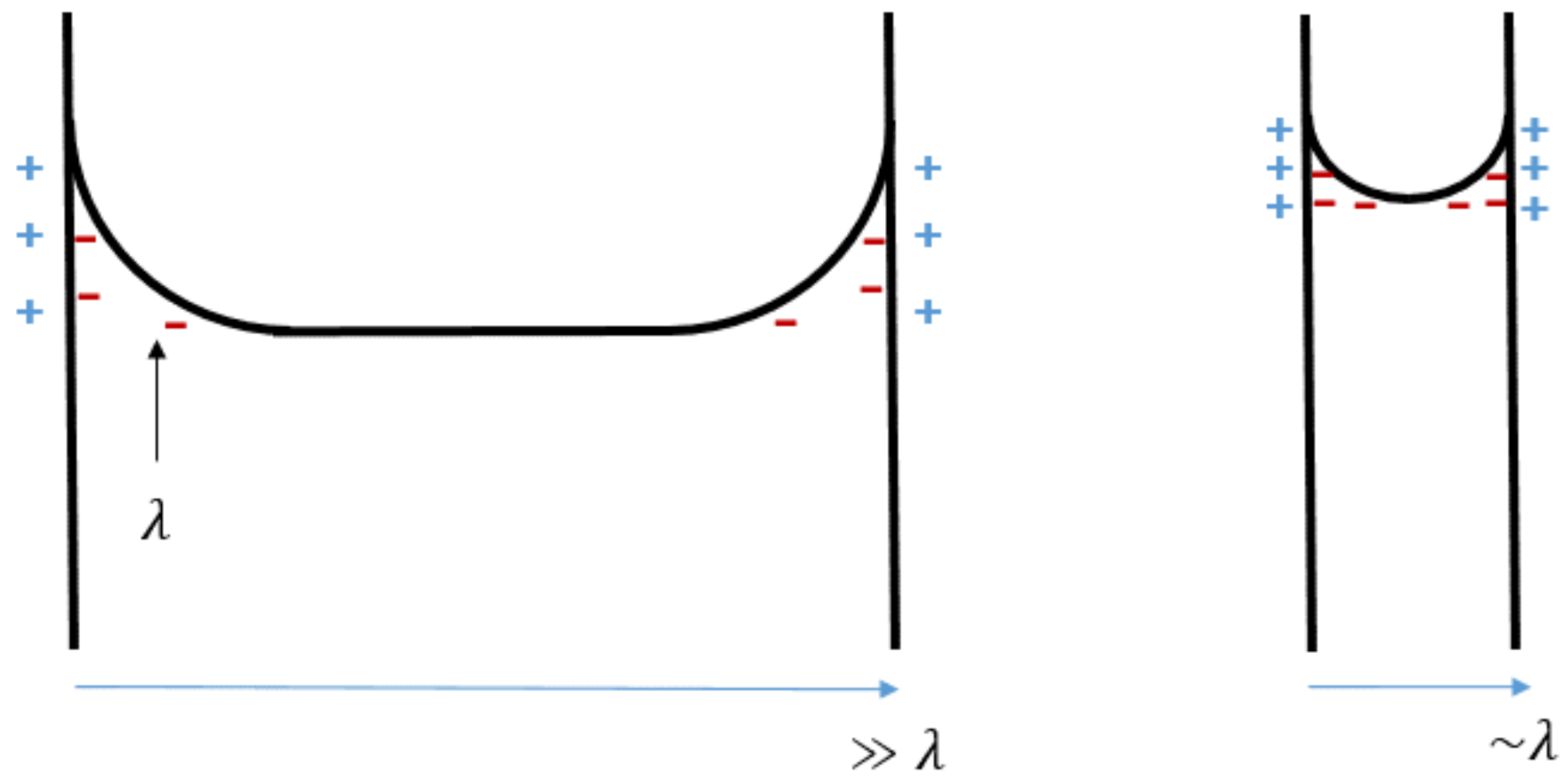

Fig. SⅤ-3 Sketch of two space charge layers.

Fig. SⅤ-3 gives the situation for a distance of two charged layers, which is far beyond the screening length $\lambda$ (left) and on the order or less than $\lambda$ (right), displayed for the case of fixed interfacial charges. On approach of very distant layers towards the situation shown at the left, certainly the energetic situation differs but also the configurational entropy for the distribution of the diffuse charges. In fact, various authors in particular Overbeek have shown that the latter contribution is dominant in the majority of cases.

In such a situation where electric fields occur, the decisive potential is the electrochemical potential (that unlike the chemical potential also includes the electric potential $\phi$)

$$\widetilde{\mu} = \mu^{\circ} + ze\phi + \mu_{\mathrm{conf}}$$

In the space charge free enthalpy ($G_{\mathrm{scl}}$), it is useful to separate the space charge enthalpy ($\mathrm{H}_{\mathrm{scl}}$) from the configurational entropy of space charge distribution, i.e. $\Delta S^{\circ}_{\mathrm{scl}}$.

For high interfacial potential Overbeek[10] gives

$$-T\Delta S^{\circ}_{\mathrm{scl}} = G_{\mathrm{scl}} \frac{\psi_0 - 3 + 6\,e^{-\frac{\psi_0}{2}}}{\psi_0 - 2 + 4\,e^{-\frac{\psi_0}{2}}} > H_{\mathrm{scl}}$$

with

$$H_{\mathrm{scl}} = G_{\mathrm{scl}} \frac{1 - 2\,e^{-\frac{\psi_0}{2}}}{\psi_0 - 2 + 4\,e^{-\frac{\psi_0}{2}}}$$

where $\psi_0 = e\varphi_0/kT$ and $\varphi_0$ = interfacial electric potential,

while at small $\psi_0$, $-T\Delta S^{\circ}_{\mathrm{scl}} \simeq \Delta H^{\circ}_{\mathrm{scl}}$.

The equations hold for a flat layer situation.

For not too small $\varphi_0$, the approach of two equally charged interfaces is primarily hindered by the loss of entropy on confining the distribution. The total G-curve of approach increases (maximum at ~ $\lambda/4$ for the example treated in Overbeek[10] and then decreases towards the value of the aggregated layers. This maximum refers to a transition state and holds irrespective of the aggregation free enthalpy being negative or positive.

Obviously $\lambda$ - in interplay with the Na-activity $a_{\mathrm{Na}}$ (OCV) - determines the staging situation. If $a_{\mathrm{Na}}$ is very small, the Na-layers will be randomly separated by distances larger than a few times $\lambda$. To be precise, variations of the surface potential and additional effects due to strain and electron correlation have to be considered.

**4) Standard free enthalpy of aggregation**

So far we have not explicitly addressed the full standard entropy $\Delta S^{\circ}$ of aggregation, i.e., the entropy connected with assembling isolated chemical bilayers to higher aggregates. There are essentially three contributions: the first is the change of the phononic entropy, the second the entropy loss as a consequence of the loss of space charge layers, the third the entropy change on the configurational change plus the chemical interaction on aggregation. As shown above, the second term is large and assumed to be dominant. Contrarily, the corresponding enthalpy change is most probably dominated by the chemical term, as the space charge enthalpy change should be small. When comparing the insertion enthalpies of Na at high and low T, we need to consider the large enthalpy change of Na metal from LT to HT, i.e., the specific heat which even includes the melting enthalpy, as well as the aggregation enthalpy when going from the HT (chemical bilayers) to LT (aggregated layers), what is more, the thermal enthalpy change owing to the specific heat of the Na-containing HOPG.

Let us consider the different contributions:

The thermal enthalpy change of Na ($\Delta H_{\mathrm{th}}$(Na)) between 700 K and 300 K is 14.7 kJ/mol[1].

The thermal enthalpy change of the Na containing HOPG can, as the Na content ($x$) is small, be approximated by $1 + x$ times the value for pure graphite, where the thermal enthalpy change of pure graphite ($\Delta H_{\mathrm{th}}$(C)) between 700 K and 300 K is 5.7 kJ/mol[1].

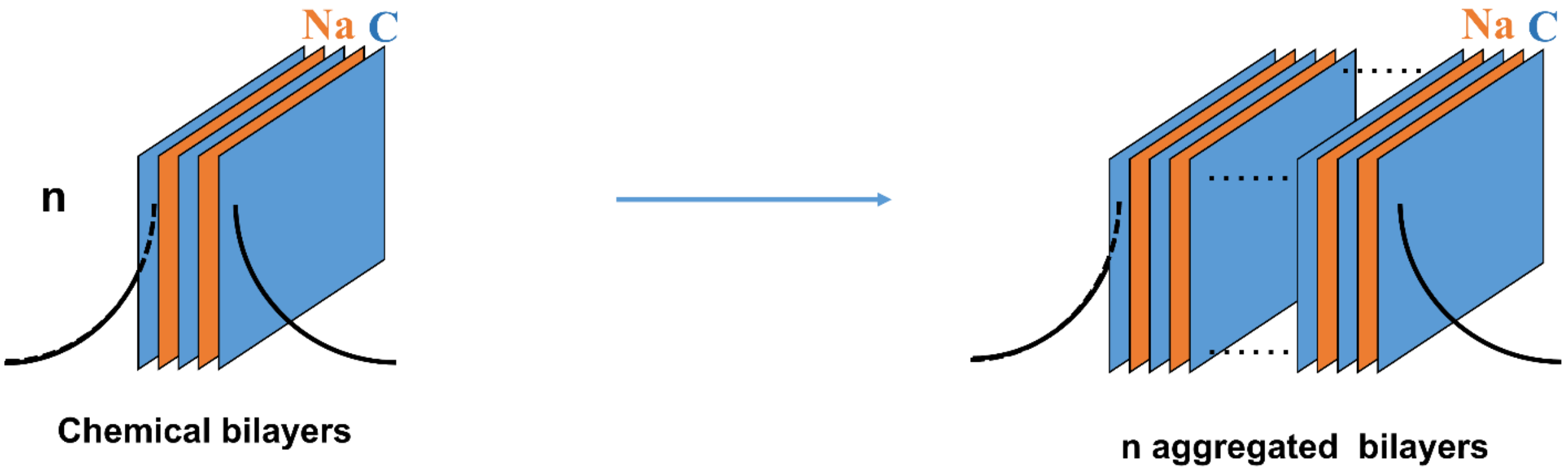

Fig. SⅤ-4 Sketch for chemical bilayer aggregation to higher aggregates, resulting in loss of space charge zones (also shown in the main text).

If changes in the phononic contributions are negligible, the (standard) aggregation entropy can be estimated from the loss of the space charge zones. As shown in Fig. SⅤ-4, for the aggregation of n isolated bilayers to a 2n-layer aggregate the loss of space charge zones is (2n-2) for n chemical bilayers. For large n-values we count 2n per n bilayers or, if we assume a 50 % occupation, per n fully occupied monolayers. This limit of 2 per full layer is taken in the further evaluation.

Let us take the text example revealing that at 700 K almost exclusively chemical bilayers are observed while at 300 K mostly aggregates (triple layers, quadruple and higher aggregates are observed). Then

$$\Delta H^{\circ}/700\ \mathrm{K}\ \ > \Delta S^{\circ} > \Delta H^{\circ}/300\ \mathrm{K}$$

where $\Delta H^{\circ}$ is standard aggregation enthalpy. When we again refer to Overbeek's work[10]. we may find expression for the free energy (free enthalpy) for losing two space charge zones, namely

$$G = \varepsilon_0 \varepsilon_r \kappa (\frac{kT}{e})^2 \left\{8 \cosh \frac{\psi_0}{2} - 8\right\}$$

where $\psi_0 = e\varphi_0/kT$ and $\kappa^{-1}(= \lambda)$ stands for the Debye length, $\varepsilon_0$ and $\varepsilon_r$ the vacuum and relative dielectric constant, respectively. The interfacial electric potential $\varphi_0$ can be obtained via

$$\varphi_0 = \frac{2kT}{e} \sinh^{-1} \frac{eq}{2\varepsilon_0 \varepsilon_r kT\kappa}$$

where $q$ is the surface charge density (charge per area).

When solely relying on space charge effects, we find the bilayer aggregation free enthalpy to be 0.065 J/m$^2$ at 500 K and 0.038 J/m$^2$ at 300 K (half occupied relative to a $NaC_8$ structure, the relative dielectric constant is assumed to be 10, and the screening length (Debye length) is taken as 1 nm[11-13]. (The free enthalpies are 0.087 J/m$^2$ at 500 K and 0.052 J/m$^2$ at 300 K when considering a half occupied $NaC_6$ structure). However, we need to take also the middle C-layer into account. It is a reasonable approach to tentatively assume that the middle layer takes up half of the total charge of the bilayer, so that each double layer contains 1/4 of the total charge, so a value that is smaller by a factor of 2 compared to the previous value. This factor becomes the smaller, the larger n, so it is even more reasonable to neglect the space charge layer contribution in the n-aggregate (Here we ignore solubility changes with temperature. Higher solubilities would weaken the last statement). (In this situation, we obtain the bilayer aggregation free enthalpy due to space charges to be 0.032 J/m$^2$ at 500 K and 0.019 J/m$^2$ at 300 K for $NaC_8$ structure; the aggregation free enthalpy values are 0.043 J/m$^2$ at 500 K and 0.026 J/m$^2$ at 300 K in terms of $NaC_6$ structure).

From these values, one can extract an entropy of bilayer aggregation due to space charges of $-1.35 \times 10^{-4}$ J/m$^2$K, and an enthalpy term of $-0.0025$ J/m$^2$. (The entropy and the enthalpy are $-1.75 \times 10^{-4}$ J/m$^2$K and $-0.0005$ J/m$^2$ for the case of $NaC_6$ structure). Then the bilayer aggregation free enthalpy due to space charges at 700 K is found to be 0.092 J/m$^2$ (23 kJ/mol). (The value is 0.122 J/m$^2$ (23 kJ/mol) in terms of $NaC_6$ structure). When we take account of the situation in which each double layer contains 1/4 of the total charge, we find the entropy of bilayer aggregation to be $-6.5 \times 10^{-5}$ J/m$^2$K, and the enthalpy term to be $-0.0005$ J/m$^2$ for $NaC_8$ structure. (The entropy and enthalpy values are $-8.5 \times 10^{-5}$ J/m$^2$K and $0.0005$ J/m$^2$ for $NaC_6$ structure). Then the bilayer aggregation free enthalpy due to space charges at 700 K is obtained to be 0.045 J/m$^2$ (23 kJ/mol). (The value is 0.06 J/m$^2$ (23 kJ/mol) in terms of $NaC_6$ structure). Even though the values in "J/m$^2$" are different for different situations (different charge densities), when we consider the charge density variation and convert the energy per area into the energy per mole ("J/m$^2$" into "kJ/mol"), all the situations give a similar value of aggregation free enthalpy.

Now we refer to the total aggregation processes (beyond space charges): As the related space charge enthalpy (see above) is negligible we have essentially the chemical term ( $\Delta H_{\mathrm{ch}}$ ) dominating $\Delta H^{\circ}$ (total aggregation process). (Such chemical term will include bonding effects and Madelung terms.) From this we can calculate free enthalpies of total aggregation for 700 K and 300 K (including $\Delta H^{\circ}$). As the first value must be positive and the second one negative, $\Delta H_{\mathrm{ch}}$ can be assessed to be in the range of -10 kJ/mol and -23 kJ/mol, probably close to the latter as we still observe a small portion of triple layer at 700 K. The heterogeneity at 300 K is to be ascribed to not having reached equilibrium.

Note that this aggregation energy is an average value for aggregating bilayers to the aggregates at room temperature (without thermal contribution). It is intelligible that the aggregation energy per particle of bilayers to quadruple layers is larger than for the higher aggregation. This is even more pronounced for the aggregation of hypothetical monolayers to bilayers what explains the fact that monolayers have not been observed for Na. (The improved Madelung energy and/or the increased electron delocalization may account for this.) In the case of Li the ionicity as major reason for the energy gain is much lower while for K etc. the strain energy seems to dominate. It is pertinent to recall that the strain energy is proportional to the square of the misfit meaning that doubling the misfit will lead to an energy twice the sum of the strain energy of two individual misfitting layers. When treating the virtual transformation of filled monolayers to half-filled bilayers by the above arguments one has also to take account of the effects of lower occupancy on misfit and on space charge potential.

For the other alkali metals, also a tendency for dissociation at higher temperature[14-16] is observed but less than for Na conforming to the more favorable situation of full aggregates for the other alkali elements (experiments and DFT[7, 17, 18]).

**5) Enthalpies and entropies of intercalation**

So far, we have been concerned with thermodynamic problems in which the overall configurational entropy did not play a significant role (the configuration problem in Section 4) still referred of individual space charge layers).

Here we consider very low Na concentrations for which we can assume random distribution, i.e., before a critical concentration is reached opening pathways of "pipe" diffusion.

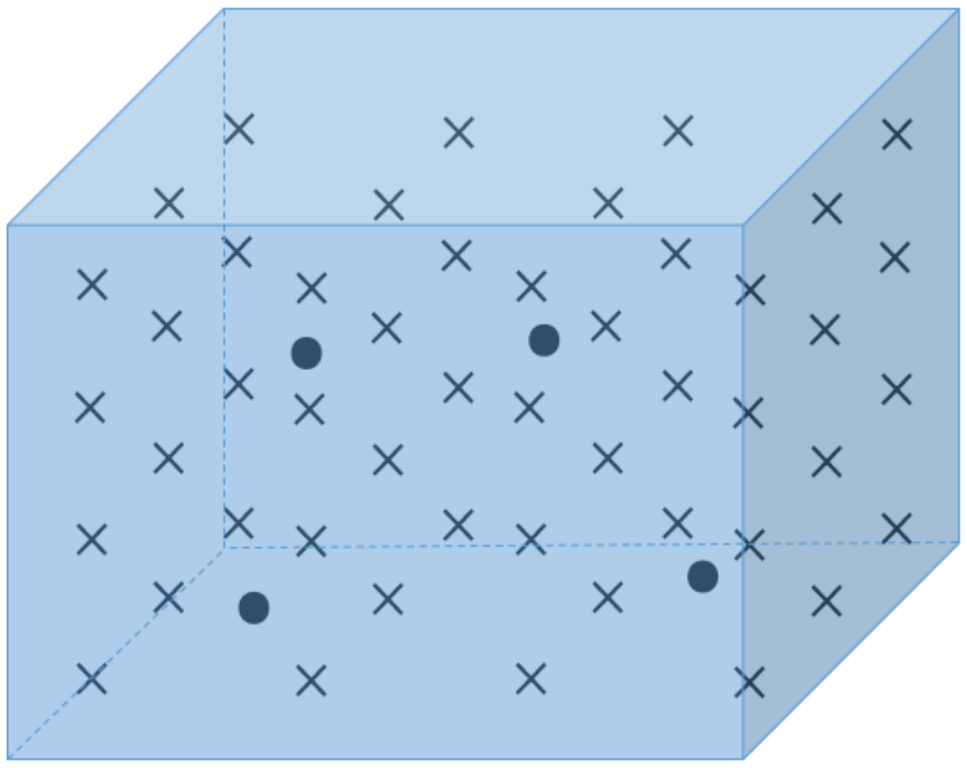

Fig. SⅤ-5 Statistics of random Na distribution in three dimension.

Then (see Figure SⅤ-5) the same treatment as in 1) but now in a three dimension-space gives us appreciable numbers for the configurational entropy. The resulting mass action law for the process

$$x\,\mathrm{Na} + \mathrm{C} \rightleftharpoons \mathrm{Na}_x\mathrm{C}$$

yields

$$[\mathrm{Na}]_{\mathrm{eq}} = a^{\circ}_{\mathrm{Na}}\exp\,(-\frac{\Delta H^{\circ}_{\mathrm{ins}}}{RT})\exp\,(\frac{\Delta S^{\circ}_{\mathrm{ins}}}{R})$$

($a^{\circ}_{\mathrm{Na}}$= activity of Na in Na, can be taken as 1).

The equilibrium Na concentration logarithmically plotted vs. *1/T* (Van't Hoff plot) delivers $\Delta H^{\circ}_{\mathrm{ins}}$ and $\Delta S^{\circ}_{\mathrm{ins}}$ (see the main text). The slope is well defined at high temperature (HT), whereas a lower slope (or even sign inversion) is indicated by the measurement at low temperature (LT). This LT value can be assessed within our model by the thermodynamic cycle to be described now. Note that we certainly overstress the random assumption. The statistical problem involving aggregation is more complicated and in view of the lack of information not tractable yet. At any rate, the activity coefficient is smaller than 1 and the concentration curve is higher than the linear $\ln a$ vs. $1/T$ plot expected from a random situation. So $\Delta H^{\circ}_{\mathrm{ins}}$ and $\Delta S^{\circ}_{\mathrm{ins}}$ have to be considered as effective values. Note also that the ab-initio calculations usually refer to $\Delta H^{\circ}_{\mathrm{ins}}$ and do not only ignore $\Delta S^{\circ}_{\mathrm{ins}}$, but also $\Delta S_{\mathrm{conf}}$.

If we assume the Na concentrations at HT and LT to be roughly the same (e.g., $y{\sim}0.01$), we can write the chemical processes as follows (see also the thermodynamic cycle as shown in Fig. 6 in the main text):

$$y\,\mathrm{Na}\ (300\ \mathrm{K}) + \mathrm{C}\ (300\ \mathrm{K}) \rightarrow \mathrm{Na}_y\mathrm{C}\ (300\ \mathrm{K})$$

$$\mathrm{Na}_y\mathrm{C}\ (300\ \mathrm{K}) \rightarrow \mathrm{Na}_y\mathrm{C}\ (700\ \mathrm{K})$$

$$\mathrm{Na}_y\mathrm{C}\ (700\ \mathrm{K}) \rightarrow y\,\mathrm{Na}\ (700\ \mathrm{K}) + \mathrm{C}\ (700\ \mathrm{K})$$

The overall reaction can be written as

$$y \text{ Na } (300 \text{ K}) + \text{C } (300 \text{ K}) \rightarrow y \text{ Na } (700 \text{ K}) + \text{C } (700 \text{ K})$$

From this set of equations, we obtain the enthalpy relation as

$y\ \Delta H_{\text{th}}(\text{Na}) + \Delta H_{\text{th}}(\text{C})$

$= y\ \Delta H_{\text{ins}}(300\ \text{K}) + \Delta H_{\text{th}}(\text{Na}_{0.01}\text{C}) + y\ \Delta H_{\text{dis}} - y\ \Delta H_{\text{ins}}(700\ \text{K})$

where $\Delta H_{\text{dis}}$ is disaggregation enthalpy, the specific heat $\Delta H_{\text{th}}$ (Na) also includes melting enthalpy of Na, and the thermal enthalpy change of the Na containing carbon is approximated by

$\Delta H_{\text{th}}(\text{Na}_y\text{C}) = (1 + y)\Delta H_{\text{th}}(\text{C})$ (see 4)).

Then we have the insertion enthalpy difference

$$\Delta H_{\text{ins}}(300\ \text{K}) - \Delta H_{\text{ins}}(700\ \text{K}) = \Delta H_{\text{th}}(\text{Na}) - \Delta H_{\text{th}}(\text{C}) - \Delta H_{\text{dis}}$$

Combining with the values given in 4) (note that here we use the upper limit of $\Delta H_{\text{dis}}(\equiv -\Delta H_{\text{ch}}) = (20 \pm 4)\ \text{kJ/mol}$, we can calculate the difference between the insertion enthalpies at 300 K and 700 K to be $(-11 \pm 4)\ \text{kJ/mol}$. Taking the latter from the HT slope of Fig. 4(a) in the main text, which is $8.5\ \text{kJ/mol}$, we estimate the LT value as $(-2.5 \pm 4)$ kJ/mol, which can explain the trend of our experimental results. We obtain from Fig. 4(a) (main text) the insertion entropy as $\Delta S^{\circ}_{\text{ins}}$ at HT to be $-26.6\ \text{J/mol K}$. We may assume that this value is essentially due to the loss of Na(s) where entropy is around $50\ \text{J/mol K}$. In fact, the values for other alkali metals are similar. Note that the values in the literature (e.g., Ref[16], even though labeled as $\Delta S^{\circ}$, contain still a configurational contribution). However, the extracted positive insertion enthalpy causes the significant discrepancy in comparison to other alkali metals, where all the others have negative insertion enthalpies (e.g., $-10\ \text{kJ/mol}$ for Li[19] and $-35\ \text{kJ/mol}$ for K[16]).

Alternatively to this mechanistic picture, we can also understand the upwards bending in Fig. 4(a) (main text) at lower temperature by introducing overall activity coefficients which are smaller than unity (interaction) and are the smaller the higher the concentration. Notwithstanding the various rough approximations, the cycle shows the internal consistency of our model.

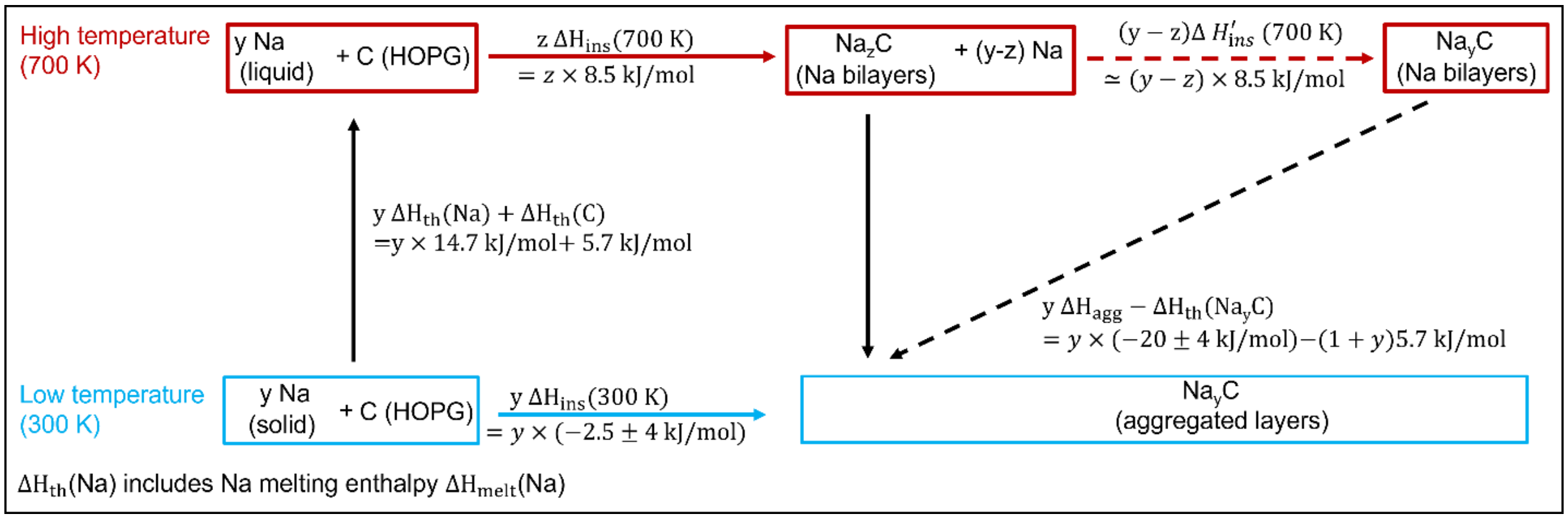

Fig. SⅤ-6. Unlike the approximate thermodynamic cycle shown as Fig. 6 in the main text, this figure is more general and includes the possibility that $\Delta H'_{\mathrm{ins}}$ differs from $\Delta H_{\mathrm{ins}}$ (see below, Refinements of the above $\Delta H$-cycle).

Refinements of the above $\Delta H$-cycle

a) Taking into account that $y\ (700\ \mathrm{K}) \neq y\ (300\ \mathrm{K})$. To simplify the notation, we use *y* for *y* (300 K) and *z* for *y* (700 K). (see Fig. SⅤ-6)

Then we have to write

(1) $y\ \mathrm{Na}\ (300\ \mathrm{K}) + \mathrm{C}\ (300\ \mathrm{K}) \rightarrow \mathrm{Na}_y\mathrm{C}\ (300\ \mathrm{K})$

(2) $\mathrm{Na}_y\mathrm{C}\ (300\ \mathrm{K}) \rightarrow \mathrm{Na}_z\mathrm{C}\ (700\ \mathrm{K}) + (y - z)\ \mathrm{Na}\ (700\ \mathrm{K})$

(3) $\mathrm{Na}_z\mathrm{C}\ (700\ \mathrm{K}) \rightarrow z\ \mathrm{Na}\ (700\ \mathrm{K}) + \mathrm{C}\ (700\ \mathrm{K})$

where the overall reaction (T) is

(T) $y\ \mathrm{Na}\ (300\ \mathrm{K}) + \mathrm{C}\ (300\ \mathrm{K}) \rightarrow z\ \mathrm{Na}\ (700\ \mathrm{K}) + (y - z)\ \mathrm{Na}\ (700\ \mathrm{K}) + \mathrm{C}\ (700\ \mathrm{K})$

Reaction (2) can be split into

(2a) $\mathrm{Na}_y\mathrm{C}\ (300\ \mathrm{K}) \rightarrow \mathrm{Na}_y\mathrm{C}\ (700\ \mathrm{K})$

(2b) $\mathrm{Na}_y\mathrm{C}\ (700\ \mathrm{K}) \rightarrow \mathrm{Na}_z\mathrm{C}\ (700\ \mathrm{K}) + (y - z)\ \mathrm{Na}\ (700\ \mathrm{K})$

$\Delta H(\mathrm{T})$ then follows as

$$\Delta H(\mathrm{T}) = \Delta H(1) + \Delta H(2\mathrm{a}) + \Delta H(2\mathrm{b}) + \Delta H(3)$$

or reformulated:

$y\ \Delta H_{\mathrm{th}}(\mathrm{Na}) + \Delta H_{\mathrm{th}}(\mathrm{C})$

$= y\,\Delta H_{\mathrm{ins}}(300\ \mathrm{K}) + \Delta H_{\mathrm{th}}(\mathrm{Na}_y\mathrm{C}) - (y-z)\,\Delta H'_{\mathrm{ins}}(700\ \mathrm{K}) + y\,\Delta H_{\mathrm{dis}} - z\,\Delta H_{\mathrm{ins}}(700\ \mathrm{K})$

where $\Delta H_{\mathrm{dis}} \equiv -\Delta H_{\mathrm{ch}}$ and $\Delta H'_{\mathrm{ins}}$ is the negative reaction enthalpy of reaction (2b) (i.e. Na insertion into Na-containing HOPG rather than in pure HOPG).

From this, it follows that

$y\,\Delta H_{\mathrm{ins}}(300\ \mathrm{K}) - z\,\Delta H_{\mathrm{ins}}(700\ \mathrm{K})$

$= y\,\Delta H_{\mathrm{th}}(\mathrm{Na}) + \Delta H_{\mathrm{th}}(\mathrm{C}) - \Delta H_{\mathrm{th}}(\mathrm{Na}_y\mathrm{C}) + (y-z)\,\Delta H'_{\mathrm{ins}}(700\ \mathrm{K}) + y\,\Delta H_{\mathrm{ch}}$

where $\Delta H_{\mathrm{th}}(\mathrm{Na}_y\mathrm{C}) = (1+y)\,\Delta H_{\mathrm{th}}(\mathrm{C})$ (see 4)), yielding

$y\,\Delta H_{\mathrm{ins}}(300\ \mathrm{K}) - z\,\Delta H_{\mathrm{ins}}(700\ \mathrm{K}) + (z-y)\,\Delta H'_{\mathrm{ins}}(700\ \mathrm{K})$

$\cong y\,\Delta H_{\mathrm{th}}(\mathrm{Na}) + \Delta H_{\mathrm{th}}(\mathrm{C}) - (1+y)\,\Delta H_{\mathrm{th}}(\mathrm{C}) + y\,\Delta H_{\mathrm{ch}}$

and finally

$$\Delta H_{\mathrm{ins}}(300\ \mathrm{K}) - \frac{z}{y}\,\Delta H_{\mathrm{ins}}(700\ \mathrm{K}) + \frac{(z-y)}{y}\,\Delta H'_{\mathrm{ins}}(700\ \mathrm{K}) \cong \Delta H_{\mathrm{th}}(\mathrm{Na}) - \Delta H_{\mathrm{th}}(\mathrm{C}) + \Delta H_{\mathrm{ch}}$$

If $y = z$, we trivially obtain the previous result

$$\Delta H_{\mathrm{ins}}(300\ \mathrm{K}) - \Delta H_{\mathrm{ins}}(700\ \mathrm{K}) \cong \Delta H_{\mathrm{th}}(\mathrm{Na}) - \Delta H_{\mathrm{th}}(\mathrm{C}) + \Delta H_{\mathrm{ch}}$$

If $y \neq z$, and considering dilute situation, the enthalpies of Na insertion into C and $\mathrm{Na}_x\mathrm{C}$ $(x \ll 1)$ are the same to a good approximation. Hence, we recover the previous result, viz

$$\Delta H_{\mathrm{ins}}(300\ \mathrm{K}) - \Delta H_{\mathrm{ins}}(700\ \mathrm{K}) \cong \Delta H_{\mathrm{th}}(\mathrm{Na}) - \Delta H_{\mathrm{th}}(\mathrm{C}) + \Delta H_{\mathrm{ch}}$$

b) Taking $\mathrm{Na}^+$ and $\mathrm{e}^-$ as independent charge carriers:

$y\ \mathrm{Na}\ (+\ \mathrm{C}) \rightarrow y\ \mathrm{Na}^+ + y\ \mathrm{e}^-$

or more precisely, $\mathrm{Na} + \mathrm{V_i} \rightarrow \mathrm{Na_i^{\cdot}} + \mathrm{e}'$

Assuming mass action laws, we could write

$$[\mathrm{Na_i^{\cdot}}][\mathrm{e'}] = a^{\circ}\exp\left(-\frac{\Delta H_{\mathrm{ins}}^{*}}{RT}\right) \qquad ([\mathrm{Na_i^{\cdot}}][\mathrm{e'}] = [\mathrm{Na_i^{\cdot}}]^2 = [\mathrm{Na}]^2)$$

Then we have

$$[\mathrm{Na}] = a^{\circ 1/2}\exp\left(-\frac{\Delta H_{\mathrm{ins}}^{*}}{2RT}\right)$$

As there is no separate impact on $\mathrm{Na^+}$ and $\mathrm{e^-}$ in our considerations, it is only a question of notation with $\Delta H_{\mathrm{ins}} = \Delta H_{\mathrm{ins}}^{*}/2$.

The above equation presupposes classic random distribution and independency, which are of course harsh approximations. Partial association can be introduced by corresponding equilibria according to defect chemistry.

**6) Special energetic role of bilayers**

In order to understand the special energetics of bilayer formation from monolayers of ionized alkali metals, we need to understand the Madelung effects as well as the delocalization effects. For simplicity we assume for the Madelung problem rock-salt structured layers and for the delocalization homogeneous nano-boxes. Note that the estimation of lattice energies of metals by considering a lattice of metal cations and electrons as anions has been used by Haber[20] and shown to be semi-quantitavely applicable investigated by Meyer[20].

Firstly we calculate the Madelung factors (*M*) of mono-, bi- and multilayers (rocksalt structure). From the literature[21, 22], we know

$M_{3D}$ = 1.748; $M_{2D\ (monolayer)}$ = 1.616; $M_{1D\ (monolayer)}$ = 2ln2 = 1.386

We consider the transition

(a) from 1D-Monolayer (1.386)

to 1D-Bilayer etc.

up to 1D-Multilayer = 2D-Monolayer (1.616)

(b) from 2D-Monolayer (1.616)

to 2D-Bilayer etc.

up to 2D-Multilayer = 3D-Crystal (1.748)

Let $M_0$ be the sum within a layer, $M_1$ the sum of a given to the next layer, $M_2$ the sum to the second next layer, and $M_3$ the one to the third next layer, then we can state:

Monolayer: $M_{\text{monolayer}} = M_0$

Bilayer: $M_{\text{bilayer}} = 2(M_0 + M_1)/2 = M_0 + M_1$

Trilayer: $M_{\text{trilayer}} = 2(M_0 + M_1 + M_2)/3 + (M_0 + 2M_1)/3 = M_0 + 4M_1/3 + 2M_2/3$

Quadruplelayer: $M_{\text{quadruplelayer}} = 2(M_0 + M_1 + M_2 + M_3)/4 + 2(M_0 + 2M_1 + M_2)/4 = M_0 + 3M_1/2 + M_2 + M_3/2$

(a) 1D monolayer (1D chain, alternating +/- charges along the x-axis)

$$M_{\text{intra}} = M_0 = 2\sum_{n=1}^{\infty}\frac{(-1)^n}{n} = 2\ln 2 \approx 1.386$$

1D bilayer (add an opposite chain at a distance of *d*)

$$M_{\text{inter}} = M_1 = \sum_{n=-\infty}^{\infty}\frac{(-1)^{n+1}}{\sqrt{n^2+d^2}}$$

Then the total Madelung constant is

$$M_{\text{1D-bilayer}} = 2\sum_{n=1}^{\infty}\frac{(-1)^n}{n} + \sum_{n=-\infty}^{\infty}\frac{(-1)^{n+1}}{\sqrt{n^2+d^2}}$$

For 1D trilayer and quadruplelayer, we have

$$M_{\text{1D-trilayer}} = 2\sum_{n=1}^{\infty}\frac{(-1)^n}{n} + \frac{4}{3}\sum_{n=-\infty}^{\infty}\frac{(-1)^{n+1}}{\sqrt{n^2+d^2}} + \frac{2}{3}\sum_{n=-\infty}^{\infty}\frac{(-1)^n}{\sqrt{n^2+4d^2}}$$

$$M_{\text{1D-quadruplelayer}} = 2\sum_{n=1}^{\infty}\frac{(-1)^n}{n} + \frac{3}{2}\sum_{n=-\infty}^{\infty}\frac{(-1)^{n+1}}{\sqrt{n^2+d^2}} + \sum_{n=-\infty}^{\infty}\frac{(-1)^n}{\sqrt{n^2+4d^2}} + \frac{1}{2}\sum_{n=-\infty}^{\infty}\frac{(-1)^{n+1}}{\sqrt{n^2+9d^2}}$$

If d=1 (ions are equally spaced), numerical evaluations give:

$$M_{\text{1D-bilayer}} \approx 1.504$$

$$M_{\text{1D-trilayer}} \approx 1.541$$

$$M_{\text{1D−quadruplelayer}} \approx 1.560$$

The Madelung factor jump from 1D monolayer to 1D bilayer is greater than that from bilayer to trilayer (and further to multilayers). It will converge slowly toward the 2D monolayer value (1.616).

(b) 2D monolayer

$$M_{\text{intra}} = \sum_{(i,j)\neq(0,0)} \frac{(-1)^{i+j}}{\sqrt{i^2+j^2}} \approx 1.616$$

2D bilayer (add a second layer at a distance of d)

$$M_{\text{inter}} = \sum_{(i,j)} \frac{(-1)^{i+j+1}}{\sqrt{i^2+j^2+d^2}}$$

Then the total Madelung constant is

$$M_{\text{2D−bilayer}} = \sum_{(i,j)\neq(0,0)} \frac{(-1)^{i+j}}{\sqrt{i^2+j^2}} + \sum_{(i,j)} \frac{(-1)^{i+j+1}}{\sqrt{i^2+j^2+d^2}}$$

For 2D trilayer and quadruplelayer, the Madelung factor can be written as

$$\mathrm{M}_{\text{2D−trilayer}} = \sum_{(i,j)\neq(0,0)} \frac{(-1)^{i+j}}{\sqrt{i^2+j^2}} + \frac{4}{3}\sum_{(i,j)} \frac{(-1)^{i+j+1}}{\sqrt{i^2+j^2+d^2}} + \frac{2}{3}\sum_{(i,j)} \frac{(-1)^{i+j}}{\sqrt{i^2+j^2+4d^2}}$$

$$\mathrm{M}_{\text{2D−quadruplelayer}} = \sum_{(i,j)\neq(0,0)} \frac{(-1)^{i+j}}{\sqrt{i^2+j^2}} + \frac{3}{2}\sum_{(i,j)} \frac{(-1)^{i+j+1}}{\sqrt{i^2+j^2+d^2}} + \sum_{(i,j)} \frac{(-1)^{i+j}}{\sqrt{i^2+j^2+4d^2}} + \frac{1}{2}\sum_{(i,j)} \frac{(-1)^{i+j+1}}{\sqrt{i^2+j^2+9d^2}}$$

If $d$=1, numerical evaluations give:

$$M_{\text{2D−bilayer}} \approx 1.683$$

$$M_{\text{2D−trilayer}} \approx 1.705$$

$$M_{\text{2D−quadruplelayer}} \approx 1.716$$

The Madelung factor jump from 2D monolayer to 2D bilayer by a greater factor than that from bilayer to trilayer (and further to multilayers). It converges slowly toward the 3D bulk value (1.748).

In order to assess the delocalization effect, we calculate-notwithstanding the weakness (but finiteness) of the delocalization perpendicular to the layers when compared to the in-plane situation - the electron-in-the-box energies for box sizes (L, length) given by mono-, bi- and multilayers. Let us define N as charge number, then the free energy change corresponding to layer evolutions,

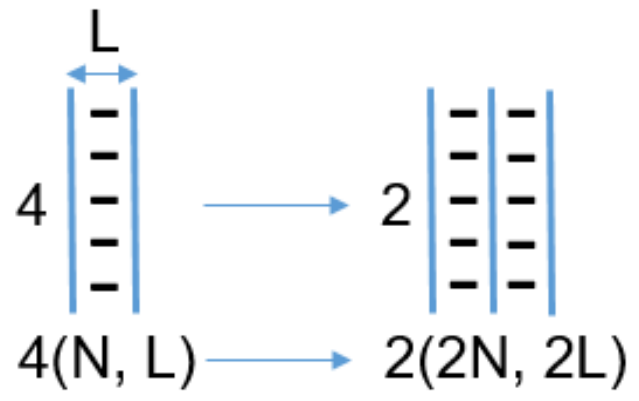

1a

$$\Delta E_{1a} = 2\frac{2N}{4L^2} - 4\frac{N}{L^2} = -3\frac{N}{L^2}$$

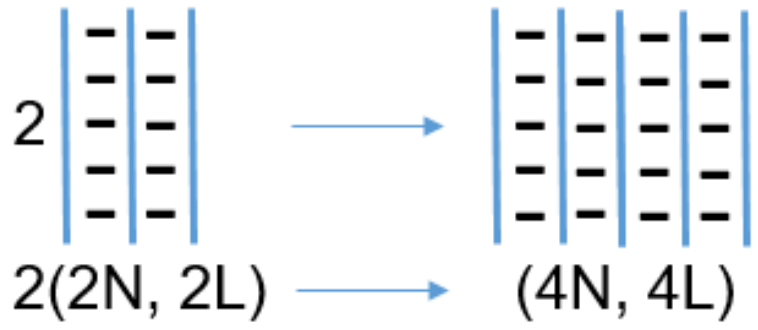

1b

$$\Delta E_{1b} = \frac{4N}{16L^2} - \frac{4N}{4L^2} = -\frac{3}{4}\frac{N}{L^2}$$

One finds

$$|\Delta E_{1b}| < |\Delta E_{1a}|$$

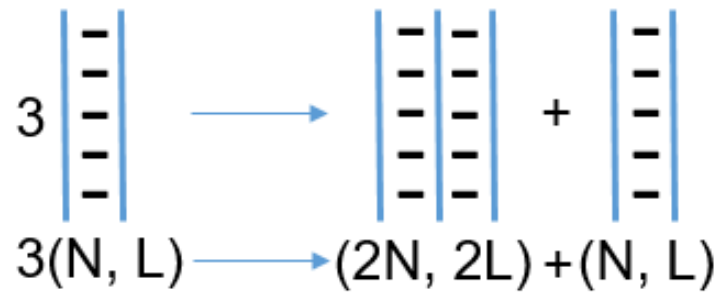

2a

$$\Delta E_{2a} = \frac{2N}{4L^2} + \frac{N}{L^2} - \frac{3N}{L^2} = -\frac{3}{2}\frac{N}{L^2}$$

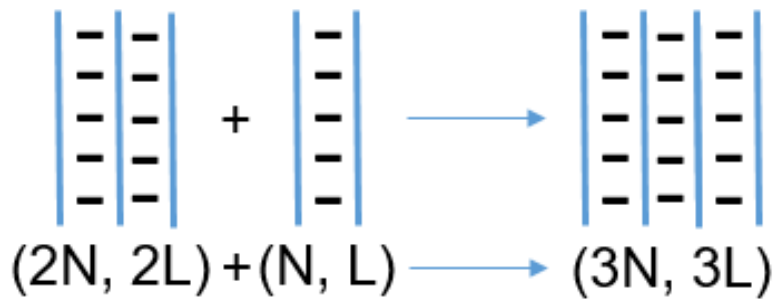

2b

$$\Delta E_{2b} = \frac{3N}{9L^2} - \frac{2N}{4L^2} - \frac{N}{L^2} = -\frac{7}{6}\frac{N}{L^2}$$

Then we have

$$|\Delta E_{2b}| < |\Delta E_{2a}|$$

Considering both the Madelung effects and the delocalization effects together, we can recognize that for Na the energy gain from monolayer to bilayer is greater than for further aggregation.

This is in striking difference to the other alkali metals. The reason, why bilayers are not favored for the higher alkali metals can be explained by size and thus by the larger strain, Why bilayers are also not found for Li, may be attributed to the higher covalency [EPR tests[23], DFT calculation[24]]. The fact that conductivities are not very different for Li-inserted materials does note prove the opposite, as they are not DC values and thus do not represent carrier densities.)

**7) Relation between intercalation free energy ($\Delta G_{\mathrm{ins}}$) and the free enthalpy measured by OCV ($\Delta G_{\mathrm{OCV}}$)**

$$\Delta G_{\mathrm{ins}} = \Delta G\{x\mathrm{Na} + \mathrm{C} \rightleftharpoons \mathrm{Na}_x\mathrm{C}\} = \mu_{\mathrm{Na}_x\mathrm{C}} - x\mu^0_{\mathrm{Na}} - \mu^0_{\mathrm{C}}$$

$$\Delta G_{\mathrm{OCV}} = \mu_{\mathrm{Na\ in\ Na}_x\mathrm{C}} - \mu^0_{\mathrm{Na}}$$

As $\mu_{\mathrm{Na}_x\mathrm{C}} = x\mu_{\mathrm{Na\ in\ Na}_x\mathrm{C}} + \mu_{\mathrm{C\ in\ Na}_x\mathrm{C}}$ it follows that

$$\Delta G_{\mathrm{ins}} = x\Delta G_{\mathrm{OCV}} + \mu_{\mathrm{C\ in\ Na}_x\mathrm{C}} - \mu^0_{\mathrm{C}}$$

where $\mu_{\mathrm{Na}_x\mathrm{C}}$ is chemical potential of Na containing-HOPG, $\mu^0_{\mathrm{Na}}$ standard chemical potential of pure Na metal, $\mu^0_{\mathrm{C}}$ standard chemical potential of pure HOPG, $\mu_{\mathrm{Na\ in\ Na}_x\mathrm{C}}$ chemical potential of Na in Na containing-HOPG, $\mu_{\mathrm{C\ in\ Na}_x\mathrm{C}}$ chemical potential of carbon in Na containing-HOPG.

(1) If Na is equilibrated with sodiated carbon (Na-containing HOPG) at the temperature of measurement:

$$\Delta G_{\mathrm{OCV}} = 0 \Leftrightarrow \Delta G_{\mathrm{int}} = \mu_{\mathrm{C\ in\ Na}_x\mathrm{C}} - \mu^0_{\mathrm{C}}$$

(2) If carbon (HOPG) is equilibrated with sodiated carbon (Na-containing HOPG) at the temperature of measurement:

$$\Delta G_{\mathrm{ins}} = x\Delta G_{\mathrm{OCV}}$$

**Supplementary References**